\documentclass[fleqn,usenatbib,useAMS]{mnras}

\usepackage{graphicx}	
\usepackage{amsmath}	
\usepackage{amssymb}	
\usepackage{multicol}        
\usepackage{bm}		
\usepackage{pdflscape}	
\usepackage{enumitem}	
\usepackage{multirow}   
\usepackage{CJK}
\usepackage{longtable}

\usepackage{etoolbox}
\makeatletter
\patchcmd\@combinedblfloats{\box\@outputbox}{\unvbox\@outputbox}{}{%
   \errmessage{\noexpand\@combinedblfloats could not be patched}%
}%
 \makeatother

\newcommand{\teff}{$T_\mathrm{eff}$}
\newcommand{\logg}{$\log g$}
\newcommand{\Feh}{$\mathrm{[Fe/H]}$}
\newcommand{\XFe}{$\mathrm{[X/Fe]}$}
\newcommand{\vmic}{$v_\mathrm{mic}$}
\newcommand{\vsini}{$v \sin i$}

\newcommand{\Gaia}{\textit{Gaia}}
\newcommand{\TF}{T_{\rm{eff}}}

\usepackage[T1]{fontenc}
\usepackage{ae,aecompl}

\usepackage{newtxtext,newtxmath}

\begin{document}
\begin{CJK*}{UTF8}{gbsn}
\label{firstpage}
\pagerange{\pageref{firstpage}--\pageref{lastpage}}


\title{The GALAH Survey: Second Data Release}

\author[Buder et al.]{Sven Buder$^{1,2}$\thanks{Contact e-mail: \href{mailto:buder@mpia.de}{buder@mpia.de}},
Martin~Asplund$^{3,4}$,
Ly~Duong$^{3}$,
Janez~Kos$^{5}$,
Karin~Lind$^{1,6}$, \newauthor
Melissa~K.~Ness$^{7,8}$, 
Sanjib~Sharma$^{5,4}$,
Joss~Bland-Hawthorn$^{5,4,9}$,
Andrew~R.~Casey$^{10,11}$, \newauthor
Gayandhi~M.~De~Silva$^{12,5}$, 
Valentina~{D'Orazi}$^{13}$,
Ken~C.~Freeman$^{3}$,
Geraint~F.~Lewis$^{5}$,\newauthor
Jane~Lin$^{3,4}$,
Sarah~L.~Martell$^{14}$, 
Katharine.~J.~Schlesinger$^{3}$, 
Jeffrey~D.~Simpson$^{12}$,\newauthor
Daniel~B.~Zucker$^{15}$,
Toma\v{z}~Zwitter$^{16}$, 
Anish~M.~Amarsi$^{1}$,
Borja~Anguiano$^{17}$,\newauthor
Daniela~Carollo$^{18}$, 
Luca~Casagrande$^{3,4}$, 
Klemen~\v{C}otar$^{16}$,
Peter~L.~Cottrell$^{11,19}$,\newauthor
Gary~Da~Costa$^{3}$,
Xudong~D.~Gao$^{1}$,
Michael~R.~Hayden$^{5,4}$, 
Jonathan~Horner$^{20}$,\newauthor
Michael~J.~Ireland$^{3}$,
Prajwal~R.~Kafle$^{21}$,
Ulisse~Munari$^{22}$,
David~M.~Nataf$^{23}$,\newauthor
Thomas~Nordlander$^{3,4}$,
Dennis~Stello$^{14,24,4}$,
Yuan-Sen~Ting (丁源森)$^{25,26,27}$, \newauthor
Gregor~Traven$^{16}$,
Fred~Watson$^{12,28}$,
Robert~A.~Wittenmyer$^{29}$,
Rosemary~F.~G.~Wyse$^{23}$, \newauthor
David~Yong$^{3}$, 
Joel~C.~Zinn$^{30}$,
Maru{\v s}a~{\v Z}erjal$^{3}$,
and~the~GALAH~collaboration
\\
\\
(Affiliations listed after the references)}

\date{Last updated 2018 April 18; in original form 2018 April 18}

\pubyear{2018}

\maketitle
\end{CJK*}

\begin{abstract}
The Galactic Archaeology with HERMES (GALAH) survey is a large-scale stellar spectroscopic survey of the Milky Way and designed to deliver chemical information complementary to a large number of stars covered by the \Gaia\ mission. We present the GALAH second public data release (GALAH DR2) containing 342,682 stars. For these stars, the GALAH collaboration provides stellar parameters and abundances for up to 23 elements to the community.
Here we present the target selection, observation, data reduction and detailed explanation of how the spectra were analysed to estimate stellar parameters and element abundances.
For the stellar analysis, we have used a multi-step approach. We use the physics-driven spectrum synthesis of \textit{Spectroscopy Made Easy} (SME) to derive stellar labels (\teff, \logg, \Feh, \XFe, \vmic, \vsini, $A_{K_S}$) for a representative training set of stars. This information is then propagated to the whole survey with the data-driven method of \textit{The Cannon}.
Special care has been exercised in the spectral synthesis to only consider spectral lines that have reliable atomic input data and are little affected by blending lines. Departures from local thermodynamic equilibrium (LTE) are considered for several key elements, including Li, O, Na, Mg, Al, Si, and Fe, using 1D {\sc marcs} stellar atmosphere models.
Validation tests including repeat observations, \Gaia\ benchmark stars, open and globular clusters, and K2 asteroseismic targets lend confidence to our methods and results.
Combining the GALAH DR2 catalogue with the kinematic information from \Gaia\ will enable a wide range of Galactic Archaeology studies, with unprecedented detail, dimensionality, and scope.
\end{abstract}





\section{Introduction} 

The last decade has seen a revolution in Galactic astronomy. This is particularly evident in the domain of spectroscopic studies where sample sizes have grown from tens or hundreds of stars to several hundred thousands of stars, enabled by the availability of multi-object spectrographs. We now live in an extraordinary era of interest and investment in stellar surveys of the Milky Way. At optical and infrared wavelengths, large-scale photometric surveys have delivered accurate photometric magnitudes and colours for several hundred millions of stars \citep[e.g.,][]{Skrutskie2006, Saito2012, Chambers2016, Wolf2018}. 

Spectroscopic surveys of stars in our Galaxy are fundamental to astrophysics because there are important measurements that can only be made in the near-field and through stellar spectroscopy. This is especially true for the measurement of detailed chemical compositions of stars. The determination of accurate stellar ages continues to be a challenge but there is cautious optimism that the situation will improve in the decades ahead. Asteroseismic surveys from NASA's \textit{Kepler} and K2 missions are providing reliable estimates of stellar surface gravities and masses \citep[e.g.][]{Stello2015, Sharma2016} and this is expected to continue with missions like TESS \cite[][]{Ricker2015,Sharma2018} and PLATO \cite[][]{Rauer2014,Miglio2017}. 
The ESA \textit{Gaia} astrometric mission will measure accurate distances for billions of stars belonging to most components of the Galaxy \citep{Perryman2001, Lindegren2016}.

The Galactic Archaeology with HERMES (GALAH) survey\footnote{\url{https://galah-survey.org/}} brings a unique perspective to the outstanding problem of understanding the Galaxy's history \citep{DeSilva2015, Martell2017}. Our study makes use of the High Efficiency and Resolution Multi-Element Spectrograph (HERMES) at the Anglo-Australian Telescope \citep{Barden2010, Sheinis2015}. This unique instrument employs the Two Degree Field (2dF) fibre positioner on the Anglo-Australian Telescope to provide multi-object ($n \sim 392$), high-resolution ($R \sim 28,000$) spectra optimized for elemental abundance studies for up to 30 elements in four optical windows. The HERMES instrument was built specifically for the GALAH survey and is largely achieving its original design goals, as we demonstrate here in our major data release (DR2).

The overarching goal of the GALAH survey is to acquire high-resolution spectra of a million stars for chemical tagging, in order to investigate the assembly history of the Galaxy \citep{FreemanBlandHawthorn2002, DeSilva2009, BlandHawthorn2010a}. The GALAH selection criteria are simple, with the baseline selection being a magnitude cut of $12 < V < 14$ with Galactic latidute $\vert b\vert > 10\,\mathrm{deg}$. As such, GALAH probes mainly the thin and thick disc of the Galaxy. However, due to the unprecedented sample size, the survey also includes a substantial number of halo stars, as well as other stellar populations serendipitously along the GALAH stars the line of sight, such as stars in the Magellanic clouds. In addition, complementary programs with the same instrument targeting the Milky Way bulge (Duong et al., in preparation) and open clusters (De Silva et al., in preparation) are underway. For the main GALAH survey, roughly two-thirds of the sample are dwarf stars, whilst the rest are predominantly red giant branch stars located at distances up to several kpc from the solar neighbourhood. 

Whilst the volume and wealth of detail contained within the GALAH dataset present a broad range of science opportunities in Galactic and stellar astrophysics, such work is beyond the scope of this publication. The key science questions that motivated the GALAH survey are as follows:
\begin{itemize}[leftmargin=0.5cm]
\item What were the conditions of star formation during early stages of Galaxy assembly?
\item When and where were the major episodes of star formation in the disc, and what drove them?
\item To what extent are the Galactic thin and thick discs composed of stars from merger events?
\item Under what conditions and in what types of systems did accreted stars form?
\item How have the stars that formed in-situ in the disc evolved dynamically since their formation?
\item Where are the solar siblings that formed together with our Sun?
\end{itemize}

The above science questions can be presented as specific delivery goals for the survey. It is these delivery goals that are the highlight of this Data Release. The goals can be summarised as follows:
\begin{itemize}[leftmargin=0.5cm]
\item To determine the primary stellar parameters: effective temperature, surface gravity, metallicity
\item To derive up to 30 individual chemical element abundances per star from Li to Eu
\item To measure radial velocities
\item To classify targets of an "unusual" nature: e.g., binaries, stellar activity, chemical peculiarities
\end{itemize}

Achieving the above delivery goals for millions of stars in an unrestricted parameter space is a major challenge; doing so consistently and in a timely manner takes us into new territory for spectral analysis methods. Teams have been required to revisit their classical approach and enter into other disciplines of data analysis to develop machinery suitable for meeting these new challenges. In recent years, we have learned to bring together many different strands to make progress \citep{Ho2017}. This has required major advances in a range of areas, including template matching \citep{Jofre2010}, automated machine learning \citep{Ness2015}, atmospheric modelling \citep{Magic2013}, spectral line formation \citep{Amarsi2016b} and internal instrument calibrations \citep{Bland-Hawthorn2017, Kos2018}. We discuss some of these advances, to the extent that they have benefited the GALAH data reduction and analysis, as part of this paper.

The GALAH survey joins a vibrant and exciting landscape of numerous Galactic spectroscopic surveys that are highly complementary in scale and scope. The \textit{Gaia}-ESO Survey on the VLT \citep{Gilmore2012, Randich2013} contains a sample of about 100,000 stars over 14 -- 19 in $V$ band, most of which are thick disc and halo stars, as well as numerous open clusters. The observations used specific spectral windows with the GIRAFFE+FLAMES fibre spectrograph system, and the typical spectral resolving power is $R\approx 20,000$, somewhat lower than that of GALAH. However, a subsample of targets observed with the UVES spectrograph have high-resolution ($R \approx 47,000$) spectra. The infrared APOGEE survey \citep{SDSSDR14} mainly targets the low-latitude Galactic disc, probing through the optically thick dust regions with a sample of 150,000 red giants at a resolving power of $R\approx 22,500$. To expand their survey, the APOGEE team have recently begun the APOGEE-South survey, using the Dupont telescope at Las Campanas Observatory, in order to study a similar number of stars in the southern sky. While the overlap in targets between the three surveys is currently only a few hundred stars, the scientific complementarity is significant. Given the different magnitude ranges and regions of the Galaxy observed, both \textit{Gaia}-ESO and APOGEE are highly complementary to the GALAH sample.
Several even larger high-resolution spectroscopic surveys of the Milky Way are expected to commence over the next decade, including WEAVE \citep{Dalton2014}, 4MOST \citep{DeJong2014},
and SDSS5 \citep{Kollmeier2017}.
 
LAMOST is a lower spectral resolution ($R \sim 1800$) survey but has observed a far greater number of stars than other surveys, with over 1.5 million stellar spectra collected to date \citep{Luo2015}. The RAVE survey \citep{Kunder2017} contains a similar number of stars as the current GALAH sample but at a lower resolving power of $R = 7,500$ and with the wavelength coverage limited to the infrared calcium triplet region. The RAVE sample, spanning magnitudes from 9 - 12 in the I band, was selected from numerous sources (Tycho-2, super-COSMOS, DENIS and 2MASS) and included a colour selection of $J-K > 0.5\,\mathrm{mag}$ for stars with $\vert b \vert < 25\degr$ to preferentially select giants \citep[see][]{Wojno2017}. The simple selection function of the GALAH sample, on the other hand, is dominated by local disc dwarfs, making it more sensitive to local substructure. As part of the second \Gaia\ data release in April 2018, radial velocities for some seven million stars brighter than $G \approx 13$ have been measured based on $R = 11,500$ spectra around the Ca infrared triplet (845-872\,nm) that also provide information on stellar parameters and some limited elemental abundances. 

In this paper we present the second major public data release of the GALAH survey including stellar parameters and individual abundances of 23 elements from Li to Eu for 342,682 stars. In Section~\ref{sec:selection_observation_reduction} we provide an overview of the target selection, observation and reduction procedures. In Section~\ref{sec:analysis} we present the details of the spectroscopic analysis, followed by a description of the validation in Section~\ref{sec:validation} and contents of the data release in Section~\ref{sec:catalog}. We highlight studies of the GALAH team  accompanying this data release in Section~\ref{sec:papers} before we conclude in in Section~\ref{sec:conclusions}.

\begin{figure*}
  \includegraphics[width=\textwidth]{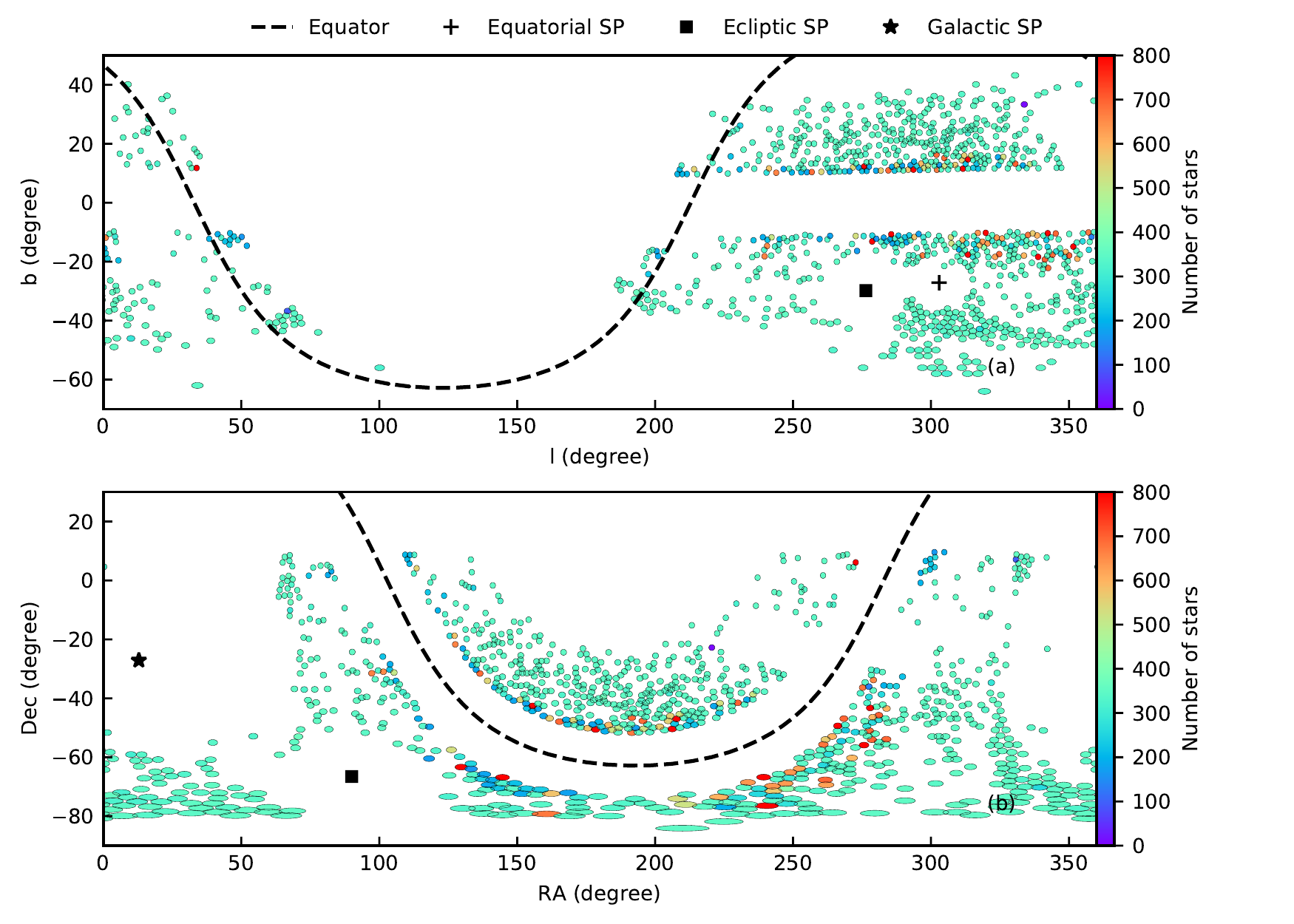}
  \caption{Distribution of observed fields in Galactic and  Equatorial coordinates. The fields are color coded by the number of observed stars. The bold dashed line corresponds to equatorial (upper panel) and  Galactic (lower panel) equators. The equatorial, the ecliptic, and the Galactic south poles are also marked on the panels.} 
  \label{fig:galah_fov}
\end{figure*}

\section{Target selection, observation, reduction} \label{sec:selection_observation_reduction}

\subsection{Target selection}

\begin{figure}
  \includegraphics[width=\columnwidth]{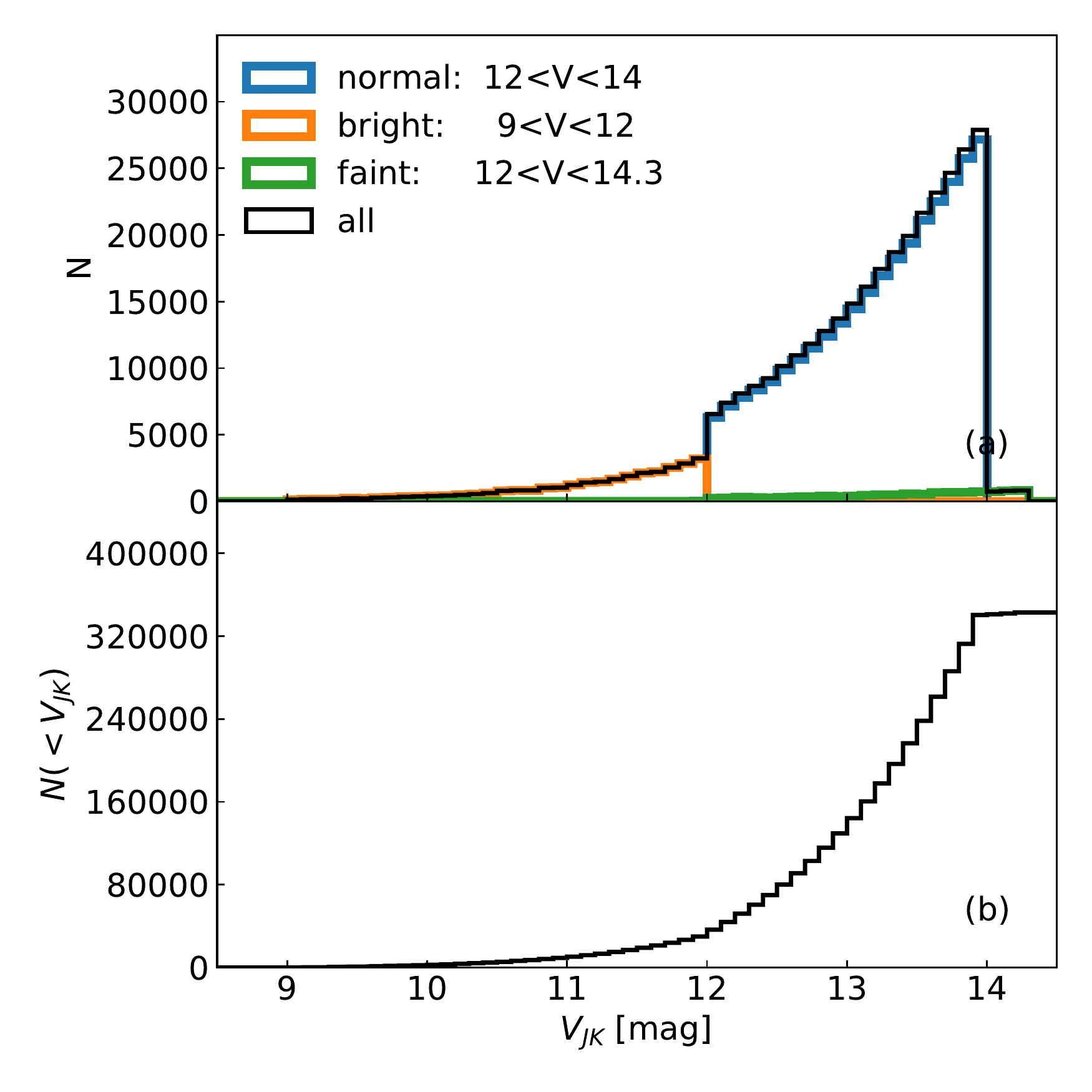}
  \caption{Distribution of V band magnitude of GALAH stars. The distributions  for normal, bright and faint fields are shown separately.} 
  \label{fig:galah_vmag_dist}
\end{figure}

\begin{figure}
  \includegraphics[width=\columnwidth]{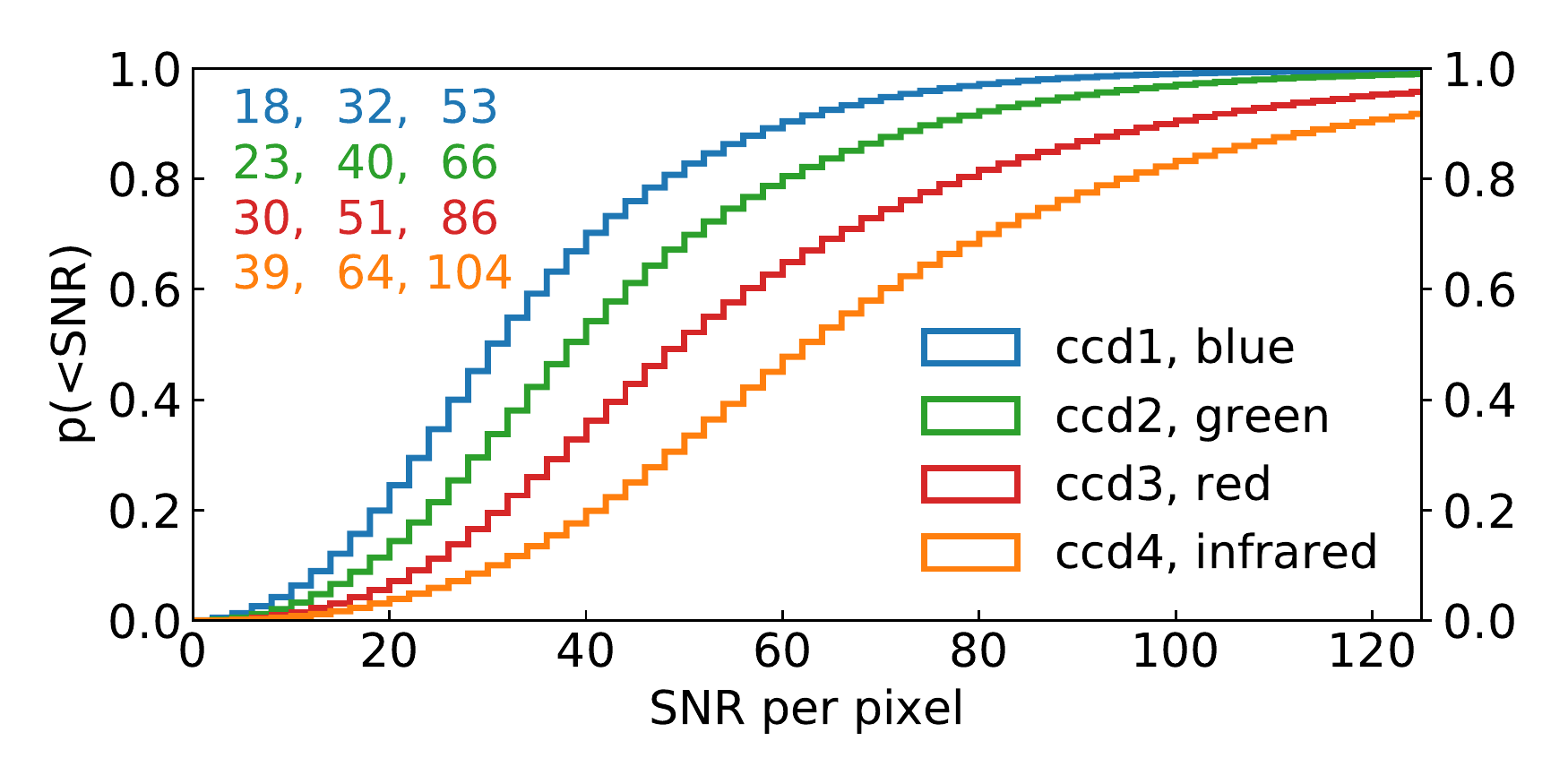}
  \caption{Distribution of SNR per pixel for the different HERMES wavelength bands. The numbers in the upper left corner denote the 16th, 50th and 84th percentile values.} 
  \label{fig:galah_snr_dist}
\end{figure}

In order to carry out a large-scale survey with broad applications across astrophysics and Galactic archaeology, the GALAH survey uses a very simple observational selection function. This makes it relatively straightforward to transform between the observed data set and the underlying Galactic populations. It also makes a clear imprint on which types of stars across which regions of the Milky Way are sampled by GALAH because of their luminosities.

The GALAH input catalogue is constructed from the union of 2MASS \citep{Skrutskie2006} and UCAC4 \citep{Zacharias2013} catalogues. Only stars with appropriate 2MASS quality flags (Q=``A'', B=``1'', C=``0'', X=``0'', A=``0'', prox$\ge 6\arcsec$) and no brighter neighbours within a radius of $V_{\rm neighbour}=(130-[10\times V_\mathrm{neighbour}])$ arcseconds were chosen. Because APASS was not complete in the Southern sky at the start of observations, we use a synthetic $V_\mathrm{JK}$ magnitude calculated from 2MASS photometry: $V_\mathrm{JK} = K+2(J-K+0.14)+0.382e^{((J-K-0.2)/0.5)}$. Using PARSEC iscochrones \citep{Marigo2017} it was  shown in \citet{Sharma2018} that this transformation is reasonably accurate across the parameter space where the majority of GALAH stars fall. 
The GALAH observations are done in following three fixed magnitude ranges of $V_{JK}$.  
\begin{itemize}
\item Normal mode $12<V_{JK}<14$: Most observations are done with this selection function.
\item Bright mode $9<V_{JK}<12$: 
This is used during twilight or  poor observing conditions.
\item Faint mode $12<V_{JK}<14.3$: This is used when fields with normal of bright configuration are not available to be observed.
\end{itemize}

As a result of this simple selection function, all stars with $-80\degr \leq \delta \leq +10\degr$ and $\mid b \mid > 10\degr$ within these magnitude limits are potentially targets for our survey. We imposed an additional requirement, namely that we only sample fields with an on-sky density of stars of at least 400 per $\pi$ square degrees, to match the number of fibres and field of view of the fibre positioner that feeds our spectrograph. The potential targets are then tiled into 6546 normal fields. The radius of each field varies inversely with the target density, to improve efficiency in dense regions. A normal field can be observed in either normal or bright mode but not faint mode. In bright configuration, the radius is set to 1 degree. In addition to normal fields, 285 faint fields were added to be observed exclusively in faint mode, and they do not overlap with the normal fields.

\subsection{Observations}
Data for the GALAH survey are taken with the HERMES spectrograph on the 3.9-metre Anglo-Australian Telescope (AAT) at Siding Spring Observatory. HERMES is a fibre-fed high-resolution ($R = 28,000$) spectrograph optimised to do Galactic archaeology from a 4m-class telescope, with four discrete optical wavelength channels covering 4713--4903\hbox{\AA}, 5648--5873\hbox{\AA}, 6478--6737\hbox{\AA} and 7585--7887\hbox{\AA} \citep{DeSilva2015,Sheinis2015}.

The 2dF top end for the AAT provides the fibre feed for HERMES \citep{Lewis2002}; its name derives from the two-degree diameter field of view. It has two metal field plates, each with 392 science fibres and 8 guide bundles. The fibres are affixed to magnetic "buttons" and placed on the field plate by a robotic fibre positioning system. While one field plate is being observed, the other is in position for setup by the positioning robot. The 2dF fibres are fed through the telescope structure to the HERMES enclosure on the fourth floor of the AAT dome, where they are arranged into two pseudoslits, such that the set of fibres corresponding to the field plate that is currently being observed is fed into the spectrograph. Light from the fibres is collimated and then sent through a series of dichroic elements to separate the four wavelength channels. It is then dispersed using volume phase holographic diffraction gratings and imaged by four independently controlled cameras. Further details of instrument design, and its as-built performance, can be found in \citet{Barden2010}, \citet{Brzeski2011}, \citet{Heijmans2012}, \citet{Farrell2014} and \citet{Sheinis2015}.

The GALAH survey observations are carried out by team members, either at the AAT or from remote observing sites in Australia. The \textsc{ObsManager} software developed by the team is used to select a field that will be near the meridian at the time of observation. This field is required to be at least 30 degrees from the Moon, to have a relatively small change in airmass over the exposure time, and to have no Solar System planets within it at the time of observation \citep[as bright sources such as Jupiter have caused trouble with scattered light contaminating spectra in previous 2dF survey observations, though HERMES has been found to suffer much less from this than AAOmega; see][]{Simpson2017}. Fiducial targets, used for field acquisition and guiding, are chosen from the stars with $11 \leq V \leq 12$ in the input catalogue that are located in the field. For observations in bright mode fiducial are selected form stars with $12 \leq V \leq 13$. Once the observer chooses a field from the options provided by \textsc{ObsManager}, it generates a list of targets in the correct input format for the \textsc{Configure} software \citep{Miszalski2006}, and it tracks which fields have been observed. 

Figure \ref{fig:galah_fov} shows maps of GALAH DR2 stars (observed between 2014 January 16 and 2018 January 29) in equatorial and Galactic coordinates. The target selection can clearly be seen in the avoidance of the Galactic plane and of fields at high Galactic latitude where the target density is too low. Figure \ref{fig:galah_vmag_dist} shows the distribution of $V_\mathrm{JK}$ in DR2. The upper panel shows the overall distribution and subdivides the data into bright, regular and faint survey fields, and the lower panel is a cumulative histogram.

\textsc{Configure} uses a simulated annealing algorithm to identify a set of target allocations for the 2dF fibres that maximises the number of science targets observed and the number of fiducial targets used for field alignment and guiding. It then places a user-defined number of fibres on sky locations, follows user-defined target priorities and obeys restrictions on fibre placement (as an example, because of the size of the 2dF buttons, the minimum fibre spacing is 30~arcsec, although the fibres themselves have a field of view of only 2~arcsec). The output of \textsc{Configure} is then passed to the 2dF fibre positioning robot, which places the fibres serially onto the currently unused field plate, checking that each is within an acceptable tolerance of its intended position. Field setup typically takes 40 minutes for a full 400-target field, and essentially no survey observing time is lost waiting for reconfiguration.
    
As discussed in \citet{Martell2017}, the exposure time is set to achieve a signal to noise ratio (SNR) of 100 per resolution element, at an apparent magnitude of 14 in the Johnson/Cousins filter closest to each bandpass. For regular survey fields, the standard procedure is to take three 1200-second exposures. If the seeing is between 2" and 2"5, this is extended to four exposures, and to six exposures, if the seeing is between 2"5 and 3". For bright fields, three $360\,\mathrm{s}$ exposures are taken. Figure \ref{fig:galah_snr_dist} shows cumulative distributions for SNR {\em per pixel} in each of the four HERMES channels. With about four pixels per resolution element, the SNR per resolution element is almost twice what is shown in Figure~\ref{fig:galah_snr_dist}.

Fibre flat fields and ThXe arc exposures, both with exposure times of 180s, are taken along with each science field. Including the readout time of $71\,\mathrm{s}$ and a slew and acquisition time of 2--$5\,\mathrm{min}$ per field, the time spent on overheads is roughly $20\%$ of the time spent acquiring science data. The typical data rate in survey fields is 4.2 stars per on-sky minute. The GALAH survey has typically been awarded 35 nights per semester since 2014 February, secured across a number of competitive time-allocation rounds. 
    
\subsection{Data reduction}\label{sec:reduction}

We use a reduction pipeline designed specifically for the GALAH survey, where steps are tailored to the observing strategy and scientific requirements of the survey. The reduction pipeline is described in detail in a separate paper \citep{Kos2017} with only a short overview given here.
    
Raw images are corrected for the bias level using a series of bias frames taken every night. One flat field exposure is taken for each science field, which is used to identify damaged columns and pixels, serves as a reference to find spectral traces on the science images, and supports the measurement of scattered light and fibre cross-talk. 
    
After cosmetic corrections, the cosmic rays are removed utilizing a modified LaCosmic algorithm \citep{Dokkum2001}. Before the extraction of the spectra, a geometric transformation is used to correct some basic optical aberrations. This reduces the variation of resolving power between different fibres and different wavelength regions. 

To ensure the most accurate analysis of our data, we have refined the previously measured literature wavelengths for the ThXe arc lamp. The wavelength fitting algorithm has been improved since the publication of \citet{Kos2017}, so it now learns the trends and instabilities in the wavelength calibration to predict a better solution when arc lamp spectra are hard to identify. We have to rely on weak lines in the ThXe arc lamp spectrum to calibrate the wavelengths. These are hard to measure in low throughput fibres, so the wavelength calibration for some fibres and some wavelength regions wassuboptimal in previous reductions.
    
The sky spectrum is modelled from $\sim25$ dedicated sky fibres over the whole $2^\circ$ focal plane. After the sky spectrum is subtracted, a telluric absorption spectrum is calculated using \textsc{Molecfit} software \citep{Kausch2015, Smette2015} and the science spectra are corrected.
    
The products of the reduction pipeline are unnormalized and normalized spectra, together with an uncertainty spectrum and a resolution map.
    
At the end of the reduction pipeline we calculate basic stellar parameters for all reduced objects. Radial velocities \texttt{rv\_synt} and their errors \texttt{e\_rv\_synt} are computed by cross-correlating reduced spectra with 15 synthetic AMBRE spectra \citep{Laverny2012}. Radial velocities are measured independently in the blue, green and red channels, which are weighted to yield one radial velocity with its uncertainty. Such radial velocities are precise enough for most of the following analysis. First they are used for a continuum normalization calculated from regions minimally affected by spectral lines. Normalized spectra are then matched with 16783 synthetic AMBRE spectra to estimate initial values of effective temperature $T_\text{eff}$, surface gravity $\log g$, and metallicity $\mathrm{[M/H]}$ for the subsequent detailed spectral analysis.

\subsection{Alternate radial velocities}

The high SNR and resolution of HERMES spectra with a wide wavelength coverage and careful data reduction make GALAH an excellent source of accurate radial velocities in the $12<V<14$ magnitude range to complement, e.g., \Gaia\ distances and proper motions. To achieve this goal we measure radial velocities in a custom three-stage process (Zwitter et al., in preparation). First, we build a library of median observed spectra with very similar stellar parameter values. In particular, we combine spectra within bins of 50~K in $T_\text{eff}$, 0.2~dex in $\log g$ and 0.1~dex in [Fe/H]. Initial values of radial velocities, as measured by the data reduction pipeline, allow us to perform this task with virtually no velocity smearing. These median spectra have a very high SNR and are used in the second stage to measure the radial velocity of an observed spectrum vs.\ its corresponding median spectrum very accurately. At this stage, the median spectrum is not guaranteed to be at zero velocity due to the combined effects of convective line shifts and gravitational redshifts \citep{Asplund2000,AllendePrieto2013}. We therefore use the grid of synthetic spectra \citep{Chiavassa2018} computed with 3D hydrodynamical stellar atmosphere models with the {\sc stagger} code \citep{Magic2013}, which include the effects of convective motions in stellar atmospheres. Finally, we incorporate the effects of gravitational redshift, to put spectra of different stellar types on the same scale and thus allows us to speak about the accuracy and not just the precision of radial velocities. This can be crucial when studying the internal dynamics of stellar clusters, associations, or streams, which generally contain different types of stars and in which relative velocities do not exceed a few km\,s$^{-1}$. 

The propagation of internal velocity errors as well as the comparison of velocities measured for multiple observations of the same object show that a typical error of derived radial velocities is close to 0.1 km\,s$^{-1}$. It can be worse for turn-off stars where accurate radii and masses of stars are unknown, so that reliable values of gravitational redshift can not be determined. Our velocity results are reported in three columns: \texttt{\texttt{rv\_obst}} is the final value of the radial velocity, including the gravitational redshift, \texttt{\texttt{e\_rv\_obst}} is its formal error, and \texttt{\texttt{rv\_nogr\_obst}} is the value of the radial velocity without gravitational redshift. The latter may be useful for two purposes: (i) to compare our results to those of other surveys, which generally do not take gravitational redshift into account, and (ii) to allow the user to use a different, more accurate value of gravitational redshift. In particular, astrometric results from \textit{Gaia} DR2 will allow an accurate estimate of the absolute magnitude of the source which, when combined with the value of \teff\ we report here, will allow the accurate calculation of the stellar radius. However this will require an estimate of extinction; we discuss prospects for measuring extinction below. Once the radius is estimated, isochrones and metallicities can be used to estimate the object's mass, and hence to determine the value of gravitational redshift. Details of the above procedure and comparison to results of other radial velocity surveys are discussed in more detail elsewhere (Zwitter et al., in preparation). 

\section{Analysis} \label{sec:analysis}

In this section, we describe the multi-step approach we use to estimate stellar parameters and element abundances from GALAH spectra.

\subsection{Analysis strategy} \label{sec:analysis_strategy}

Because of the large volume of data, we are using a new data analysis approach, which has proven successful when dealing with very large datasets: train a data-driven approach on physics-driven input to connect data (spectra) with labels (in our case stellar parameters and element abundances) and then propagate this information onto the whole sample.
We first create such a training set of 10605 stars and estimate stellar labels through detailed spectrum synthesis using the code Spectroscopy Made Easy (Section~\ref{sec:analysis_step1}), investing much effort ensuring that the inferred stellar parameters and abundances are trustworthy (e.g. line selection, atomic/molecular data, blends, non-LTE effects). We then create spectral models with \textit{The Cannon} and use these models to propagate the information from the training set on to the whole survey (Section~\ref{sec:analysis_step2}). This approach makes the implementation of a flagging algorithm vital, because \textit{The Cannon} in its current form will always produce label estimates, which then have to be vetted as we describe in Section~\ref{sec:analysis_step3}.

\subsection{Analysis step 1: The training set analysis with \textit{Spectroscopy Made Easy}} \label{sec:analysis_step1}
For the model-driven analysis, we use the spectral synthesis code SME (\emph{Spectroscopy Made Easy}) v360~\citep{Valenti1996,Piskunov2017}.
SME performs spectrum synthesis for 1D stellar atmosphere models, which in our case for DR2 consist of {\sc marcs} theoretical 1D hydrostatic models \citep{Gustafsson2008}; we use spherical symmetric stellar atmosphere models for $\log g \leq 3.5$ assuming 1$M_\odot$ and plane parallel models otherwise. While the radiative transfer in SME is typically carried out under the assumption of LTE, it is possible to provide departure coefficients for level populations calculated elsewhere; we make use of this feature in the GALAH survey in order to derive accurate stellar parameters and elemental abundances as free of systematic errors as possible. 
For DR2 We incorporate non-LTE line formation for several key elements, including  Li \citep{Lind2009}, O \citep{Amarsi2016b}, Na \citep{Lind2011}, Mg \citep{Osorio2015}, Al \citep{Nordlander2017}, Si \citep{Amarsi2017}, and Fe \citep{Amarsi2016a}, mostly with additional dedicated calculations besides those previously published. 
In all cases the non-LTE computations have been performed using exactly the same grid of 1D {\sc marcs} model atmospheres. 
Future GALAH data releases will have additional elements treated in non-LTE. 

In addition to providing a formal solution of the radiative transfer, SME attempts to find the optimal solution for various free parameters specified by the user; we use this feature to estimate the stellar parameters of the GALAH targets, allowing $T_{\rm eff}$, $\log g$, [Fe/H], [X/H], $V_{\rm broad}$ (spectral line broadening, consisting of the combined effects of macroturbulence and rotation), $v_{\rm rad}$ and continuum normalisation to vary during the optimisation process.  

We have carefully selected the most reliable atomic lines within the HERMES wavelength regions  to ensure accurate determination of the stellar parameter and abundances for the analysis of late-type stars. The line list selection was originally done in conjunction with the corresponding compilation for the  
\textit{Gaia}-ESO survey \citep{Heiter2015b}. Our guiding principle has been to include only spectral lines that both have reliable atomic data and be as little affected by blending lines as possible in order to enable also absolute abundances to be inferred. Naturally this dramatically limits the number of spectral features to be used in late-type stellar spectra, since the majority of lines are either blended to various extent and/or lack good atomic data, especially transition probabilities. The selection of lines to employ was initially based on a detailed comparison of the predicted spectrum against observations for the Sun and Arcturus but also tested for other benchmark stars. Whenever possible, experimental oscillator strengths are used if thrustworthy, but for some lines used as elemental abundance diagnostics we have had to resort to more or less uncertain theoretical transitional probabilities in the absence of better alternatives. In addition to the primary line list for abundance purposes we have included background blending lines, which have largely been taken from the \textit{Gaia}-ESO linelist; in several cases we have updated the oscillator strengths $\left(\log gf \right)$ empirically compared to those in the \textit{Gaia}-ESO master linelist in order to provide better agreements between observed HERMES spectra and the predicted stellar spectrum for stars. 
The list of the primary spectral lines used for the determination of stellar parameters and abundances is given in Table \ref{tab:linelist}.

To determine stellar parameters, we make use of the \teff -sensitive H$\alpha$ and H$\beta$ lines \citep[e.g.][]{Amarsi2018} and neutral/ionized lines of Sc, Ti, and Fe; the latter elements provide constraints on the effective temperature and surface gravity \logg through excitation and ionisation balance as well as metallicity. SME first synthesises the initial model based on radial velocities from the reduction process and a set of initial stellar parameters. If available and not flagged, stellar parameters from a prior version of \textit{The Cannon} (version 1.3) are chosen. This version was also used for previous data releases of GALAH \citep{Martell2017}, TESS-HERMES \citep{Sharma2018}, and K2-HERMES \citep{Wittenmyer2018}. If these parameters are flagged or not available, the synthesis commences with the stellar parameter estimates from the reduction process. If these are also flagged, we start from an arbitrary set of stellar parameters ($T_\text{eff} = 5000\,\mathrm{K}$, $\log g = 3.5\,\mathrm{dex}$, and $\mathrm{[M/H]} = -0.5\,\mathrm{dex}$). 

We then use two main iteration cycles in SME to optimise the parameters, unless we have to use the arbitrary set of stellar parameters, in which case the second iteration cycle is repeated. In the first cycle, each wavelength segment of typically 10\,\AA\ is normalized using a linear function, which is adequate for the short wavelength intervals used here. SME then computes synthetic spectra in 46 selected (masked) regions. The free parameters $T_\textrm{eff}$, $\log g$, [M/H]\footnote{SME returns the iron abundance of the model atmosphere during the parameter determination stage, which is called metallicity, or [M/H].}, $\xi_t$ (micro turbulence), $v\sin i$ (rotational velocity) are simultaneously determined. We also solve for radial velocity ($v_\mathrm{rad}$) at each iteration to correct for local variations due to uncertainties in wavelength calibration. SME uses the Levenberg-Marquardt algorithm to find parameters that correspond to the near-minimum~$\chi^2$.

The final parameters from the first cycle are used to synthesise the initial model in the second cycle and re-normalize each segment. SME goes through the same iteration process, optimising $\chi^2$ until convergence is achieved (when $\chi^2$ changes by less than $0.1\%$). The number of iterations necessary to reach convergence varies from star to star. Typically, more metal-rich and cooler stars take longer to converge, but normally still do so in less than 20 iterations. During the parameter determination, we implement non-LTE corrections from~\citet{Amarsi2016a} for Fe lines. 

While the nominal resolving power of HERMES is $R \approx$ 28000, it is known to vary from fibre to fibre, and as a function of wavelength~\citep{Kos2017}. This issue is resolved by interpolating the observed spectrum with pre-computed resolution maps from~\citet{Kos2017} to estimate a median resolution for each segment. The GALAH survey is currently implementing a photonic comb, which will map the aberrations and point-spread-function across the full CCD images \citep{Bland-Hawthorn2017}.

\subsubsection{Constraints on spectral line broadening}

The resolving power and SNR of GALAH spectra are not adequate to separate the projected rotational velocity ($v \sin i$) and macroturbulence ($v_{\mathrm{mac}}$) due to degeneracies in their line broadening. When both $v \sin i$ and $v_{\mathrm{mac}}$ are allowed as free parameters the results show greater scatter and poorer convergence performance. Therefore, we solve for $v\sin i$ and set all $v_{\mathrm{mac}}$ values to zero. This effectively incorporates $v_{\mathrm{mac}}$ into our $v\sin i$ estimates, hereafter used as $v_\mathrm{broad}$. 

Similarly, setting micro-turbulence ($\xi_t$) as a free parameter causes additional scatter in the results. While micro-turbulence is updated in every iteration, it is dictated by the empirical formulae that have been calibrated for the LUMBA node of the \textit{Gaia}-ESO survey~\citep{Smiljanic2014}, which uses a similar SME-based analysis pipeline as ours.	

\noindent For cool main sequence stars ($T_\textrm{eff} \leq 5500$K; $\log g \geq 4.2$):
\begin{equation}
\xi_t = 1.1 + 1.0 \times 10^{-4} \times (T_\textrm{eff}-5500)+ 4 \times 10^{-7} \times (T_\textrm{eff}-5500)^2
\end{equation}
\noindent For evolved and hotter stars ($T_\textrm{eff} \geq 5500$K; $\log g \leq 4.2$):
\begin{equation}
\xi_t = 1.1 + 1.6 \times 10^{-4} \times (T_\textrm{eff}-5500),
\end{equation}
where $T_\textrm{eff}$ in K.

Since macroturbulence (and microturbulence) is reflecting convective motions and oscillations 
in the stellar atmosphere \citep{Asplund2000} those velocities are typically limited to $<10$ km/s for late-type stars \citep{Gray2008}. For greater $v_{\rm broad}$, the broadening is dominated by rotation, which is normally the case for $T_{\rm eff} \ga 7000$\,K.

\subsubsection{Constraints on surface gravity}
\label{extlogg}

There are few unblended ionized lines of suitable strength in HERMES spectra, making a fully spectroscopic surface gravity determination a challenge; we note that the HERMES spectrograph and GALAH survey were designed with the expectation that parallaxes from \Gaia\ would provide superior surface gravities in general. For dwarf stars cooler than about 4500\,K the purely spectroscopic surface gravities are underestimated, causing an `up-turn' in the lower stellar main sequence.
This is a common shortcoming in stellar spectroscopic studies relying on ionisation balance
in the framework of LTE spectral line formation in 1D stellar atmosphere models \citep[e.g.][]{Yong2004, Bensby2014, Aleo2017}. The cause for this breakdown of 1D LTE ionisation balance has not yet been identified. 

To help improve the accuracy of the $\log g$ determination, the GALAH survey observed fields that are in the \textsc{Hipparcos} \citep{Perryman1997,vanLeeuwen2007} and \textit{Tycho}-\textit{Gaia} Astrometric Solution (TGAS) catalogues and within the K2 footprint, providing spectra with parallax and asteroseismic information~\citep{Perryman1997,Brown2016,Stello2017}. These are used to determine $\log g$ during the parameter optimisation process.

For stars with asteroseismic information, surface gravity is not strictly a free parameter, but is determined at each SME iteration with respect to solar values using the scaling relation~\citep{Kjeldsen1995}:
\begin{equation}
\nu_\mathrm{max} = \nu_\mathrm{max,\odot} \frac{g/g_\odot}{\sqrt{T_\mathrm{eff}/T_\mathrm{eff,\odot}}}
\label{eq.numax}
\end{equation}
Here $\nu_\mathrm{max}$ is the measured frequency at maximum power. 

For stars with reliable parallax information, the surface gravity is updated at each SME iteration using the fundamental relation~\citep{Nissen1997,Zhang2005}:
\begin{equation}
\log \frac{g}{g_\odot} = \log \frac{\mathcal{M}}{\mathcal{M_\odot}} - 4 \log \frac{T_\mathrm{eff}}{T_\mathrm{eff,\odot}} + 0.4\left(M_\mathrm{bol}-M_\mathrm{bol,\odot}\right)
\end{equation}

\noindent where 

\begin{equation}
M_\mathrm{bol} = K_S+ BC_{K_S} - 5 \log D_\varpi + 5.
\end{equation}

\noindent 
Here the mass $\mathcal{M}$ of each star is estimated with the age estimation code \textsc{Elli} \citep{Lin2018} using 2MASS photometry and parallaxes from Hipparcos or \Gaia . For the absolute bolometric magnitude $M_\mathrm{bol}$, bolometric corrections $\left(BC\right)$ from \citet{Casagrande2014} are applied to the 2MASS $K$ magnitude. For all stars from the \textsc{Hipparcos} catalogue, the distance $D_\varpi$ is computed by the transformation $ D = {1}/{\varpi} $ with $\varpi$ being the parallax. For all stars from the TGAS catalogue, Bayesian distances from \citet{Astraatmadja2016} are used. For more details on the use of astrometry for the GALAH+TGAS overlap, we refer the reader to \citet{Buder2018a}.

\subsubsection{Elemental abundances}\label{sec:sme_abundances}

\begin{figure}
	\includegraphics[width=\columnwidth]{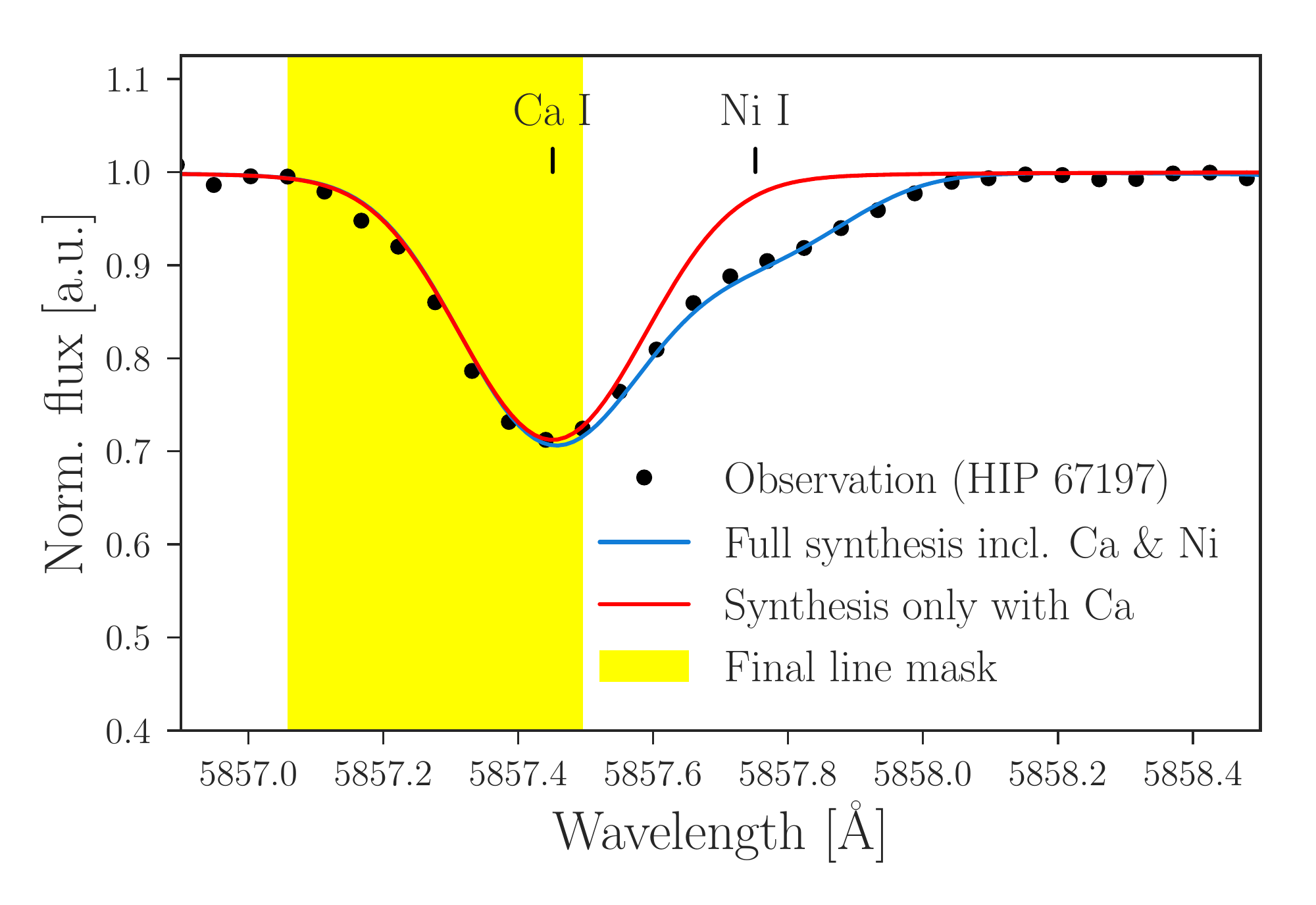}
	\caption{Example for the blending test as part of the abundance estimation for the Ca~I at 5857~\AA\ line in HIP 67197. The black dots show the observed spectrum with a Ca~I line, partially blended with a Ni~I line in the observed spectrum (black dots). Syntheses with all lines (blue) and only calcium (red) show that the red wing of the Ca~I line for this particular star is strongly blended. Hence the line mask (yellow) is adjusted to only include this region for the calcium abundance determination.}
	\label{fig:blendtest}
\end{figure}

After the atmospheric parameters have been established, we apply corrections of the biases estimated in Section~\ref{sec:GBS} (shift of $+0.15\,\mathrm{dex}$ for purely spectroscopic $\log g$ and $+0.1\,\mathrm{dex}$ for all metallicities) and then fix them for abundance determination. The lines of each element are synthesised, and line blending is modelled using the atomic and molecular information provided. The blended wavelength points are excluded from the line mask (see Figure~\ref{fig:blendtest}). 

\begin{figure}\includegraphics[width=\columnwidth]{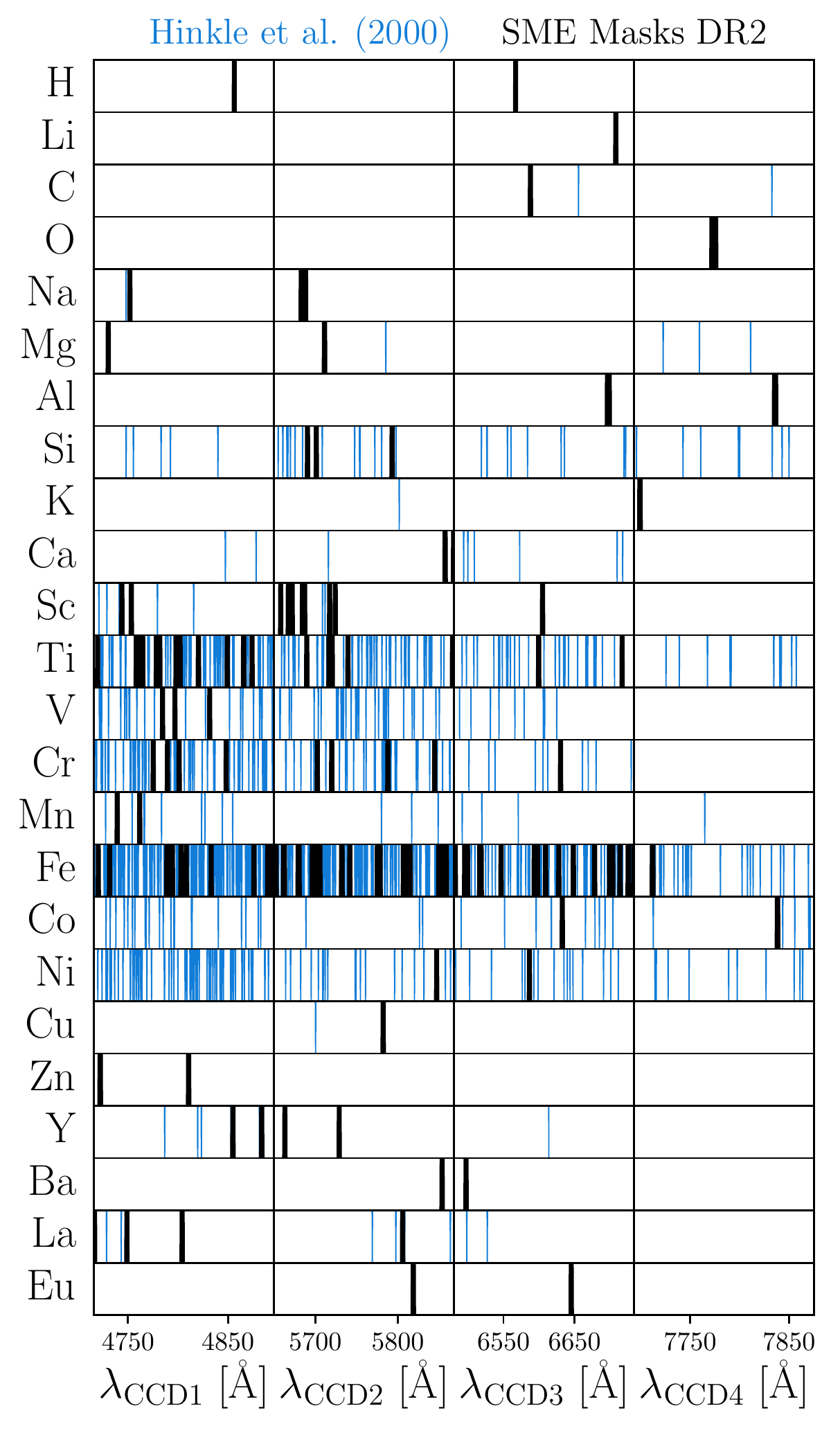}
	\caption{Visualisation of the element lines within the GALAH wavelength range. Blue regions indicate the lines identified by \citet{Hinkle2000} in the Sun and Arcturus, including weak and blended lines. The subset of these regions used for the SME and \textit{The Cannon} analysis are indicated as black regions.}
	\label{fig:line_masks}
\end{figure}

The element lines used for this data release were initially selected from the lines identified by \citet{Hinkle2000} within spectra of the Sun and Arcturus, but carefully vetted in order to be strong enough across the parameter range, have line data based on laboratory measurements, and blend-free, if possible. Therefore, we only use a subset (marked in black in the line region overview of Figure~\ref{fig:line_masks}) of the lines from \citet{Hinkle2000}, indicated as blue regions, to measure element abundances.

\begin{figure*}
	\includegraphics[width=\textwidth]{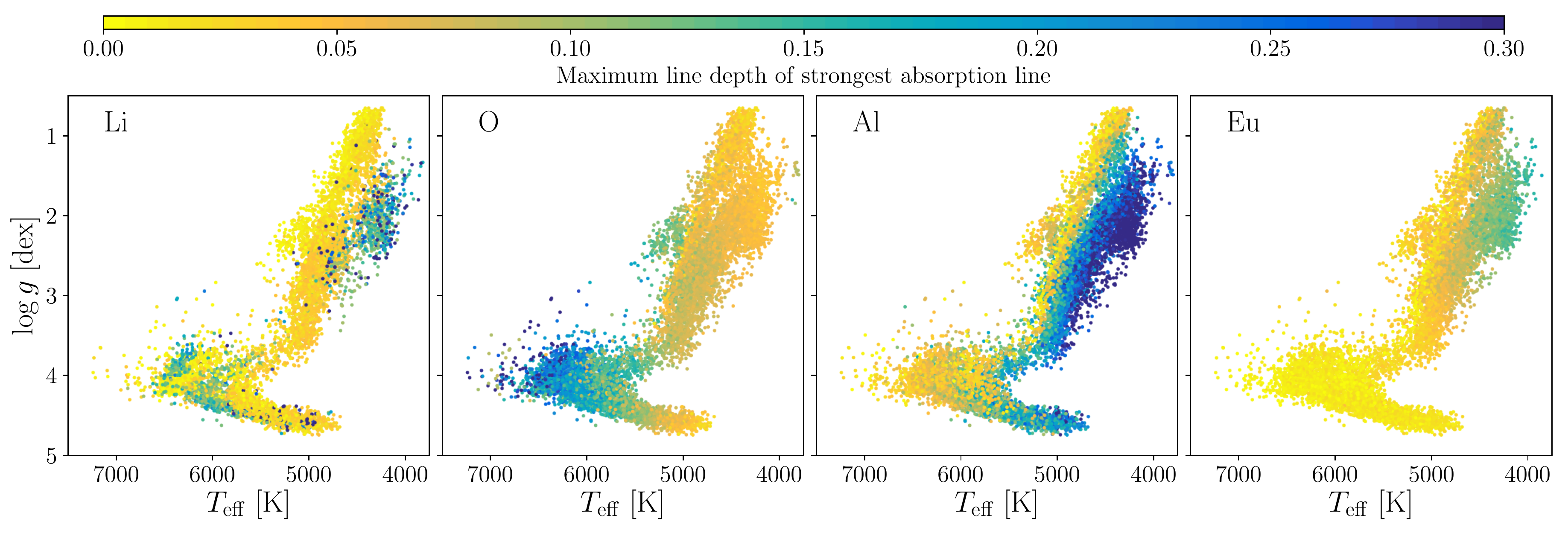}
	\caption{Visualisation of the parameter dependence of line strengths for the four elements Li, O, Al, and Eu. Shown are Kiel diagrams coloured by the maximum normalized absorption line depth within the used masks of the four elements, ranging from 0 to 0.3. For clarity, we truncated the plotted line strength at 0.3. These panels show that Li can not be measured in most stars (except hot dwarfs and especially Li-rich stars), while O has strong lines in hotter dwarfs and Al lines are very strong in cool, metal-rich giants. Eu is, similar to most neutron-capture elements within the GALAH range, almost exclusively detectable in giants.}
	\label{fig:linedepth}
\end{figure*}

For the measurements, we define lines as detected, when the line is deeper than 3$\sigma$ of the flux error within the line mask, and at least 5\% below the continuum flux, otherwise the measurement is considered an upper limit. Additionally, we require the measurement error to be less than $0.3\,\mathrm{dex}$, be based on at least 3 data points, and neglect the measurements if it is above an empirically calibrated SNR-dependent $\chi^2$-limit.

Abundance ratios are given in bracket notation as [X/Fe]. To minimize systematic errors and to calibrate the solar zero point, we analysed HERMES twilight spectra with the same SME set-up as other spectra and compute solar-relative abundances, 
\begin{equation}
\mathrm{[X/H] = A(X) - A(X)_{\odot}}
\end{equation}
where $\mathrm{A(X)_{\odot}}$ is the measured abundance from our solar spectra. Elemental abundance ratios are then defined as
\begin{equation}
\mathrm{[X/Fe] = \mathrm{[X/H]} - \mathrm{[Fe/H]}.}
\end{equation}
These may be converted back to absolute values ($\mathrm{A(X)^{A+09}}$) on the absolute abundance scale of \citet{Asplund2009} by computing
\begin{equation}
\mathrm{A(X)^{A+09} = [X/Fe] + [Fe/H] + A(X)_{\odot}^{A+09}.}
\end{equation}
The values of the solar abundances $\mathrm{A(X)_{\odot}}$ measured with GALAH as well as the reference values from \citet{Grevesse2007} that are adopted in the MARCS atmosphere grids and the  solar composition $\mathrm{A(X)_{\odot}^{A+09}}$ from \citet{Asplund2009} are given in Table~\ref{tab:solar}. We note that both references compositions are very similar with the latter being more recent and commonly used.

We report the individual abundances of $\alpha$-elements, but also include an error-weighted combination of unflagged abundances of the elements Mg, Si, Ca, and Ti reported as
\begin{equation}
\mathrm{[\alpha/Fe]} = \frac{\sum_\mathrm{X} \frac{\mathrm{[X/Fe]}}{\left( \mathrm{e\_[X/Fe]} \right)^2}}{\sum_\mathrm{X} \left( \mathrm{e\_[X/Fe]} \right)^{-2}}\text{, where X = Mg, Si, Ca, Ti}
\end{equation}
We recommend this definition, because the different elements are estimated with different precisions and a simple average would hence lead to a less precisely estimated [$\alpha$/Fe]. We note, however, that because Ti is typically the most precisely measured element among these four $\alpha$-elements, the combined [$\alpha$/Fe] is mainly tracing Ti. [$\alpha$/Fe] is reported also for stars where one or more of Mg, Si, Ca, and Ti is not available. 

\begin{table}
 \caption{Comparison of solar abundances ($\mathrm{A(X)}$) with respect to the standard composition of MARCS model atmospheres \citep{Grevesse2007} and the solar photospheric abundances by \citet{Asplund2009}.}
 \label{tab:solar}
 \begin{tabular}{cccc}
  \hline
X & $\mathrm{A(X)_{\odot}}$ & \citet{Grevesse2007} & \citet{Asplund2009} \\
\hline
Li & $0.95 \pm 1.00$ & $1.05 \pm 0.10$ & $1.05 \pm 0.10$ \\ 
C & $8.50 \pm 0.23$ & $8.39 \pm 0.05$ & $8.43 \pm 0.05$ \\ 
O & $8.85 \pm 0.04$ & $8.66 \pm 0.05$ & $8.69 \pm 0.05$ \\ 
Na & $6.10 \pm 0.04$ & $6.17 \pm 0.04$ & $6.24 \pm 0.04$ \\ 
Mg & $7.54 \pm 0.03$ & $7.53 \pm 0.09$ & $7.60 \pm 0.04$ \\ 
Al & $6.45 \pm 0.03$ & $6.37 \pm 0.06$ & $6.45 \pm 0.03$ \\ 
Si & $7.45 \pm 0.04$ & $7.51 \pm 0.04$ & $7.51 \pm 0.03$ \\ 
K & $5.50 \pm 0.05$ & $5.08 \pm 0.07$ & $5.03 \pm 0.09$ \\ 
Ca & $6.36 \pm 0.06$ & $6.31 \pm 0.04$ & $6.34 \pm 0.04$ \\ 
Sc & $3.12 \pm 0.04$ & $3.17 \pm 0.10$ & $3.15 \pm 0.04$ \\ 
Ti & $4.89 \pm 0.02$ & $4.90 \pm 0.06$ & $4.95 \pm 0.05$ \\ 
V & $3.93 \pm 0.55$ & $4.00 \pm 0.02$ & $3.93 \pm 0.08$ \\ 
Cr & $5.62 \pm 0.04$ & $5.64 \pm 0.10$ & $5.64 \pm 0.04$ \\ 
Mn & $5.31 \pm 0.03$ & $5.39 \pm 0.03$ & $5.43 \pm 0.04$ \\ 
Fe & $7.40 \pm 0.01$ & $7.45 \pm 0.05$ & $7.50 \pm 0.04$ \\ 
Co & $4.91 \pm 0.48$ & $4.92 \pm 0.08$ & $4.99 \pm 0.07$ \\ 
Ni & $6.21 \pm 0.04$ & $6.23 \pm 0.04$ & $6.22 \pm 0.04$ \\ 
Cu & $4.03 \pm 0.07$ & $4.21 \pm 0.04$ & $4.19 \pm 0.04$ \\ 
Zn & $4.43 \pm 0.04$ & $4.60 \pm 0.03$ & $4.56 \pm 0.05$ \\ 
Y & $1.89 \pm 0.09$ & $2.21 \pm 0.02$ & $2.21 \pm 0.05$ \\ 
Ba & $2.18 \pm 0.21$ & $2.17 \pm 0.07$ & $2.18 \pm 0.09$ \\ 
La & $1.10 \pm 0.11$ & $1.13 \pm 0.05$ & $1.10 \pm 0.04$ \\ 
Eu & $0.58 \pm 0.28$ & $0.52 \pm 0.06$ & $0.52 \pm 0.04$ \\ 
  \hline
 \end{tabular}
\end{table}

Although HERMES spectra in principle cover a large variety of element lines, their strength varies with the abundance of the element itself, but also the line properties and the stellar parameters. For this reason, we can not detect all elements equally well in all stars from the GALAH spectra. To visualise this, we show Kiel diagrams for the four elements Li, O, Al, and Eu in  Figure~\ref{fig:linedepth}, where we color each point by the depth of the strongest line of the respective element. For Li, the majority of stars only show weak Li lines. However both in warm dwarfs and several cool giants, it can be detected. For this element, we have added stars to the training set with projected high Li by the spectrum classification algorithm t-SNE (see Section~\ref{sec:tsne}), which show up as blue dots in the left panel of Figure~\ref{fig:linedepth}. The O triplet shows strongest lines for hot dwarf and turn-off stars due to its high excitation potential, but is in general detectable across the whole parameter space. Al is usually detectable in cooler and metal-rich stars across the parameter space, but not always in warmer stars. Eu lines are in general not detectable for dwarfs with the GALAH setup, but only for giants. The training set hence contains stellar parameters for all stars, but not all stars have abundance measurements for all elements.

\subsection{Analysis step 2: \textit{The Cannon}} \label{sec:analysis_step2}

We implement \textit{The Cannon} as described in \citet{Ness2015}, adopting a simple quadratic model with coefficients $\boldsymbol{\theta}_\lambda$, which describes the flux $f_{n,\lambda}$ of a given spectrum $n$ with stellar labels $\boldsymbol{\ell}_n$:
\begin{equation} \label{eq:cannon}
f_{n,\lambda} = \boldsymbol{\theta}_\lambda^T \cdot \boldsymbol{\ell}_n + \mathrm{noise}
\end{equation}
We augment this procedure with a number of additional processing steps and derive many more labels than the original implementation. We interpolate all spectra of the survey on to a common wavelength grid of 14304 pixels and use the normalized spectra at rest from our reduction pipeline, see Section~\ref{sec:reduction}.

\begin{figure*}
  \includegraphics[width=0.98\textwidth]{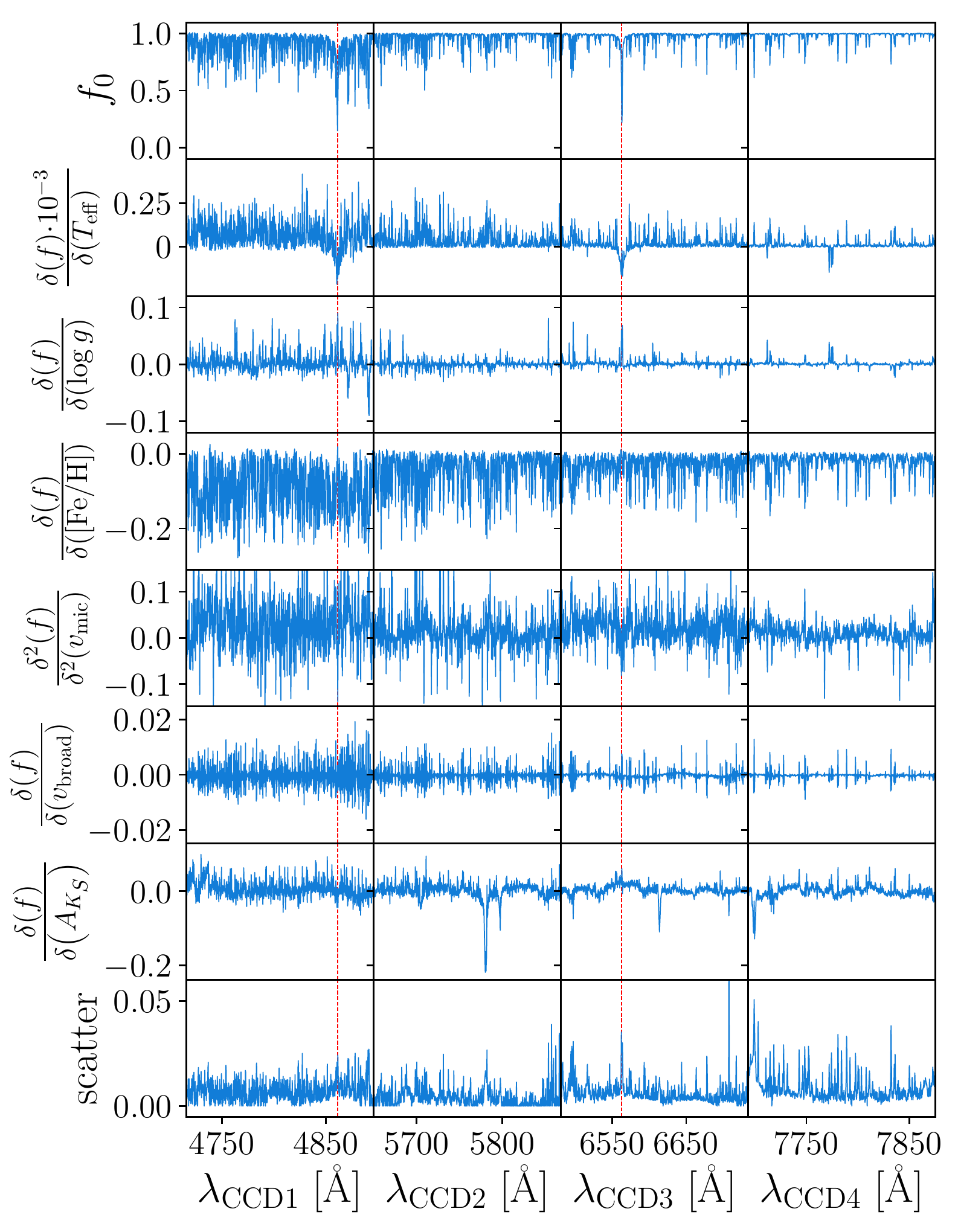}
  \caption{Linear coefficients of \textit{The Cannon} model as a function of wavelength across the four HERMES wavelength regions. We show the spectrum $f_0$ with labels equivalent to the median of the training set and linear coefficients for five stellar labels. We also include the quadratic coefficient for microturbulence velocity, showing the same pattern as the molecular absorption bands and the scatter of the model. Red dashed lines indicate the Balmer lines.}
  \label{fig:coefficients}
\end{figure*}

A limitation of the currently published versions of \textit{The Cannon} is that all labels have to be known for each training set spectrum. For our reference objects, we have many stars that have a subset of the full number of the possible abundances measured. There are relatively fewer stars in the set of reference objects with all individual abundances measured. While efforts are being made to extend \textit{The Cannon} in order to handle label errors or partially missing labels (Eilers et al., in preparation), we still have to rely on a different approach in this data release. We handle this issue of partial labels in the training set by creating an ensemble of models; one for each element [X/Fe]. We start with a training set for which all stellar parameters are known and train a model using these parameters only. We then use this model on the training set itself and re-derive the stellar parameter labels.  We then exchange the labels of the initial training set with the re-derived labels. Then, for each element, X, we create a new training set by adding one more label, the element abundance [X/Fe], using only the subset of stars in the training set with this measured abundance. We therefore train a new model based on the six stellar parameter labels, plus one additional element label, and do this step for each element. In each case, for each model and each corresponding element, we restrict (mask) all coefficients with [X/Fe]-terms to be zero outside of pre-selected line regions of that element.

\begin{figure*}
  \includegraphics[width=\textwidth]{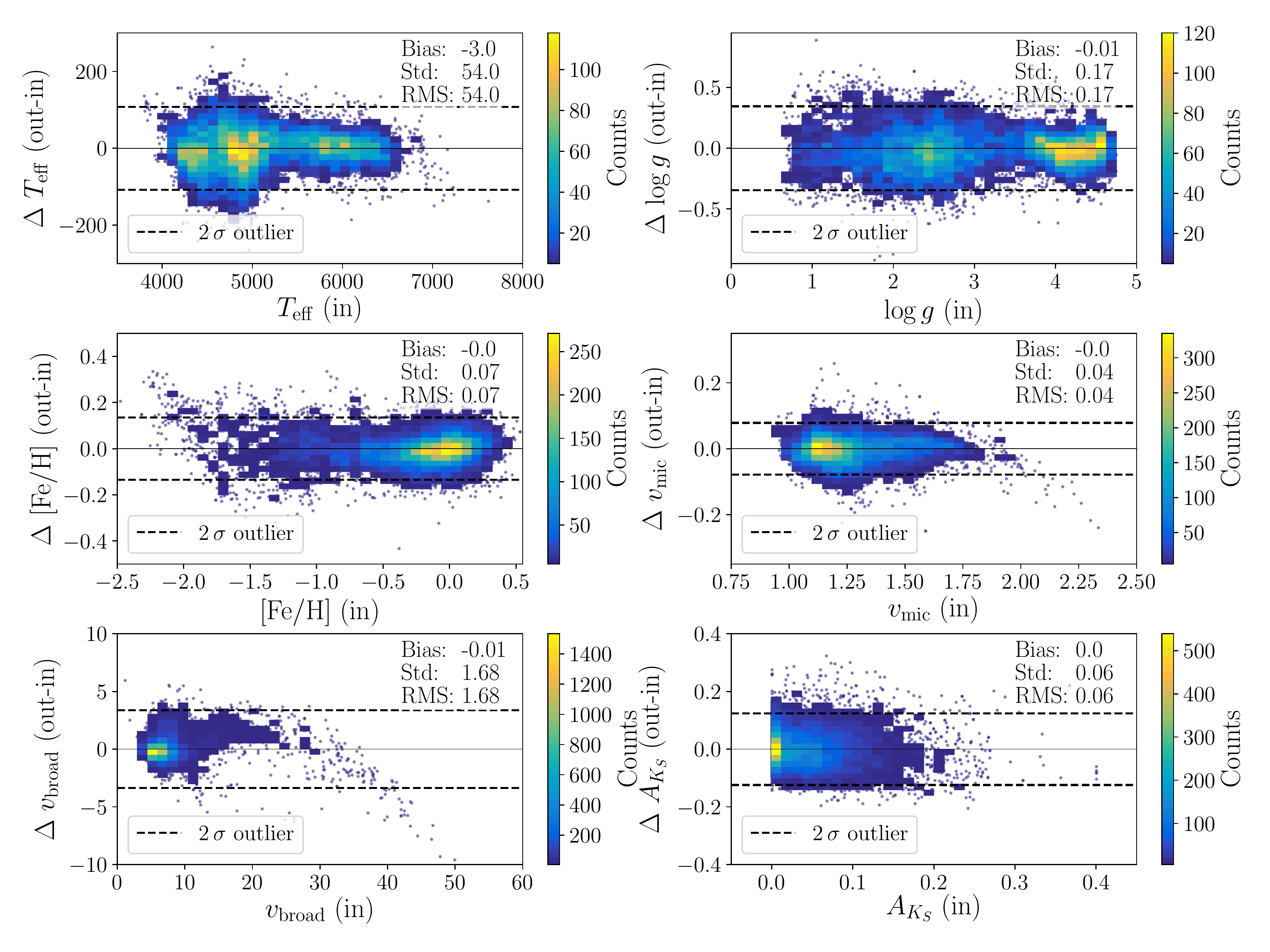}
  \caption{Comparison of training set labels from SME (input) and \textit{The Cannon} interpolation (output) for the labels of the stellar parameter model, i.e. $T_\text{eff}$, $\log g$, [Fe/H], $v_\text{mic}$, $v_\text{broad}$, and $A_{K_S}$.}
  \label{fig:selftest}
\end{figure*}

\begin{figure*}
  \includegraphics[width=\textwidth]{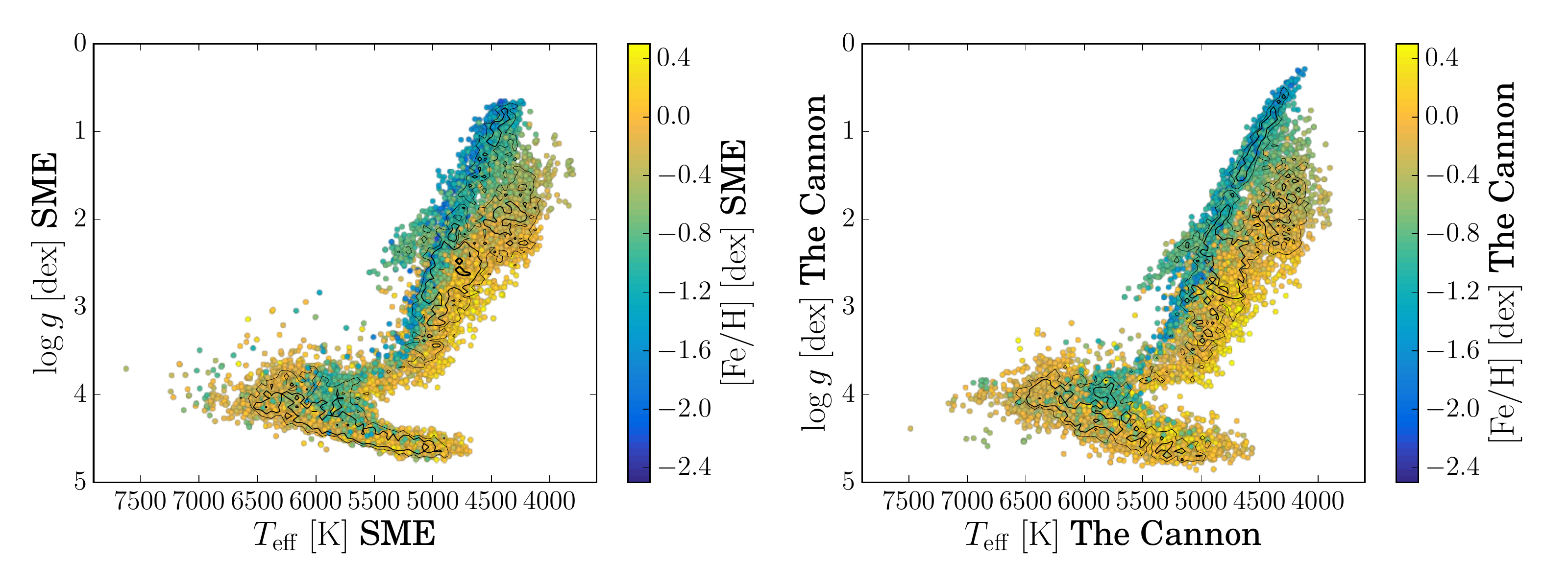}
  \caption{Kiel diagram for the GALAH DR2 training set as determined with SME (left; as input into \textit{The Cannon}) and as reproduced by \textit{The Cannon} (right).}
  \label{fig:sme_cannon}
\end{figure*}

\subsubsection{\textit{The Cannon} model for stellar parameters}

Our training data are a high-fidelity set of stars with labels generated using SME, as described in Section~\ref{sec:analysis_step1}. The final training set for stellar parameters of 10605 spectra consists of 21 \Gaia\ benchmark stars, 12 stars overlapping with \citet{Bensby2014}, 77 stars with \textsc{Hipparcos} parallaxes, 3807 stars with TGAS parallaxes, 915 stars with asteroseismic information, 669 open or globular cluster stars as well as 1805 stars already included in previous training sets \citep{Martell2017, Sharma2018, Wittenmyer2018}. To ensure a sufficient coverage of parameter space for the training step, we expand the set with 1057 selected stars in the parameter range of $\mathrm{[Fe/H]} < -1.0$, 1055 additional stars with $-1.0 < \mathrm{[Fe/H]} < -0.5$, 654 additional giants with $T_\text{eff} < 5000\,\mathrm{K}$, $\log g < 2.0\,\mathrm{dex}$, and $\mathrm{SNR} > 125$ in the green channel, 388 stars with projected high Li abundances based on the work by \citet{Traven2017}, and 145 stars with $\mathrm{SNR} > 100$ or $\mathrm{SNR} > 50$ at $\log g < 2$ overlapping with APOGEE DR14. We stress that we excluded spectroscopic binaries from the training set, either based on previous automated stellar classifications, see Section~\ref{sec:tsne} or via visual inspection of the training set spectra.

We emphasise that we include more stars and labels than \citet{Ness2015} because our parameter space covers a much larger part of the HR-diagram than the RGB. The lines of the majority of the GALAH survey main sequence and turn-off stars, are in general significantly more broadened than in giants. Therefore, the model needs to be flexible enough to track these changes in the lines of the spectrum. We build our reference labels on similar parameters as the free parameters of the SME optimisation, i.e. $T_\text{eff}$, $\log g$, [Fe/H], $v_\text{mic}$, and $v_{\rm broad}$. 

Additionally, we found that it was important to include interstellar extinction $A_{\rm K}$, similarly to \citet{Ho2017}, as diffuse interstellar bands are present in the spectra. There is a degeneracy between abundance and extinction that is present if this label is not included. The values for $A_{\rm K}$ are estimated from the RJCE method \citep{Majewski2011}.
We note that \citet{KosZwitter2013} 
have shown that the ratio of the strength of diffuse interstellar bands to extinction is a function of the ultraviolet radiation field in the interstellar medium while \citet{Nataf2016} 
have demonstrated that this ratio also depends on the shape of interstellar extinction curve. In the future, we will calibrate these additional parameters to unprecedented precision with GALAH data. 

For the stellar parameter estimation we use a quadratic model with six stellar labels, resulting in 22 coefficients for each of the 14304 pixels of \textit{The Cannon} wavelength range, on to which we interpolate each spectrum.

\begin{landscape}
\begin{figure}
\includegraphics[width=\columnwidth]{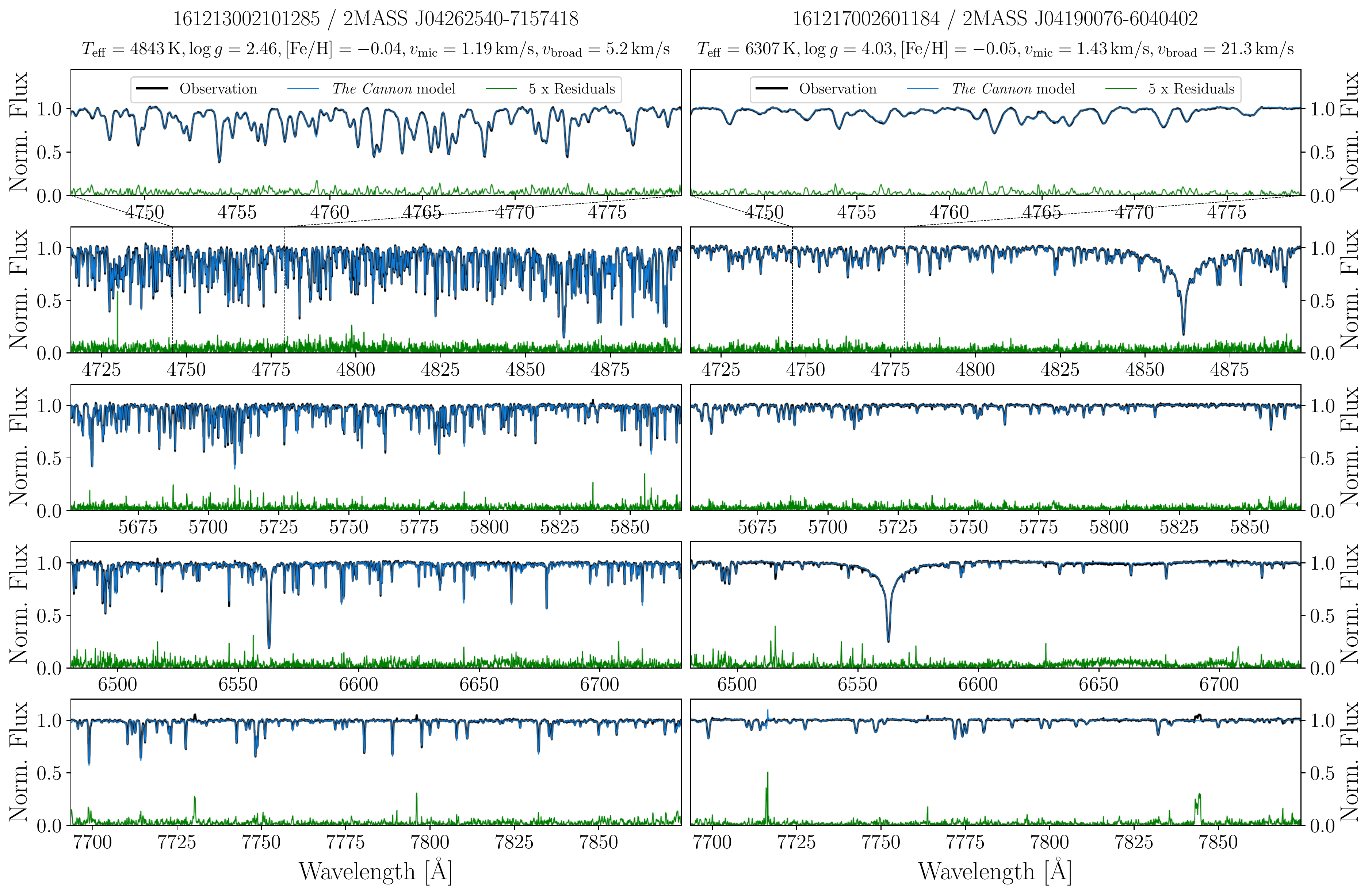}
\caption{Spectra of the giant stars 2MASS J04262540-7157418 (left panels) and the dwarf star 2MASS J04190076-6040402 (right panels). Shown are the observed spectrum (black) as well as the model spectra from \textit{The Cannon} (blue) and the magnified residuals (5 times larger in green) for different parts of the GALAH spectral range. The top panels show a magnification of a part of the blue arm (CCD 1), while the other panels show the spectral range of the four bands (CCD 1 to 4). \textit{The Cannon} is only applied to the second half of the IR-arm. The agreement between the normalised observation and model are overall very good with mean residuals around 0.005. The largest residuals can typically be seen in regions with imperfect skyline and telluric line corrections in the observed spectrum.}
\label{fig:example_spectra}
\end{figure}
\end{landscape}

The linear coefficients of this model for the labels are shown in Figure~\ref{fig:coefficients} and indicate, how the median flux $f_0$ of the training set (with median labels $T_\text{eff} = 5114\,\mathrm{K}$, $\log g = 3.0\,\mathrm{dex}$, $\mathrm{[Fe/H]} = -0.33$, $v_\text{mic} = 1.28$~km\,s$^{-1}$, $v_\text{broad} = 7.7$~km\,s$^{-1}$, and $A_{K_S} = 0.05$, respectively) changes with each of these labels. These linear coefficients are a good first diagnostic for the sensitivity of \textit{The Cannon} regarding certain labels within the GALAH range, but the quadratic terms have to be taken into account as well for the full model spectrum. In Figure~\ref{fig:example_spectra}, we show two examples of GALAH observations and the model spectrum from \textit{The Cannon}.

The panels in Figure~\ref{fig:coefficients} show that the effective temperature $T_\text{eff}$ is strongly correlated with the two hydrogen Balmer lines (indicated by red dashed lines). We note also a strong connection of the O triplet in the IR arm with $T_\text{eff}$. Many ionized lines show positive correlations with the linear coefficient for $\log g$, while for example the Fe \textsc{i} lines around $4872$ and $4891$\,\AA\ show strong negative correlations, as expected due to their pressure broadened wings. The coefficient of [Fe/H] shows not only correlations with Fe, but is a tracer of metallicity itself and consequently responds to all lines since the abundances of all elements track each other to first-order; especially the blue channel is sensitive to the metallicity due to the preponderance of lines there. The linear coefficient for $v_\text{mic}$ shows the strongest sensitivity in the blue channel. Because of the empirical, temperature-dependent relation used for this label with SME, $v_\text{mic}$ is most sensitive to changes at the hot and cool end of the parameter space. To visualise this, we show the quadratic coefficient of $v_\text{mic}$, which shows the influence of molecular absorption bands in the spectrum for the coolest stars in the training set. The broadening label $v_\text{broad}$ indicates positive correlations in the core of lines and negative ones in the wings, i.e., lines become broader with larger $v_\text{broad}$ while maintaining the overall line strength. Similar to \cite{Ho2017}, our linear coefficient for $A_{K_S}$ correlates strongly with the diffuse interstellar bands within the GALAH range \citep[][]{DeSilva2015}. We also show the scatter term of the model, which corresponds to regions not well described by the stellar labels, including telluric lines from imperfect corrections as well as the regions of the diffuse interstellar bands and the interstellar component of K \textsc{i} 7699. However, the scatter is in general very low (with a median around 0.01), suggesting that our model fits the data well.

We apply this best-fitting model to the training set spectra as a self-test and subsequently compare the stellar labels from SME with those estimated by \textit{The Cannon} in Figure~\ref{fig:selftest}. \textit{The Cannon} reproduces the labels of the training set with negligible biases and within a scatter of $\sigma \left( T_\text{eff} \right) = 71\,\mathrm{K}$, $\sigma \left( \log g \right) = 0.25\,\mathrm{dex}$, $\sigma \left( \mathrm{[Fe/H]} \right) = 0.1\,\mathrm{dex}$, $\sigma \left( v_\mathrm{mic}  \right) = 0.06$~km\,s$^{-1}$, $\sigma \left( v_\text{broad} \right) = 3.1$~km\,s$^{-1}$, and $\sigma \left( A_{K_S} \right) = 0.08\,\mathrm{mag}$. We note that although the Kiel diagrams of the input and output labels look very similar (see Figure~\ref{fig:sme_cannon}), the scatter values are slightly larger than those of previous analyses \citep{Martell2017, Sharma2018}. While we have not yet found the reason for this, an explanation could be the expansion of the training set to cover a larger (and hence different) parameter space, including fast rotators ($v_\text{broad} > 30$~km\,s$^{-1}$) and metal-poor stars, which stretches the flexibility of the quadratic model to its limits. Contrary to the previous models estimated with \textit{The Cannon} for the GALAH survey, we do not fit [$\alpha$/Fe] as part of the stellar parameters, as it would interfere with the later estimated individual $\alpha$-element abundances, and because it did not significantly decrease the scatter of the label validation.

\subsubsection{\textit{The Cannon} models for element abundances}

\begin{figure*}
  \includegraphics[width=\textwidth]{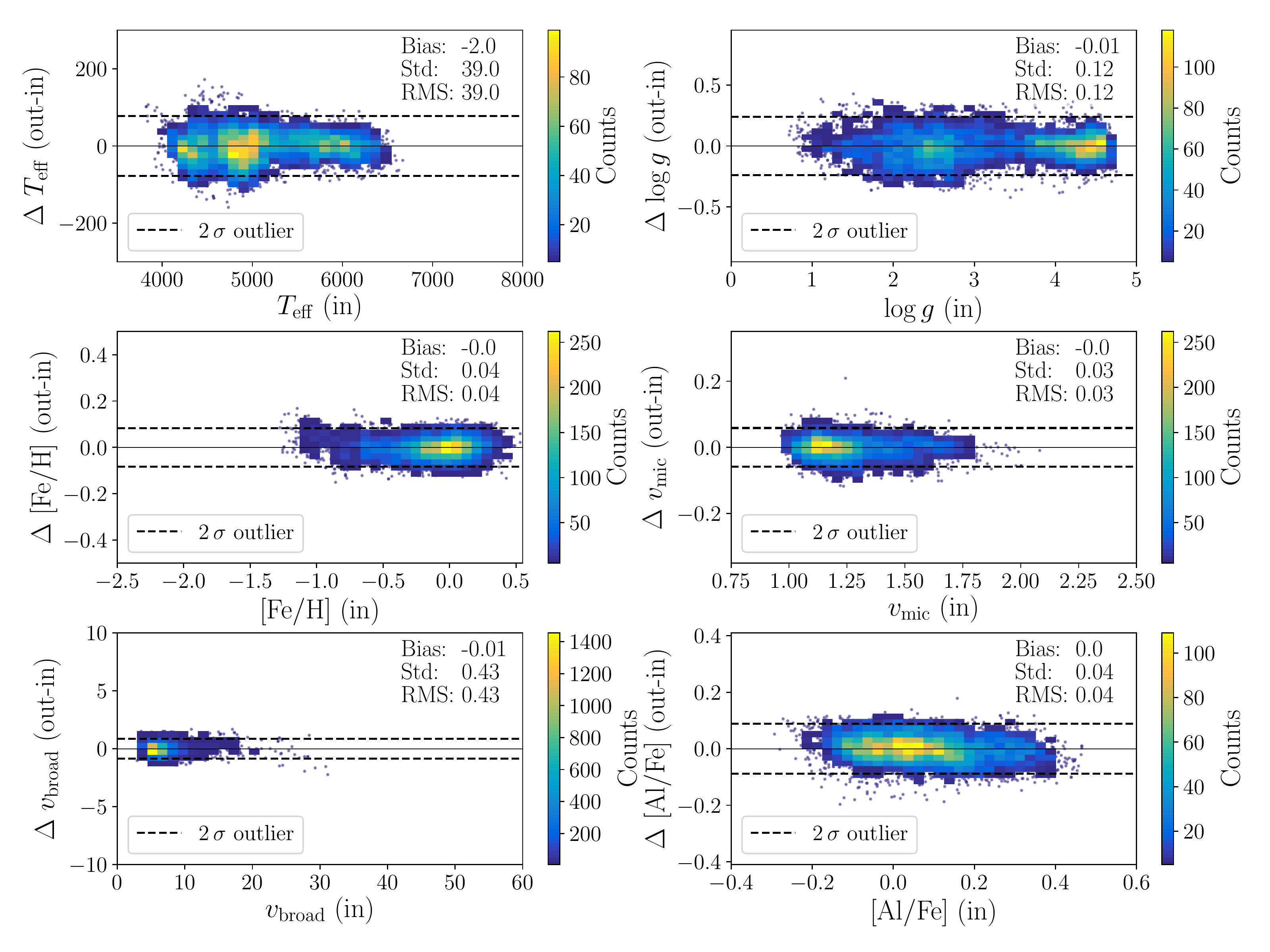}
  \caption{Self validation of \textit{The Cannon} model for Al.}
  \label{fig:Al_selftest}
\end{figure*}

We reiterate that we use an ensemble of models to infer our elements, based on the six stellar parameters plus one additional element, for all elements. For each training set for each model, we only include those stars in the individual models which have abundance detections for the respective element.  Table~\ref{tab:ts_size} summarises the relative fractions of stars in the training set with each element measurement. For each model, we exchange the stellar parameters of SME with those from the self test, before using this model at test time to estimate the individual element abundance. This introduces a minor perturbation to the model (and the stellar labels are slightly different, within the error of the inference). We confirm that the perturbation is minor; the scatter term of the labels from the self test is significantly smaller than before the exchange of stellar parameter labels, see for example the self validation for Al in Figure~\ref{fig:Al_selftest}.

We emphasise that we have chosen to restrict the abundance determination to use only lines of each element being inferred. We therefore restrict \textit{The Cannon's} model to use only certain wavelength regions for each element inference (e.g. for [O/Fe] we use the pixels of the O\,{\sc i} 7771-5\,\AA\ triplet). This masking is achieved by setting the element-dependent coefficients of the model to zero outside specified regions. 

The masks we use for the individual elements are the same as for SME, see Section~\ref{sec:sme_abundances}. This differs from the regularised approach of \citet{Casey2016}, where the model decides which coefficients of the model to force to 0, but can still learn strongly correlated features. It may be legitimate to learn these correlations and take advantage of this information \citep[e.g.][]{Ting2018}. Based on previous analyses, which showed a strong correlation of [$\alpha$/Fe] and telluric lines for the GALAH survey, we have chosen to use the most conservative approach for this analysis.

\begin{table}
 \caption{Training set size for individual \textit{The Cannon} abundance models, compared to the stellar parameter model with 10605 spectra.}
 \label{tab:ts_size} 
 \begin{tabular}{cccc}
  \hline
Model & Nr. of spectra (\%) & Model & Nr. of spectra (\%) \\
\hline
Li & \,~1652 (16\%) & V & \,~3495 (33\%) \\ 
C & \,~1204 (11\%) & Cr & 10015 (94\%) \\ 
O & \,~8538 (81\%) & Mn & 10204 (96\%) \\ 
Na & 10222 (96\%) & Co & \,~5574 (53\%) \\ 
Mg & 10470 (99\%) & Ni & \,~6388 (60\%) \\ 
Al & \,~8529 (80\%) & Cu & \,~9312 (88\%) \\ 
Si & \,~6422 (61\%) & Zn & 10012 (94\%) \\ 
K & 10237 (97\%) & Y & \,~9354 (88\%) \\ 
Ca & \,~9217 (87\%) & Ba & 10417 (98\%) \\  
Sc & 10438 (98\%) & La & \,~5215 (49\%) \\ 
Ti & 10335 (97\%) & Eu & \,~4419 (42\%) \\ 
\hline
 \end{tabular}
\end{table}

\subsubsection{\textit{The Cannon} errors}\label{sec:cannon_errors}

The errors reported by \textit{The Cannon} are based on the formal covariance errors, which are typically very small. Our total error estimate for each label is based on an additional SNR-dependent performance test of the training set for each label. This is estimated by comparing the difference of the SME input and \textit{The Cannon} output as a function of the training set SNR and fitting an exponential function to the mean values within bins of SNR of 25, 50, 75, and 100. These errors are then summed in quadrature with \textit{The Cannon} covariance errors. We note that this performance test only includes SNR $>$ 25 (as a result of our requirement for the training set spectra to be of high-fidelity). At SNR $<$ 25, the performance test is an extrapolation and tends to underestimate the errors when compared with the scatter between repeat observations, see Section~\ref{sec:repeats}.

\subsection{Analysis step 3: Flagging} \label{sec:analysis_step3}

To report the quality of both spectra and the spectroscopic analysis, we are employing several flags, which are summarised in Table~\ref{tab:catalog} and explained subsequently.

\subsubsection{Flagging of stellar parameter and element abundance estimates}

For a given spectrum $m_{DR2}$ of the GALAH Data Release 2 with labels $\ell_{m_{DR2}}$, we estimate the label distance to the labels $\ell_{n_TS}$ of the training set points $n_{TS}$ similarly to \citet[][see their Equation~7]{Ho2017}:
\begin{equation}
D  = \sum_\ell  \sum_{n_{TS}} \frac{ \left( \ell_{m_{DR2}} - \ell_{n_{TS}} \right)^2}{K_\ell^2}
\end{equation}
For the stellar parameters we use $$\ell~\in~[T_\text{eff}, \log g, \mathrm{[Fe/H]}, v_\text{broad}]$$ and for the abundance of element X we use $$\ell~\in~[T_\text{eff}, \log g, \mathrm{[Fe/H]}, v_\text{broad}, \mathrm{[X/Fe]}]$$. The uncertainties $K_\ell$ used to estimate the label distances are based on the RMS of the self validation as listed in Table~\ref{tab:self_validation}. Subsequently, we estimate the mean distance of the 10 smallest label-distances, i.e. closest training set points and raise the $\texttt{flag\_cannon}$ bitmask by 1, if this distance is larger than 8 for the stellar parameters (a mean of $2\sigma$ for 4 stellar labels) or 10 for the individual abundances (a mean of $2\sigma$ for 5 stellar labels). We report however also the mean label distance to the 10 closest training set points as \texttt{sp\_label\_distance} to allow the exploration of this flag.

\begin{table}
 \caption{Biases and RMS of the self validation test.}
\label{tab:self_validation}
 \begin{tabular}{cccccc}
  \hline
$\ell$	&	Bias	&	RMS & $K_\ell$	&	Bias	&	RMS	\\
\hline
$T_\text{eff}$ [K]	&	3	&		54	& 	$\mathrm{[Ca/Fe]}$	&	0.00	&		0.05	\\
$\log g$ [dex]		&	0.01	&		0.17	&	$\mathrm{[Sc/Fe]}$	&	0.00	&		0.04	\\
$\mathrm{[Fe/H]}$	&	 0.00	&		0.07	&	$\mathrm{[Ti/Fe]}$	&	0.00	&		0.04	\\
$v_\text{mic}$ [km\,s$^{-1}$]	&	0.00	&		0.04	&	$\mathrm{[V/Fe]}$	&	0.00	&		0.06	\\
$v \sin i$ [km\,s$^{-1}$]		&	0.0	&		1.7	&	$\mathrm{[Cr/Fe]}$	&	0.01	&		0.05	\\
$A_{K_S}$ [mag]	&	0.00	&		0.06	&	$\mathrm{[Mn/Fe]}$	&	0.00	&		0.06	\\
$\mathrm{[Li/Fe]}$	&	0.00	&		0.08	&	$\mathrm{[Co/Fe]}$	&	0.00	&		0.06	\\
$\mathrm{[C/Fe]}$	&	0.00	&		0.05	&	$\mathrm{[Ni/Fe]}$	&	0.00	&		0.07	\\
$\mathrm{[O/Fe]}$	&	0.00	&		0.11	&	$\mathrm{[Cu/Fe]}$	&	0.00	&		0.06	\\
$\mathrm{[Na/Fe]}$	&	0.00	&		0.05	&	$\mathrm{[Zn/Fe]}$	&	0.00	&		0.08	\\
$\mathrm{[Mg/Fe]}$	&	0.00	&		0.08	&	$\mathrm{[Y/Fe]}$	&	0.00	&		0.08	\\
$\mathrm{[Al/Fe]}$	&	0.00	&		0.04	&	$\mathrm{[Ba/Fe]}$	&	0.00	&		0.10	\\
$\mathrm{[Si/Fe]}$	&	0.01	&		0.08	&	$\mathrm{[La/Fe]}$	&	0.00	&		0.06	\\
$\mathrm{[K/Fe]}$	&	0.00	&		0.09	&	$\mathrm{[Eu/Fe}$]	&	0.00	&		0.07	\\  
\hline
 \end{tabular}
\end{table}

Analogous to the analysis with SME, see Section~\ref{sec:sme_abundances}, we estimate if the measured line is a detection of only an upper limit and raise the bitmask by 2, if the line is $<3\sigma$ of the flux error, but at least 5\% below the continuum flux.

Additionally, we make use of the $\chi^2$ fit statistic and raise the $\texttt{flag\_cannon}$ bitmask by 4, if the mean $\chi^2$ per pixel (with an expectation of 1 for a perfect fit and perfectly known errors) is either below 0.5 or above 10, both indicating issues with the spectra or that \textit{The Cannon} model can not describe the given spectrum.

\begin{figure*}
		\centering
		\includegraphics[trim = 40mm 10mm 40mm 0mm, clip, width=0.95\textwidth]{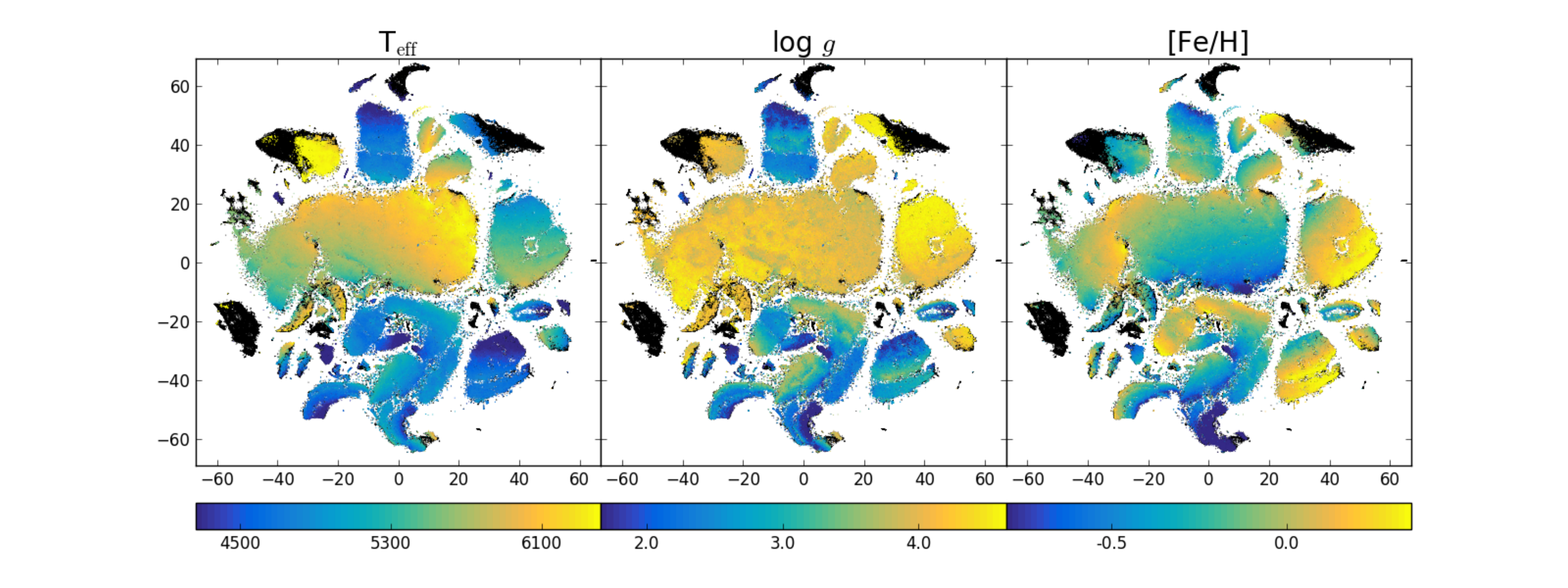}
  		\caption{t-SNE projection map of 587\,153 spectra. 413\,920 of them have reliable stellar parameters derived by \textit{The Cannon} pipeline, others are plotted as black points. The parameters are missing especially for hot and cool stars, but also for some binary stars and stars with emission lines, as seen by comparing this figure to Figure \ref{fig:tsne_flags} where the classification categories are marked. The colour-scale is done with 2.5 \% at both extremes of parameter values truncated for better contrast.}
 		\label{fig:tsne_basic}
	\end{figure*}

\begin{figure*}
		\includegraphics[trim = 40mm 45mm 45mm 45mm, clip, width=0.7\textwidth]{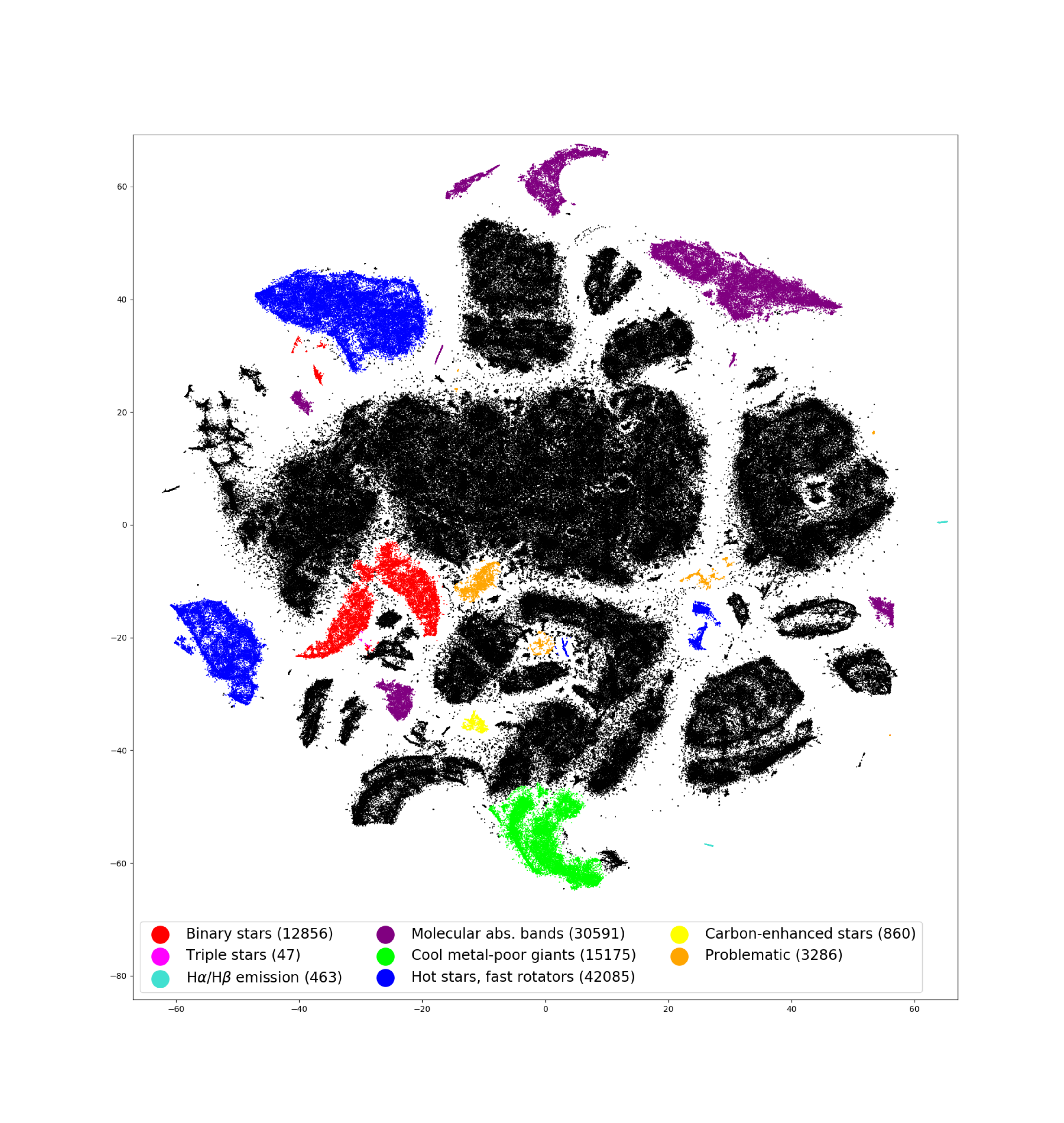}
  		\caption{t-SNE projection map, same as Figure  \ref{fig:tsne_basic}, except that the points (spectra) are color-coded by classification category. The majority of stars do not show peculiarities and are shown as black dots. The flagged triple stars are few and hardly seen next to the lower left group of binary stars, whereas the H$\alpha$/H$\beta$ emission stars are on the far right and bottom right in the map. The count of spectra for each category is given in the legend.}
 		\label{fig:tsne_flags}
	\end{figure*}
	
\subsubsection{Automated stellar classification with t-SNE}
\label{sec:tsne}

Classification of data is one of the most important steps in any kind of automatic data reduction and analysis. This is especially true in the case where the sheer quantity of collected information prevents us from manually inspecting the data as it comes in, and also when it is not possible to determine all sorts of outliers and unexpected issues {\it a priori}. Because the GALAH sample of observed spectra fits this description, it was necessary to develop a semi-automatic classification procedure, which has been presented in \citet{Traven2017}. Here we briefly outline the classification scheme of the spectra at rest wavelength and its recent improvements. 

In order to facilitate the discovery and determination of diverse morphological groups of spectra, we make use of two mathematical techniques, an increasingly more popular dimensionality reduction method t-SNE \citep{vanderMaaten2008}, and the well-established clustering algorithm DBSCAN \citep{Ester1996}. These techniques enable us to condense the information contained in our entire dataset into a two-dimensional map (Figure \ref{fig:tsne_basic}), where the spectra are arranged in such a way, that similar ones are grouped together, while there is a clear separation between distinct groups. The strongest features in the data -- indicating stellar physical parameters -- clearly dictate its global structure, hence the map can be clearly divided between dwarfs and giants or hot and cool stars. Inside these larger groups there is rich local structure, usually driven by the chemical composition of stars or by any other slowly changing spectral feature. 
In addition to the influence of stellar parameters, the shape of the projection is also driven by the presence of strong emission lines, or multiple lines from binary and higher order systems. 

With the help of the above described map we are able to manually inspect the structure of our dataset and assign classification categories to groups of spectra of interest. To each group in the map, one assigns a category by either looking at the average spectrum of the subset of the group or by the help of previous classification and other existing information/labels. Therefore, with every subsequent classification of the growing dataset, it is easier to assign categories; however, new and unexpected features can add to the complexity of the map. 

We no longer use the two-step procedure described in \citet{Traven2017} for producing a t-SNE map of peculiar spectra, since we have overcome the computational difficulties and can now produce the projection map using all available spectral information of the whole dataset in less than a day on a 24-core Xeon node. This became possible by using the parallel multicore t-SNE code \citep{Ulyanov2016} and modifying it to remove the limit of overall information that we put into t-SNE. Additionally, we exclude the infrared band as it still suffers from the presence of strong spikes (see \citealt{Traven2017}), which are now understood and accounted for. The inclusion of this band would hamper our classification significantly compared to what could be gained from information contained therein. 

Figure \ref{fig:tsne_flags} presents classification results based on all spectra of this data release which passed the basic reduction as explained in Section~\ref{sec:reduction}. We flag all groups of spectra for which a sensible physical category can be determined, however our results are not exhaustive. The population of binary stars represents $\sim 2.2$ \% of the whole dataset, down to a separation of $\sim 10$ km\,s$^{-1}$ in the case of most blended double lines. We see a morphological distinction between two larger groups of binaries, which is due to the stronger component being redshifted in one group and blueshifted in the other. 

The flagged H$\alpha$/H$\beta$ emission stars are very few in our dataset, which is partly because our observations are not focused on young open clusters, but also because a weak emission signature is sometimes not enough for those spectra to stand out in the map. Our observations and data reduction still introduce some issues that manifest in different features in the spectra and are flagged as problematic, however they are relatively few. 

The classification presented in this section serves both as a source of intrinsically interesting objects which can be studied separately (Traven et al., in preparation), and also as a guide to the development and improvement of the reduction and analysis pipeline.

\section{Validation}\label{sec:validation}

To validate the results by both SME and \textit{The Cannon}, we use a variety of tests, including the comparisons with fundamental parameters of the \Gaia\ benchmark stars, repeated observations,  photometric temperatures, asteroseismic surface gravities, open and globular clusters, and other spectroscopic surveys.

\subsection{\Gaia\ benchmark stars}
\label{sec:GBS}

\begin{figure}
\includegraphics[width=\columnwidth]{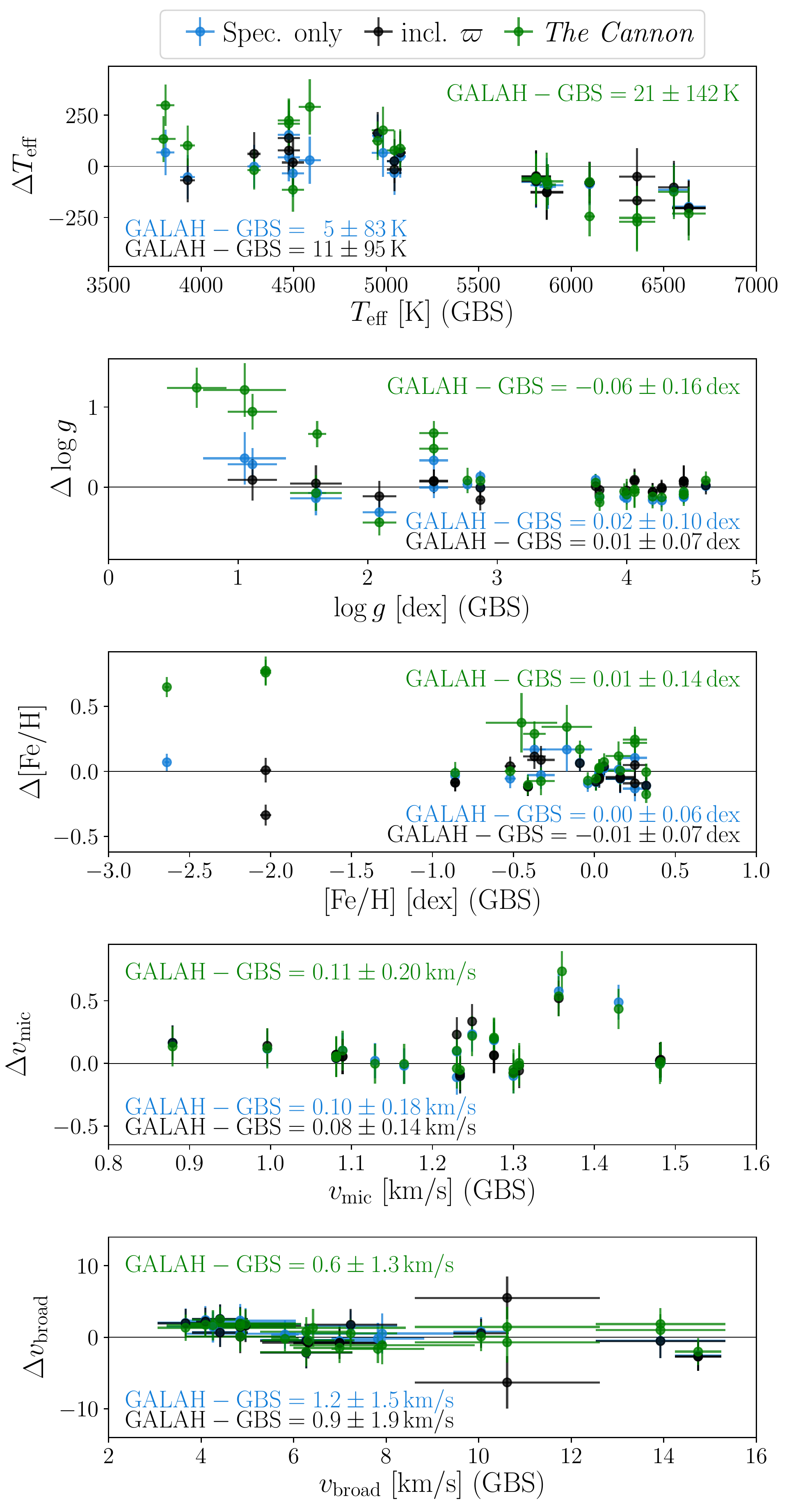}
  \caption{Comparison of GALAH stellar parameters with \Gaia\ benchmark stars (GBS). Shown are differences for the training set (intended as GALAH-Gaia benchmark stars) only based on spectral information (blue), training set including astrometric information (black), and \textit{The Cannon} output (green).}
  \label{fig:gbs_comparison}
\end{figure}

We use \Gaia\ benchmark stars \citep{Jofre2014, Heiter2015} as one way to validate the accuracy of our stellar parameters. Because of their independently estimated effective temperatures and surface gravities through interferometry and bolometric flux estimations, respectively, these parameters are less model-dependent than our spectroscopically derived ones.
In Figure~\ref{fig:gbs_comparison}, we compare results with the SME analysis (only based on spectroscopy, as well as with astrometric information) and with the \text{The Cannon}-based parameters. 

For the fundamental parameter $T_\text{eff}$ we find no significant bias (see offset and dispersion in top panel of Figure~\ref{fig:gbs_comparison}). For the warmest \Gaia\ benchmark stars, we see systematic trends of under-estimated temperatures for both SME analyses, which are propagated by \textit{The Cannon}. For the luminous giants, we see a rather good agreement for SME within $200\,\mathrm{K}$, while \textit{The Cannon} overestimates the temperature of the most luminous giants by around $250\,\mathrm{K}$. For these stars, \textit{The Cannon} also overestimates the surface gravity significantly by around $1\,\mathrm{dex}$, as the second panel of Figure~\ref{fig:gbs_comparison} shows. While the agreement for surface gravity is good by construction when including parallax information in the SME analysis, there is an offset of $0.15\,\mathrm{dex}$ to both the \Gaia\ benchmark stars and asteroseismically inferred $\log g$ when using a purely spectroscopic approach. As a result, we have opted to shift the spectroscopic SME results by this amount. 

The iron abundances as well as the iron abundances of the best-fitting atmosphere (\textit{sme.feh}) have both shown systematic biases of $0.1\,\mathrm{dex}$. We have hence increased the iron abundance globally by $0.1\,\mathrm{dex}$. and find good agreement along the metallicity range for the SME results. \textit{The Cannon}, however, shows biases of overestimated iron abundances for the most metal-poor stars (below [Fe/H] of $-1$) of around $0.7\,\mathrm{dex}$, and $0.35\,\mathrm{dex}$ for the coolest giants.

The microturbulence velocity, $v_\text{mic}$, agrees in general with those presented in \citep{Heiter2015, Jofre2014} except for luminous giants where our chosen relation differ by $0.5$ to $0.7\,\mathrm{km\, s^{-1}}$. We note that the microturbulence velocity is dependent on the adopted stellar parameters, model atmospheres and line selection among other things and it is thus not given that our values are inferior when the two sources differ. Similarly, our line broadening parameter $v_{\rm broad}$ is similar to the literature values when summing the published $v_\text{mac}$ and $v \sin i$ in quadrature. 

\subsection{Repeat observations}\label{sec:repeats}

About 50,000 spectra have independent repeated observations\footnote{As discussed in Section~\ref{sec:catalog}, in the final GALAH DR2 catalogue, if stars were observed multiple times, we only report the highest SNR observation and remove duplicates.}. These repeats are useful for characterizing the uncertainty in the stellar parameters as well as the intrinsic variability of the stellar properties. The repeats span multiple programs with HERMES, including commissioning, and the pilot and main GALAH surveys, as well as the K2-HERMES, and TESS-HERMES surveys. About 50\% of the repeats are due to bad observational conditions, which were repeated to boost the signal to noise ratio. These repeats were done using the same plate and fibre configuration as the original observation. The other 50\% had different plate and fibre configuration. Of these, 25\% are serendipitous observations of the GALAH pilot and commissioning programs, which did not have a well defined selection function and hence were not tracked when the main survey started its operation. To minimize the time spent on repeats, a significant number of the deliberate repeats were carried out for the bright fields, which require less than half of the exposure time.

\begin{figure*}\includegraphics[width=\textwidth]{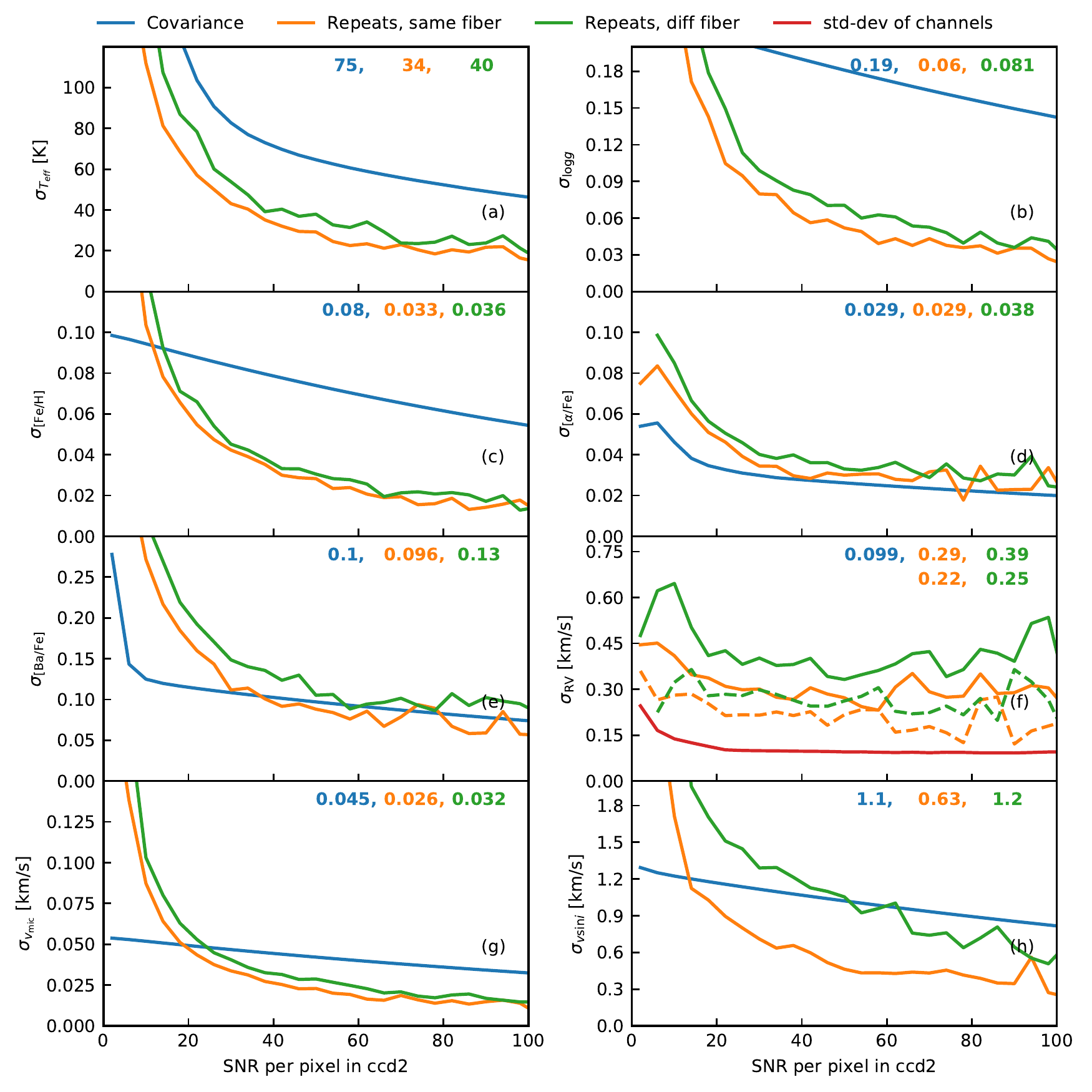}
  \caption{Uncertainty in spectroscopic stellar parameters, Ba abundance, and radial velocity as a function of SNR. The blue curve shows the uncertainty as predicted by \textit{The Cannon} (based on covariance errors and the SNR-dependent RMS within the leave-out-test). The other two curves are for uncertainties estimated using repeat observations, where the repeated stars are on the same fibre (orange curve) and different fibres (green curve). The numbers given in each panel are the uncertainties estimated from \textit{The Cannon} covariance and repeat observations for stars on the same and on different fibres at SNR=40, which is the median value of the SNR. For radial velocity, the solid lines are for \texttt{rv\_synt} while the dashed lines are for \texttt{rv\_obst}.}
  \label{fig:galah_errors_snr}
\end{figure*}

In \autoref{fig:galah_errors_snr} we make use of the repeats to estimate the uncertainty of the spectroscopic stellar parameters as a function of SNR. The uncertainty is estimated by computing, in each bin of SNR, the 16th and 84th percentile range of differences between repeats and dividing it by $2\sqrt{2}$.
The uncertainty estimated from repeats (green and orange curves) is compared with uncertainty estimated using \text{The Cannon} (blue curves) as described in Section~\ref{sec:analysis_step2}. The repeats confirm the overall trend of uncertainties with SNR, namely a strong degradation for SNR $<20$. Except for [$\alpha$/Fe], [Ba/Fe] and $v \sin i$, the covariance based method in general overestimates the uncertainty compared to the repeat-based method. Below SNR$=15$, for $\log g$, [Fe/H], $v_{\rm mic}$, and $v \sin i$, \textit{The Cannon} based method significantly underestimates the uncertainties as a result of the error-analysis described in Section~\ref{sec:cannon_errors}.

The estimated repeat-based uncertainties for RV and $v_{{\rm sin}i}$, and to a lesser degree $\log g$, differ when based on observations using different fibres as opposed to using the same fibre. This is because of the systematic differences in wavelength calibration and point spread function of each fibre. The uncertainties for other parameters are not affected by this, because a small change in point spread function does not directly translate into a variation of those parameters.

\subsection{Infrared flux method temperatures}\label{sec:irfm}

We apply the infrared flux method \citep[IRFM][]{Casagrande2010,Casagrande2014b} to over 280,000 GALAH stars having SkyMapper photometry \citep{Wolf2018}. Further details on the implementation of SkyMapper filters into the IRFM can be found in Casagrande et al. (in prep). Apart from a few pointings along the plane, nearly all of the SkyMapper targets overlapping with GALAH targets have Galactic latitudes above $|10^{\circ}|$, meaning that the most obscured and patchy regions of the Galactic plane are avoided. Yet, reddening can have a non-negligible contribution, which we account for with the same procedure used when implementing the IRFM for RAVE stars \citep{Kunder2017}. 

\begin{figure*}
  \includegraphics[width=\textwidth]{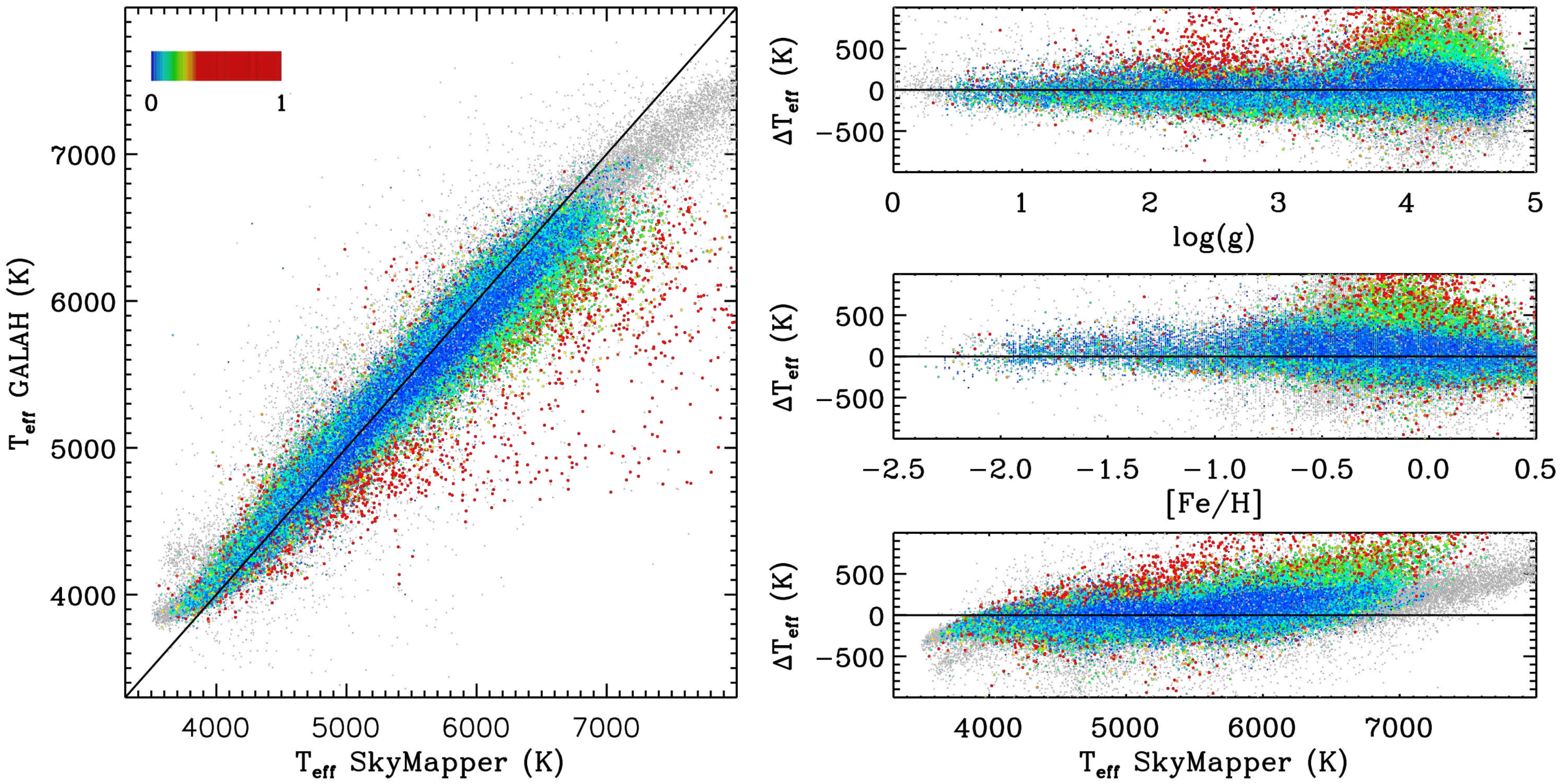}
  \caption{Comparison of GALAH effective temperatures $T_\mathrm{eff}$ with IRFM temperatures derived from SkyMapper photometry. Residuals (SkyMapper-GALAH) are shown as function of stellar parameters and colour coded by E(B-V) according to the scale on the top left. Grey points are stars flagged as unreliable in GALAH.}
  \label{fig:irfm}
\end{figure*}

Figure \ref{fig:irfm} shows the comparison between $T_\mathrm{eff}$ from the IRFM and GALAH, colour-coded by the adopted $E(B-V)$. For low reddening regions, the agreement is usually excellent across the entire stellar parameter range, although it deteriorates for regions of high extinction. Stars labelled as unreliable in GALAH ($\texttt{flag\_cannon} \neq 0$) are plotted in grey. Effectively all of the stars above $7000$~K are flagged in GALAH, because of the lack of stars in the training set in this regime, forcing the data-driven approach to extrapolate the determination of stellar parameters. The IRFM indicates that the GALAH pipeline underestimates effective temperatures in this regime, similar to what was seen from the \Gaia\ benchmark star comparison, saturating at $8000$~K which is the limit of the grid of model atmospheres used for the training set; we checked that this trend is not an artefact from stars affected by high values of extinction. 

After removing flagged stars, the SkyMapper$-$GALAH mean (median) $\Delta\TF$ is $61$~K ($49$~K) with a scatter of $183$~K when stars are considered irrespectively of their reddening. The above numbers reduce to $\Delta\TF = 51$~K ($50$~K) with a scatter of $132$~K when restricting to $E(B-V)<0.10$, and $\Delta\TF=12$~K ($12$~K) with a scatter of $123$~K for $E(B-V)<0.01$. Since we expect typical uncertainties of order 100~K for the IRFM in low extinction regions, the above scatters suggest that a slightly smaller uncertainty applies to the GALAH spectroscopic temperatures, in line with the conclusions discussed above.

\subsection{Asteroseismology}

Although GALAH does not overlap with the \textit{Kepler} field, there is a significant overlap with several observing campaigns of the K2 mission, in particular with the asteroseismic-based Galactic Archaeology Program (GAP) \citep{Stello2015}. The GAP has so far published seismic results for K2 Campaign 1 (C1) \citep{Stello2017}, and is in the process of releasing results for additional campaigns, including C4, C6, and C7 (Zinn et al., in preparation). Here, we use the results from the Bayesian Asteroseismology data Modeling Pipeline \citep[][see also Zinn et al., in preparation]{Stello2017} of stars in C1, C4, C6, and C7 to verify $\log g$ from \textit{The Cannon}.

\begin{figure}\includegraphics[width=\columnwidth]{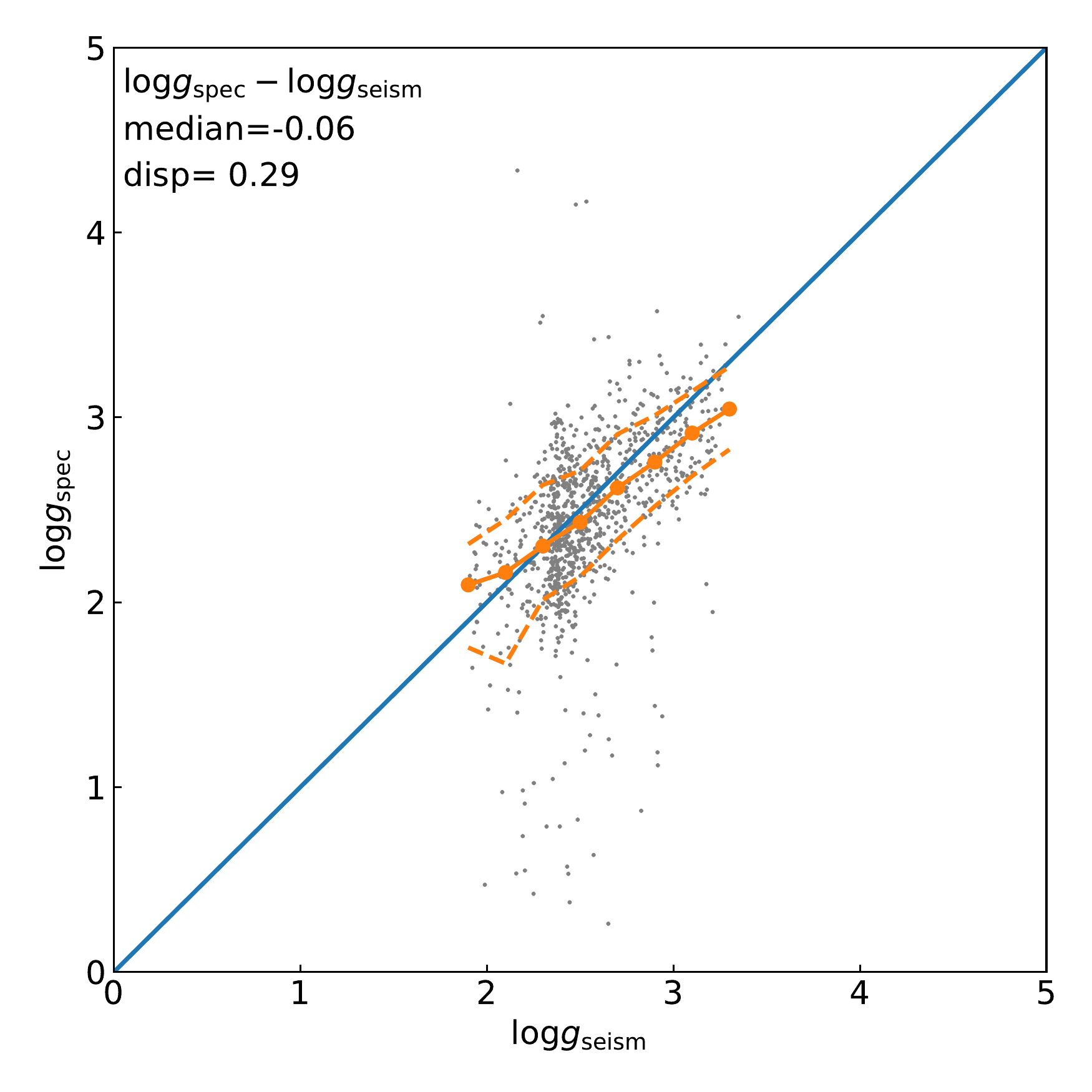}
  \caption{Comparison of GALAH surface gravity with asteroseismic surface gravity derived using data from the K2 mission (Campaign C1, C4, C6, and C7). Median and the confidence-interval-based dispersion of the difference between the two gravities are shown in the top left corner. The solid blue line shows the one-to-one relation. The solid orange line shows the 50th percentile, while the dotted lines show the 16th and 84th percentiles, derived in 0.2 dex wide bins in $\log g_\mathrm{seism}$. Stars shown are with  \texttt{flag\_cannon=0} and \texttt{snr\_c2>30}.}
  \label{fig:logg_seism_spec}
\end{figure}

In Figure~\ref{fig:logg_seism_spec} we show the comparison between the spectroscopic $\log g$ from GALAH and the seismic $\log g$ derived using Equation~\ref{eq.numax}, with data being from the K2-HERMES survey. The $\log g$  coverage of the seismic results is limited by the length ($\sim 80\,$days) and sampling ($\sim 30\,$min) of the K2 time series. To reach lower $\log g$ stars would require longer time series, while higher $\log g$ stars requires faster sampling. The dispersion seen in Figure~\ref{fig:logg_seism_spec} is dominated by the uncertainty in the spectroscopic values, which is evident from the narrow seismic $\log g$ range around 2.4 for the red clump stars, as expected from stellar evolution theory, compared to the larger spread in spectroscopic $\log g$ values .   Generally, we see a good agreement between the spectroscopic and seismic results, but with a slight bias for low luminosity red giant branch stars ($2.8<\log g<3.2$), where the spectroscopic values may be underestimated; see binned data in orange relative to the blue one-to-one line in Figure~\ref{fig:logg_seism_spec}. This bias is also observed for each campaign separately. We also note that the dispersion (dashed orange curves) of about 0.3 dex, is larger than would be expected either from the spectroscopic $\log g$ uncertainty estimated by the GALAH pipeline or from the scatter between repeat observations, which in both cases is below 0.2 dex for the majority of stars (Figure~\ref{fig:galah_errors_snr}). This may suggest that the asteroseismically inferred surface gravities also have unresolved issues. We note that stars with extreme differences between seismic and spectroscopic $\log g$ values can potentially be blends (in the K2 data) or mis-identifications.

With the TESS launch scheduled for April 2018, there will be further opportunity in the near future to train and test the GALAH $\log g$ values using seismology. In particular, stars in TESS's southern continuous viewing zone will be observed for up to one year, pushing the boundaries of the $\log g_\mathrm{seism}$ range GALAH can access.

\subsection{Globular and open clusters}

\begin{figure*}
  \includegraphics[width=\textwidth]{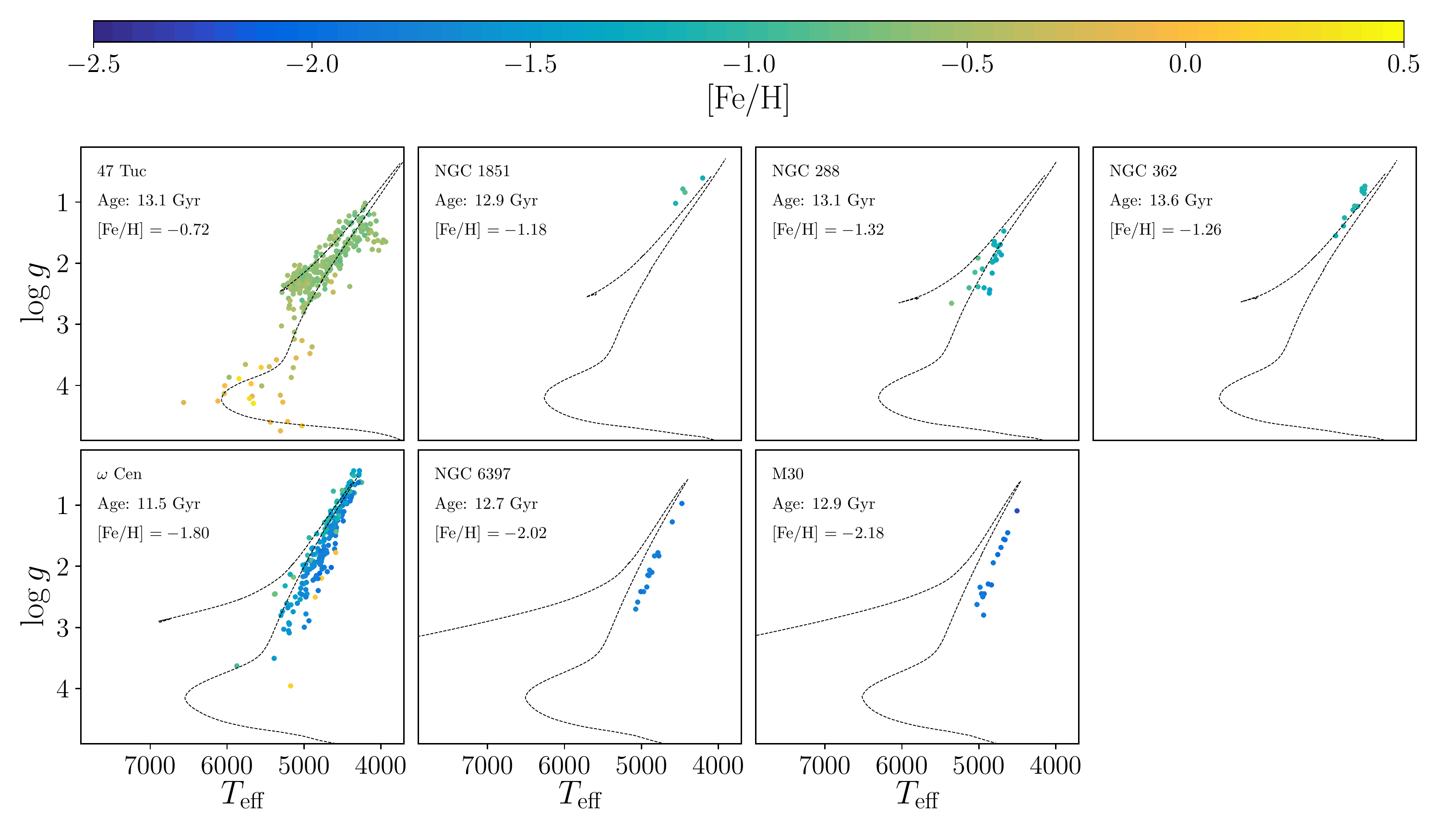}
  \caption{Kiel diagrams ($T_\text{eff}$ [K] vs. $\log g$ [dex]) of globular cluster stars observed by GALAH. Cluster members were identified from our own radial velocity measurements, proper motions from UCAC5 \citep{Zacharias2017}, and parameters from \citet{Kharchenko2013}. We stress that isochrones have not been fitted to the data. For each cluster, we plot PARSEC+COLIBRI isochrones \citep{Marigo2017} with ages from \citet{Forbes2010} and metallicities from \citet{Harris1996}.}
  \label{fig:gc_iso}
\end{figure*}

\begin{figure*}
  \includegraphics[width=\textwidth]{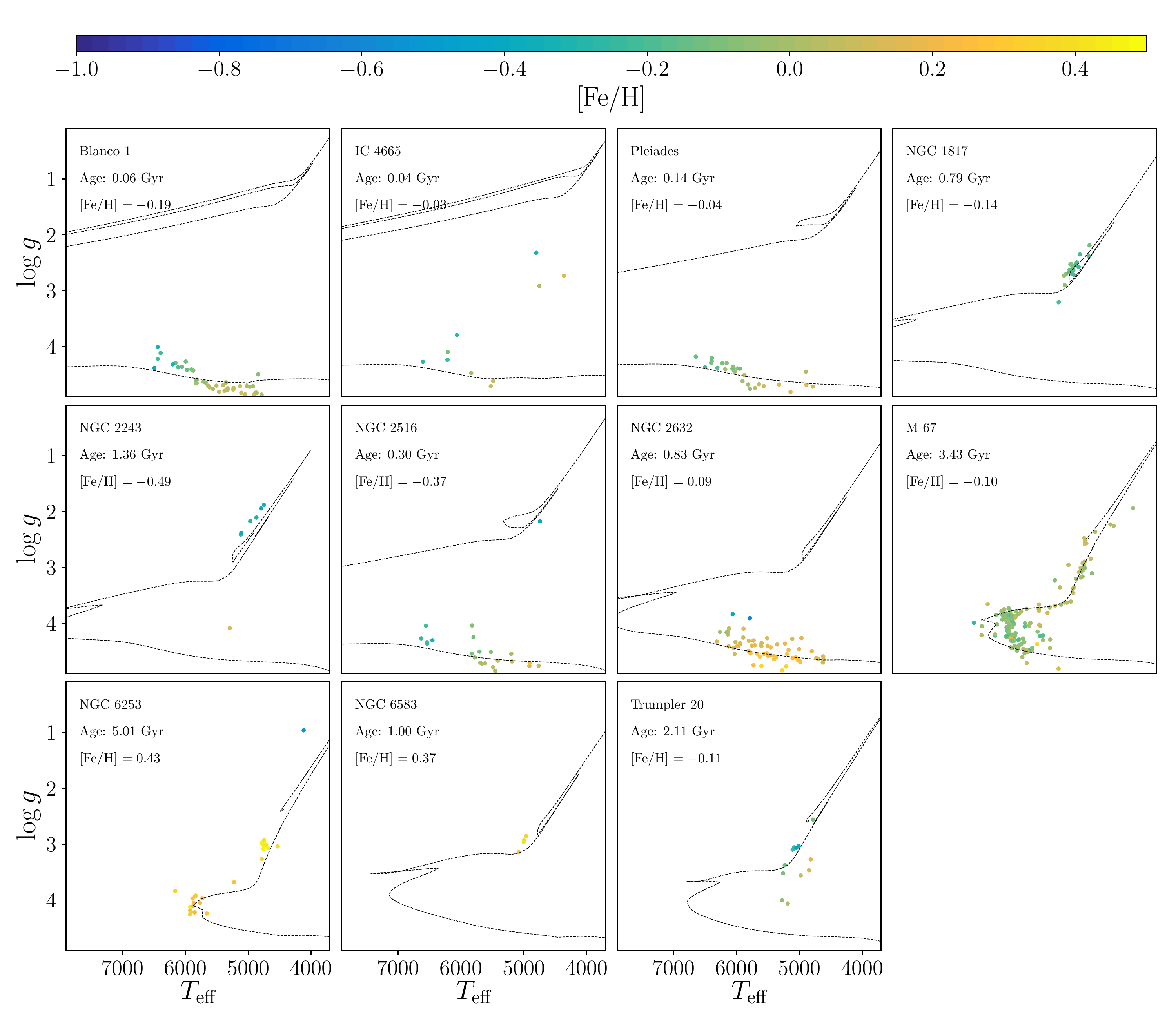}
  \caption{Kiel diagrams ($T_\text{eff}$ [K] vs. $\log g$ [dex]) of open cluster stars observed by GALAH. Cluster members were identified from our own radial velocity measurements, proper motions from UCAC5 \citep{Zacharias2017}, and parameters from \citet{Kharchenko2013}. We stress that isochrones have not been fitted to the data. For each cluster, we plot PARSEC+COLIBRI isochrones \citep{Marigo2017} with ages from and metallicities from \citet{Kharchenko2013}.}
  \label{fig:oc_iso}
\end{figure*}

\begin{figure}
  \includegraphics[width=\columnwidth]{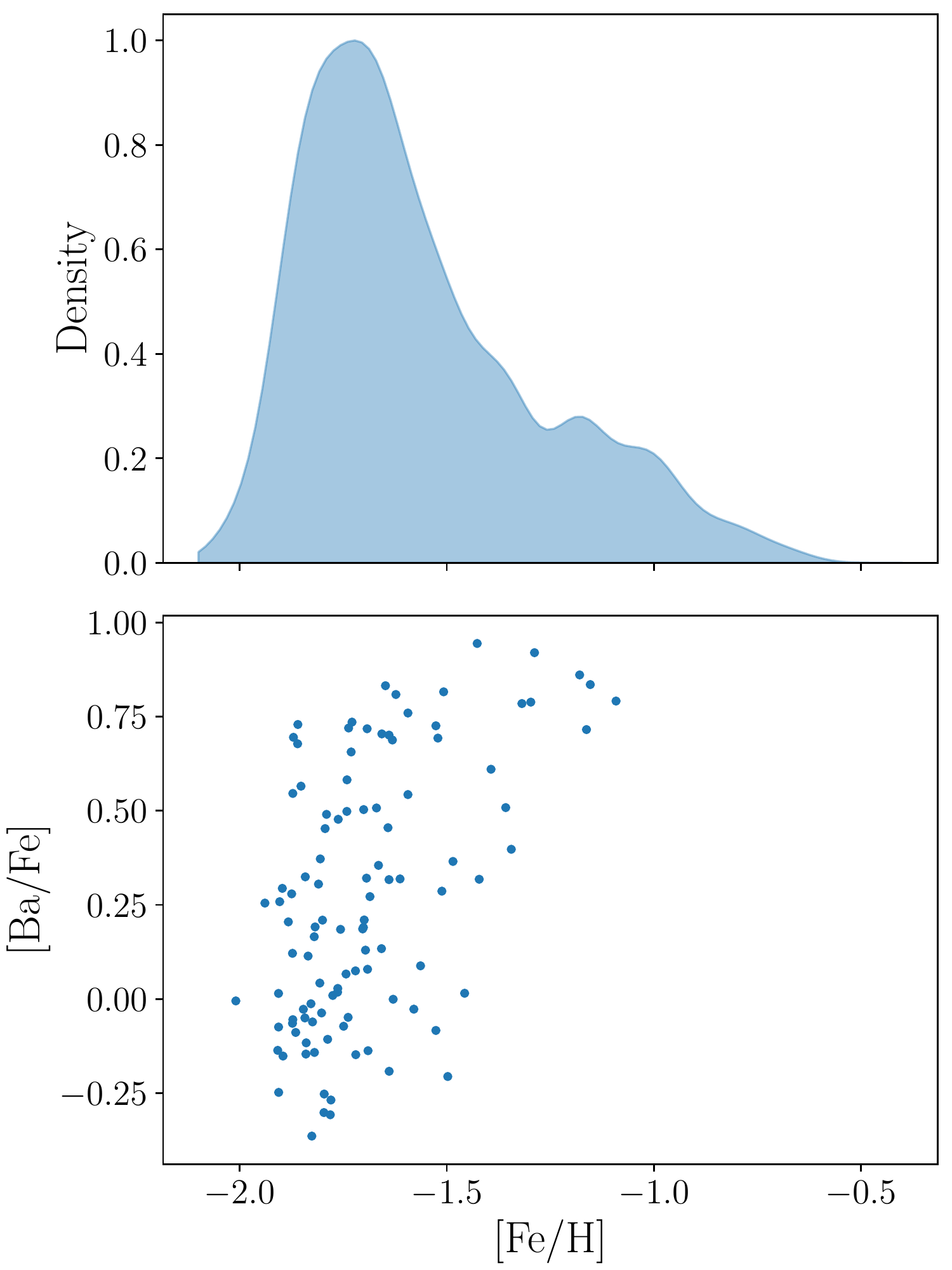}
  \caption{Top: Metallicity distribution of the members of $\omega$~Centauri observed. We find the main population at $\mathrm{[Fe/H]}\approx-1.8$, an intermediate metallicity population at about $\mathrm{[Fe/H]}\approx-1.4$, and a further population at $\mathrm{[Fe/H]}\approx-1.2$. There is a tail of stars that extends to $\mathrm{[Fe/H]}\approx-0.5$. Bottom: The distribution of [Fe/H] with [Ba/Fe] in $\omega$~Centauri. We find the expected rapid increase of s-process elemental abundances previously observed in $\omega$~Centauri.}
  \label{fig:omega_cen}
\end{figure}

Stellar clusters provide an excellent sample of stars with which to validate the GALAH results. Many clusters have been extensively studied \citep[e.g., see the review of globular cluster abundances by][]{Bastian2017}, and they span a large range of metallicities, ages, and other parameters. GALAH has targeted several globular and open clusters for validation purposes during its pilot observing campaign, and several more fell within our survey fields and members were serendipitously observed. In this section we discuss clusters that were not only observed by the GALAH survey, but also by related surveys: TESS-HERMES \citep{Sharma2018}, K2-HERMES \citep{Wittenmyer2018} and the HERMES Open Cluster Project (De Silva et al., in preparation). As such, some of the clusters and members discussed in this work are not available in GALAH DR2.

Cluster members were identified by cuts in radial velocity around the literature values \citep{Kharchenko2013} and proper motions from UCAC5 \citep{Zacharias2017}. Only members within the cluster radius given in \citet{Kharchenko2013} were considered.  We found reliable members of 7 globular and 11 open clusters. In Figure \ref{fig:gc_iso} we show the Kiel diagrams for the globular clusters and Figure \ref{fig:oc_iso} is the same for the open clusters, overplotted with PARSEC+COLIBRI isochrones \citep{Marigo2017}. The isochrones have not been fitted to the data; rather, we are simply using the literature values for each cluster. For the globular clusters we use metallicities from \citet{Harris1996} and ages from \citet{Forbes2010}. For the open clusters we use the values in \citet{Kharchenko2013}. We note that these references are compilations of literature data.

As would be expected for the magnitude range, the globular cluster members have been recovered as giant stars. They are usually found on narrow sequences, though these are sometimes offset from the isochrone. Most of the open clusters are young and most of the observed targets are main sequence stars. For the youngest clusters (e.g., Blanco 1 and Pleiades) there is evidence for a temperature dependence of their metallicities due to fast rotation causing a degeneracy within the spectra and hence systematics in the stellar parameters. We stress that these also translate into abundance trends we see.

In agreement with many previous studies \citep[e.g.,][]{Johnson2010}, we find that a large metallicity spread in $\omega$~Cen: $-2.0<\mathrm{[Fe/H]}<-0.5$ (top panel of Figure \ref{fig:omega_cen}) with four peaks in the metallicity distribution. We also confirm the distinctive distribution of s-process element abundance with metallicity, here demonstrated with [Ba/Fe] (bottom panel of Figure \ref{fig:omega_cen}). [Ba/Fe]  increases with metallicity until $\mathrm{[Fe/H]}\approx-1.5$ and appears to become roughly constant thereafter.

Open clusters are particularly valuable validation objects for our derived element abundances, because we expect stars from the same open cluster at the same evolutionary phase to show intrinsically very similar abundance patterns and they have metallicities similar to the majority of the GALAH main sample. As a result, these clusters are ideal benchmarks to validate abundance determinations.

\begin{figure*}
  \includegraphics[width=\textwidth]{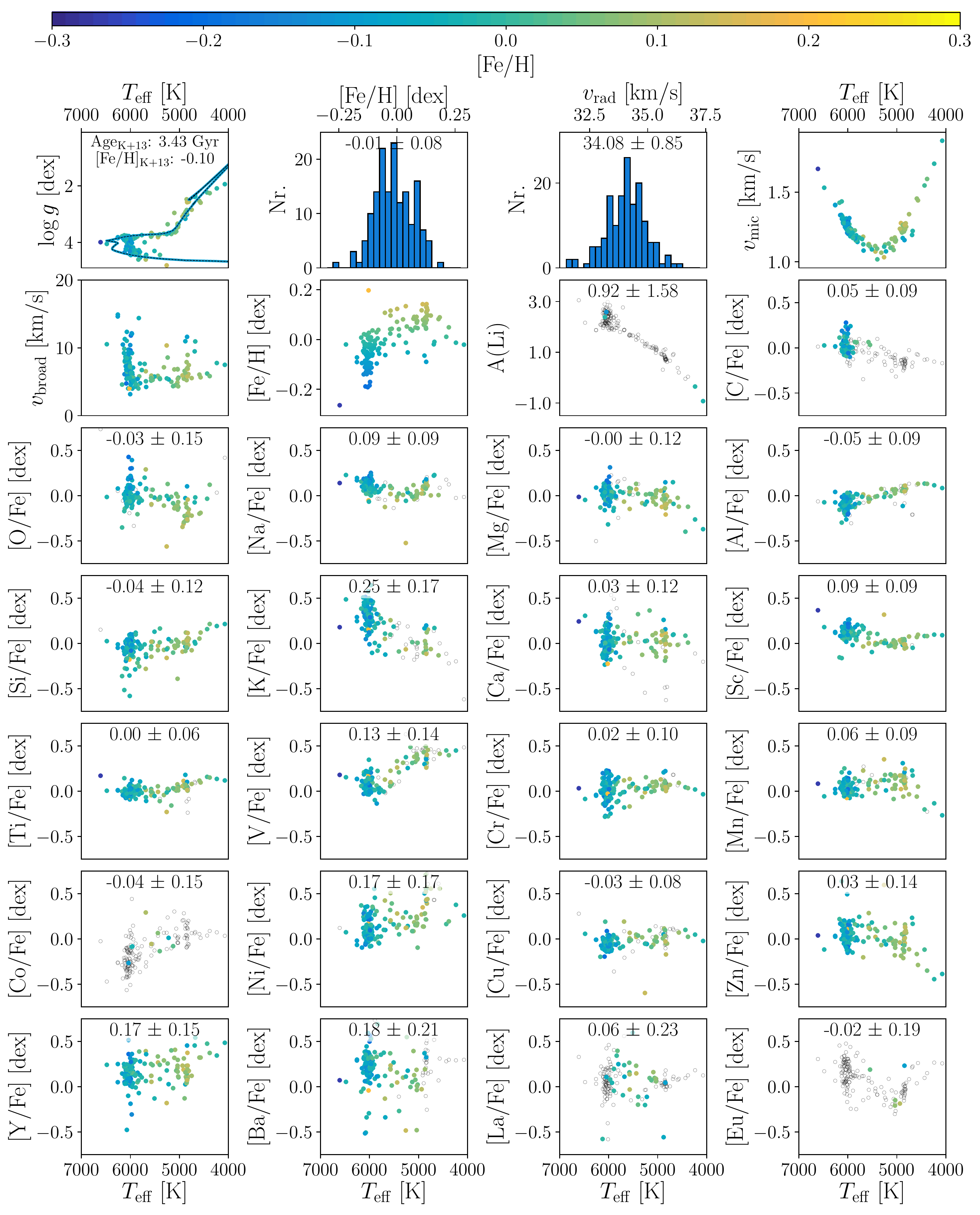}
  \caption{Overview of stellar parameters and element abundances for the open cluster M67. The top left panel shows the M67 Kiel diagram, followed horizontally by the metallicity and radial velocity distributions. The top right panel show the microturbulence velocity $v_\text{mic}$ as a function of effective temperature $T_\text{eff}$. The other panels show two more stellar parameters ($v_\text{broad}$ and [Fe/H]) as well as element abundances. Colour indicates the iron abundances and black circles indicate flagged measurements. Text annotates mean values and dispersion, which can also be found in Table~\ref{tab:m67_abundances} together with individual lists for dwarfs and giants.}
  \label{fig:m67}
\end{figure*}

For this data release, we are concentrating on the open cluster M67, for which we have observed turn-off, subgiant, red giant branch, and red clump stars with a mean iron abundance of $\mathrm{[Fe/H]} = -0.01\pm0.08$ and a radial velocity of $34.08\pm0.85\,\mathrm{km\, s^{-1}}$. We stress that the measured iron abundance is different from the value by \citet{Kharchenko2013} used in Figure~\ref{fig:oc_iso}. In addition to the Kiel diagram of the cluster members and histograms of both iron abundance and radial velocity, we show stellar parameters and element abundances as a function of effective temperature $T_\text{eff}$ in Figure~\ref{fig:m67}. We note that the turn-off stars cover a larger range of broadening velocities, whereas the evolved stars are usually slow rotators. The larger broadening is coincident with lower iron abundance estimates of the pipeline and also manifested in the iron abundance distribution, showing an increase of iron abundance towards lower temperatures and lower broadening. 

When we look at the difference between the abundances of dwarfs (defined as $T_\text{eff} > 5500\,\mathrm{K}$ or $\log g > 4.0$) and giants ($T_\text{eff} \leq 5500\,\mathrm{K}$ and $\log g \leq 4.0$) as part of the GALAH observations of M67, we see trends with effective temperature (above $1\sigma$ difference) for Al, K, Sc, and V in addition to Fe. We see no significant trends with $T_\text{eff}$ (less than $1\sigma$ difference) for O, Ti, Ni, Zn. No trends (less than $0.5\sigma$ difference) can be seen for Na, Mg, Si, Ca, Cr, Mn, Cu, Y, Ba, La. For the elements Li, C, Co, and Eu we can not assess the trends due to the low number of stars in either category. None of these trends with temperature are corrected for and hence increase the overall abundance scatter within the cluster (see Table~\ref{tab:m67_abundances}). We stress that some of these trends have also been found in other studies. For O and Al similar trends have been found by \citet{Gao2018}, when analysing higher resolution data with a very similar analysis. For K, the trend with temperature is expected because of strong non-LTE effects. As outlined in Section~\ref{sec:sme_abundances}, the lines of some elements are not strong enough in dwarf or giant spectra within the GALAH range to be detected precisely or at all. When assessing the scatter within dwarfs and giants of the cluster separately, we hence find a lower scatter for several elements, which are a more appropriate measure of our internal precision. These range from the highest precisions of $0.04-0.08\,\mathrm{dex}$ (Fe, Al, Sc, Ti, V, and Cu) over high precision of $0.08-0.12\,\mathrm{dex}$ (C, Na, Si, Cr, and Mn) and intermediate precision of $0.12-0.16\,\mathrm{dex}$ (O, Mg, K, Ca, Co, Ni, Zn, and Y) to low precision above $0.16\,\mathrm{dex}$ (Li, Ba, La, Eu) for the stars observed in M67.

\begin{table*}
 \caption{Abundances of the open cluster M67. For this compilation, we define dwarfs as stars with $T_\text{eff} > 5500\,\mathrm{K}$ or $\log g > 4.0$, and giants as stars with $T_\text{eff} \leq 5500\,\mathrm{K}$ and $\log g \leq 4.0$.} \label{tab:m67_abundances}
 \begin{tabular}{ccccccc}
  \hline
Abundance	& Nr. stars 	&  Mean & Nr. dwarfs & Mean & Nr. Giants  & Mean \\
\hline
$\mathrm{[Fe/H]}$ & 156 & $-0.01 \pm 0.08$ & 107 & $-0.05 \pm 0.07$ & 49 & $~~0.06 \pm 0.05$ \\
$\mathrm{[Li/Fe]}$ & 4 & $~~0.92 \pm 1.58$ & 2 & $~~2.48 \pm 0.10$ & 2 & $-0.64 \pm 0.29$ \\
$\mathrm{[C/Fe]}$ & 49 & $~~0.05 \pm 0.09$ & 49 & $~~0.05 \pm 0.09$ & 0 & $~~~-$ \\
$\mathrm{[O/Fe]}$ & 131 & $-0.03 \pm 0.15$ & 84 & $~~0.02 \pm 0.14$ & 47 & $-0.13 \pm 0.12$ \\
$\mathrm{[Na/Fe]}$ & 134 & $~~0.09 \pm 0.09$ & 96 & $~~0.10 \pm 0.09$ & 38 & $~~0.06 \pm 0.07$ \\
$\mathrm{[Mg/Fe]}$ & 143 & $-0.00 \pm 0.12$ & 94 & $~~0.02 \pm 0.12$ & 49 & $-0.05 \pm 0.11$ \\
$\mathrm{[Al/Fe]}$ & 122 & $-0.05 \pm 0.09$ & 93 & $-0.08 \pm 0.07$ & 29 & $~~0.05 \pm 0.05$ \\
$\mathrm{[Si/Fe]}$ & 155 & $-0.04 \pm 0.12$ & 106 & $-0.06 \pm 0.12$ & 49 & $~~0.01 \pm 0.09$ \\
$\mathrm{[K/Fe]}$ & 113 & $~~0.25 \pm 0.17$ & 97 & $~~0.28 \pm 0.15$ & 16 & $~~0.01 \pm 0.11$ \\
$\mathrm{[Ca/Fe]}$ & 137 & $~~0.03 \pm 0.12$ & 96 & $~~0.03 \pm 0.12$ & 41 & $~~0.04 \pm 0.12$ \\
$\mathrm{[Sc/Fe]}$ & 153 & $~~0.09 \pm 0.09$ & 107 & $~~0.12 \pm 0.08$ & 46 & $~~0.01 \pm 0.04$ \\
$\mathrm{[Ti/Fe]}$ & 142 & $~~0.00 \pm 0.06$ & 106 & $-0.01 \pm 0.05$ & 36 & $~~0.05 \pm 0.07$ \\
$\mathrm{[V/Fe]}$ & 129 & $~~0.13 \pm 0.14$ & 101 & $~~0.07 \pm 0.08$ & 28 & $~~0.37 \pm 0.07$ \\
$\mathrm{[Cr/Fe]}$ & 144 & $~~0.02 \pm 0.10$ & 107 & $~~0.01 \pm 0.11$ & 37 & $~~0.06 \pm 0.06$ \\
$\mathrm{[Mn/Fe]}$ & 141 & $~~0.06 \pm 0.09$ & 95 & $~~0.05 \pm 0.07$ & 46 & $~~0.08 \pm 0.11$ \\
$\mathrm{[Co/Fe]}$ & 9 & $-0.04 \pm 0.15$ & 5 & $-0.04 \pm 0.18$ & 4 & $-0.03 \pm 0.08$ \\
$\mathrm{[Ni/Fe]}$ & 147 & $~~0.17 \pm 0.17$ & 106 & $~~0.13 \pm 0.14$ & 41 & $~~0.29 \pm 0.17$ \\
$\mathrm{[Cu/Fe]}$ & 121 & $-0.03 \pm 0.08$ & 94 & $-0.05 \pm 0.09$ & 27 & $~~0.01 \pm 0.06$ \\
$\mathrm{[Zn/Fe]}$ & 154 & $~~0.03 \pm 0.14$ & 105 & $~~0.06 \pm 0.13$ & 49 & $-0.04 \pm 0.14$ \\
$\mathrm{[Y/Fe]}$ & 156 & $~~0.17 \pm 0.15$ & 107 & $~~0.14 \pm 0.15$ & 49 & $~~0.22 \pm 0.14$ \\
$\mathrm{[Ba/Fe]}$ & 122 & $~~0.18 \pm 0.21$ & 102 & $~~0.19 \pm 0.21$ & 20 & $~~0.12 \pm 0.20$ \\
$\mathrm{[La/Fe]}$ & 31 & $~~0.06 \pm 0.23$ & 18 & $~~0.11 \pm 0.24$ & 13 & $-0.01 \pm 0.20$ \\
$\mathrm{[Eu/Fe]}$ & 5 & $-0.02 \pm 0.19$ & 1 & $0.19~~~~~~~~~~$ & 4 & $-0.08 \pm 0.18$ \\
\hline
 \end{tabular}
\end{table*}

\subsection{Comparison with other studies and surveys}

Because of the variety of high-resolution, high-quality, and large-scale spectroscopic studies and surveys to compare with, we have decided to only choose two homogeneously analysed samples for comparison in this study:
\citet{Bensby2014} and its follow-up studies \citep{Battistini2015, Battistini2016, Bensby2018} for dwarf stars and APOGEE DR14 \citep{SDSSDR14} for giants. While we only use those stars with $\texttt{flag\_cannon} = 0$ and $\texttt{flag\_X\_Fe} = 0$ (see Figures~\ref{fig:Dwarf_Giant_comparison1_flag0}-\ref{fig:Dwarf_Giant_comparison3_flag0}), we append the same comparisons including extrapolated abundances ($\texttt{flag\_X\_Fe} \leq 1$) in Section~\ref{sec:behind_the_flags}. 

\subsubsection{GALAH DR2 vs. 714 dwarf spectra from \citet{Bensby2014}}

In this section, we compare the general trends of the stars with similar parameters ($T_\text{eff} > 5500\,\mathrm{K}$ or $\log g > 3.3$) to the stars analysed by \citet{Bensby2014}. Their first study included the elements O, Na, Mg, Al, Si, Ca, Ti, Cr, Ni, Zn, Y, and Ba. \citet{Battistini2015} added a similar, homogeneous follow-up analysis of Sc, V, Mn, and Co. The follow-up study of neutron-capture elements by \citet{Battistini2016} includes the elements La and Eu, while \citet{Bensby2018} have published Li measurements for the 714 dwarfs. We here abbreviate these studies together as "Bensby studies". Similar to the study by \cite{Buder2018a}, the majority of dwarfs in GALAH DR2 covers metallicity / iron abundance ranges of $-0.7 < \mathrm{[Fe/H]} < 0.5$. We hence refer the reader to the detailed discussion by \citet{Buder2018a} and only briefly discuss the trends we see for each element represented in the Bensby studies. From Figures~\ref{fig:Dwarf_Giant_comparison1_flag0}, \ref{fig:Dwarf_Giant_comparison2_flag0}, and \ref{fig:Dwarf_Giant_comparison3_flag0} we see:
\begin{itemize}
\item For several element abundances (O, Si, Ca, Ti, Mn, and Zn), measured in the Bensby studies and our Data Release 2, we find a strong agreement of the trends, manifested in the overlap of the majority of our stars (with highest density in the yellow regions in these figures) with the high-quality data points of the Bensby studies. 
\item For Na, Al, Sc, Cr, Ni, Y, and Ba we see good agreement of the general trends, but minor shifts of the mean abundance around iron abundances of $-0.2\,\mathrm{dex}$, which might originate in zero-point offsets or target selection differences. For Y and Ba, the different mean abundances might be a manifestation of the different ages and hence different s-process enhancements.
\item In GALAH we see a rather flat but not decreasing Mg abundance in the super-solar regime. Contrary to \citet{Bensby2014}, this trend is also found by APOGEE (see neighboring panel in Figure~\ref{fig:Dwarf_Giant_comparison1_flag0}).
\item Because of the detection limits and flagging algorithm, we can not measure a large number of Li, C, Co, La, and Eu abundances. 
\item  In the case of Li, we can usually only estimate the abundance for Li-rich stars. However, we have also been able to identify numerous metal-poor dwarfs following the Spite plateau \cite{Spite1982} and giants with lower Li around the solar photospheric value, which show strong enough lines due to their cool atmospheres \citep{Buder2018a}. For Co and Eu, the metal-rich end of GALAH overlaps with stars of the Bensby studies. For La, however, we see a disagreement. Zero-point offsets of around $0.25\,\mathrm{dex}$ would however lead to a agreement for dwarfs (but La under-abundances in giants). Additionally, we can not estimate C abundances in almost any giants (due to the highly excited line we can use in the GALAH range).
\end{itemize}

\subsubsection{GALAH DR2 vs. APOGEE DR14}

In the right hand panels of Figures~\ref{fig:Dwarf_Giant_comparison1_flag0} to \ref{fig:Dwarf_Giant_comparison3_flag0}, we compare GALAH DR2 giants with those of APOGEE. The GALAH DR2 giants and their abundances are based on the selection of stars with $T_\text{eff} < 5500\,\mathrm{K}$ and $\log g < 4$, covering the majority of stars from the calibrated APOGEE sample. While we can see a good agreement of the APOGEE-based contours along the red giant branch, we note that the red clump region of APOGEE is more elongated and GALAH DR2 shows a rather local overdensity. We note stars identified as red clump stars have been shifted based on calibrations with asteroseismic values. For GALAH DR2, we include a large number of stars with asteroseismic surface gravities, but to span the parameter space in this region, we have to also include stars without this precise gravity information. We also want to stress, that the parameter space for which the GALAH pipeline is optimised, does not include the very luminous giants and therefore we can not reliably provide stellar parameters for stars below $4000\,\mathrm{K}$.

\begin{itemize}
\item When comparing our results with APOGEE DR14, we see good agreement of the abundance trends for Mg, Al, and Mn.
\item The [O/Fe] vs. [Fe/H] trends exhibits a steeper slope for giants in GALAH than for dwarfs or for giant stars in APOGEE DR14. Although the calculation of the abundance includes departures from LTE, this disagreement could be caused by the 1D model atmosphere (i.e., neglecting 3D effects in the analysis). The missing flattening of [O/Fe] for the metal-rich regime is however consistent with the dwarf studies and a validation to not treat oxygen as a regular $\alpha$-element.
\item Na agrees for the highest density of stars, but in the GALAH giant abundances, the upturn of [Na/Fe] in the metal-rich regime is not seen, contrary to the results in dwarfs and APOGEE.
\item For Si in giants, we note a disagreement at the metal-rich regime ($\mathrm{[Fe/H]} > 0$) where we measure an increase of [Si/Fe] contrary to measurements for dwarfs and APOGEE DR14. We note however a significant bimodality in [Si/Fe] as one would expect from an overlap of thin and thick disc stars. 
\item K is affected by non-LTE effects in giants as well as by interstellar absorption, which makes the comparison with APOGEE more difficult.
\item For [Ca/Fe] there is a zero point difference but otherwise a good agreement for this element.
\item We note strong disagreement for Ti, where GALAH follow the expected $\alpha$-element behaviour but APOGEE does not; for further discussions see e.g. \cite{SDSSDR13,Hawkins2016b}. 
\item For V we see a significant disagreement with both the dwarf abundances and APOGEE DR14. We report this element anyway, because of the useful abundances in dwarfs, but advise not to use of [V/Fe] in giants.
\item For Cr, we see a good agreement, but the Cr abundance of APOGEE DR14 giants follows Fe even closer than for GALAH DR2 stars.
\item Even though we have not been able to measure many Co abundances due to weak and blended lines, the number densities of the two surveys seem to be consistent. 
\item Our Ni trend for giants with $\mathrm{[Fe/H]} > -0.5$ is higher than the one seen APOGEE DR14, where Ni tracks Fe closely, as expected on nucleosynthetic grounds. We note that our dwarf-based [Ni/Fe] results are significantly better. 
\item In addition, in this data release we include some elements not covered by APOGEE DR14, including Sc, Cu, Zn, Y, Ba, La, and Eu. In general the GALAH results for giants and dwarfs are in reasonably good agreement but with giants we can trace those abundance trends to lower metallicities. 
\end{itemize}

\begin{figure*}
  \includegraphics[width=0.975\textwidth]{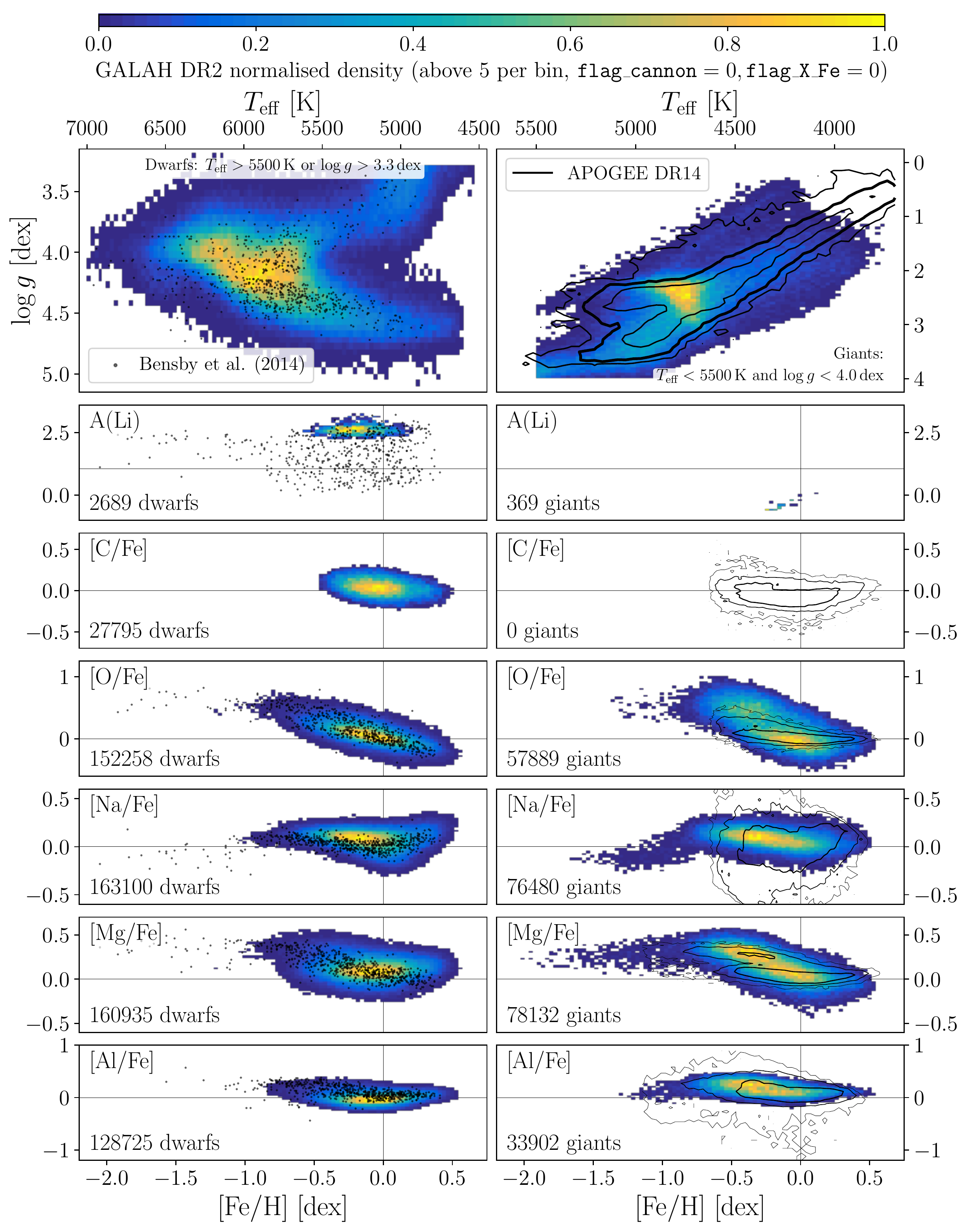}
  \caption{Comparison of the Kiel diagrams (top panel) and individual abundances (Li through Al) for the sample of 714 dwarfs \citep{Bensby2014, Battistini2015, Battistini2016, Bensby2018} on the left hand side as well as APOGEE DR14 giants \citep{SDSSDR14} on the right hand side. GALAH DR2 data (with $\texttt{flag\_cannon} = 0$ for stellar parameters and $\texttt{flag\_X\_fe} = 0$ for the respective element X) are plotted as colored density with a minimum of 5 stars per bin. The literature values for dwarfs are overplotted as black dots, while the APOGEE giants (with finite values, $\texttt{ASPCAPFLAG} = 0$, and $\texttt{STARFLAG} = 0$ for stellar parameters as well as $\texttt{X\_FE\_FLAG} = 0$ for the respective element X) are shown as contours.}
  \label{fig:Dwarf_Giant_comparison1_flag0}
\end{figure*}

\begin{figure*}
  \includegraphics[width=0.975\textwidth]{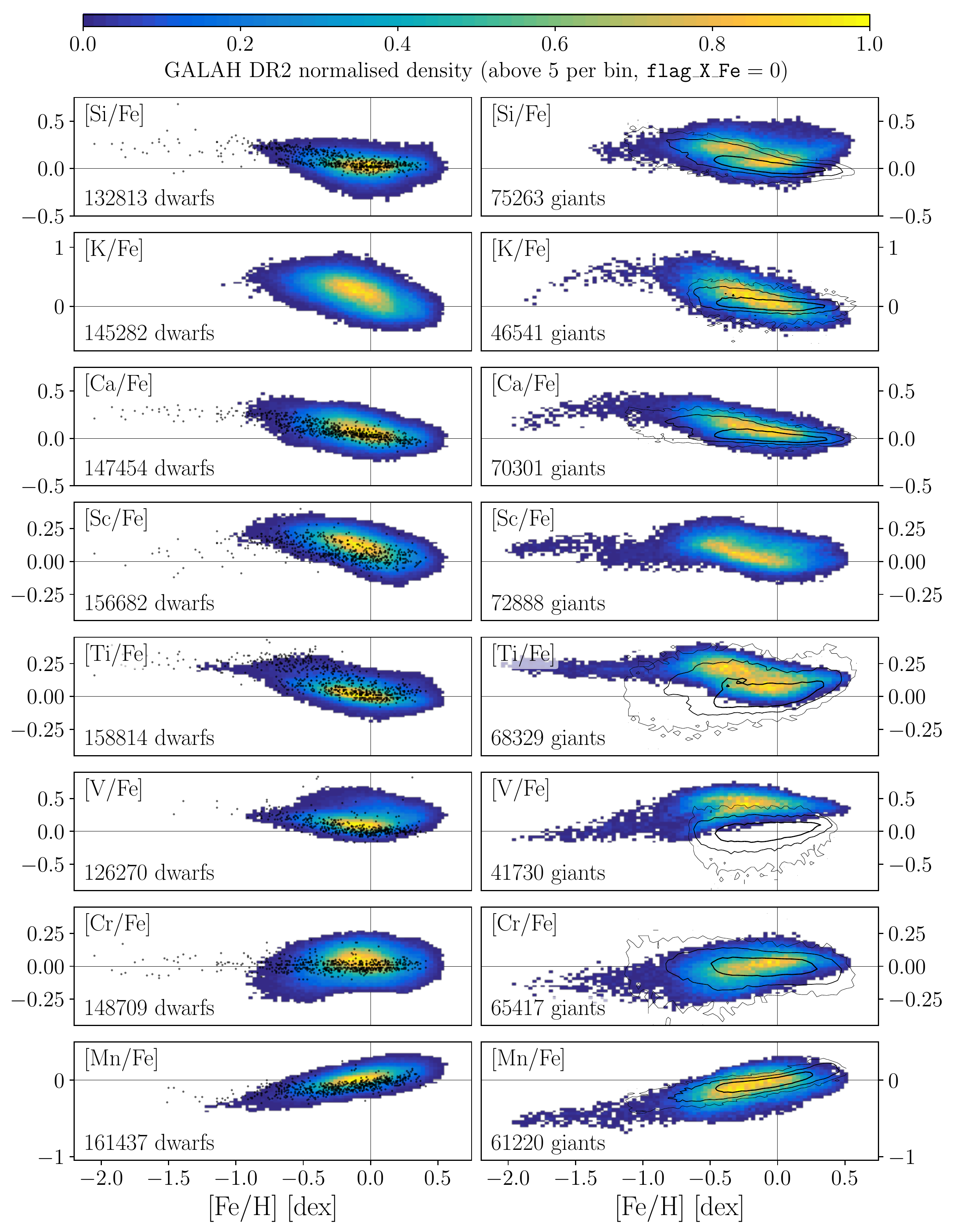}
\caption{Continuation of Figure~\ref{fig:Dwarf_Giant_comparison1_flag0} for elements Si through Mn.}
  \label{fig:Dwarf_Giant_comparison2_flag0}
\end{figure*}

\begin{figure*}
  \includegraphics[width=0.975\textwidth]{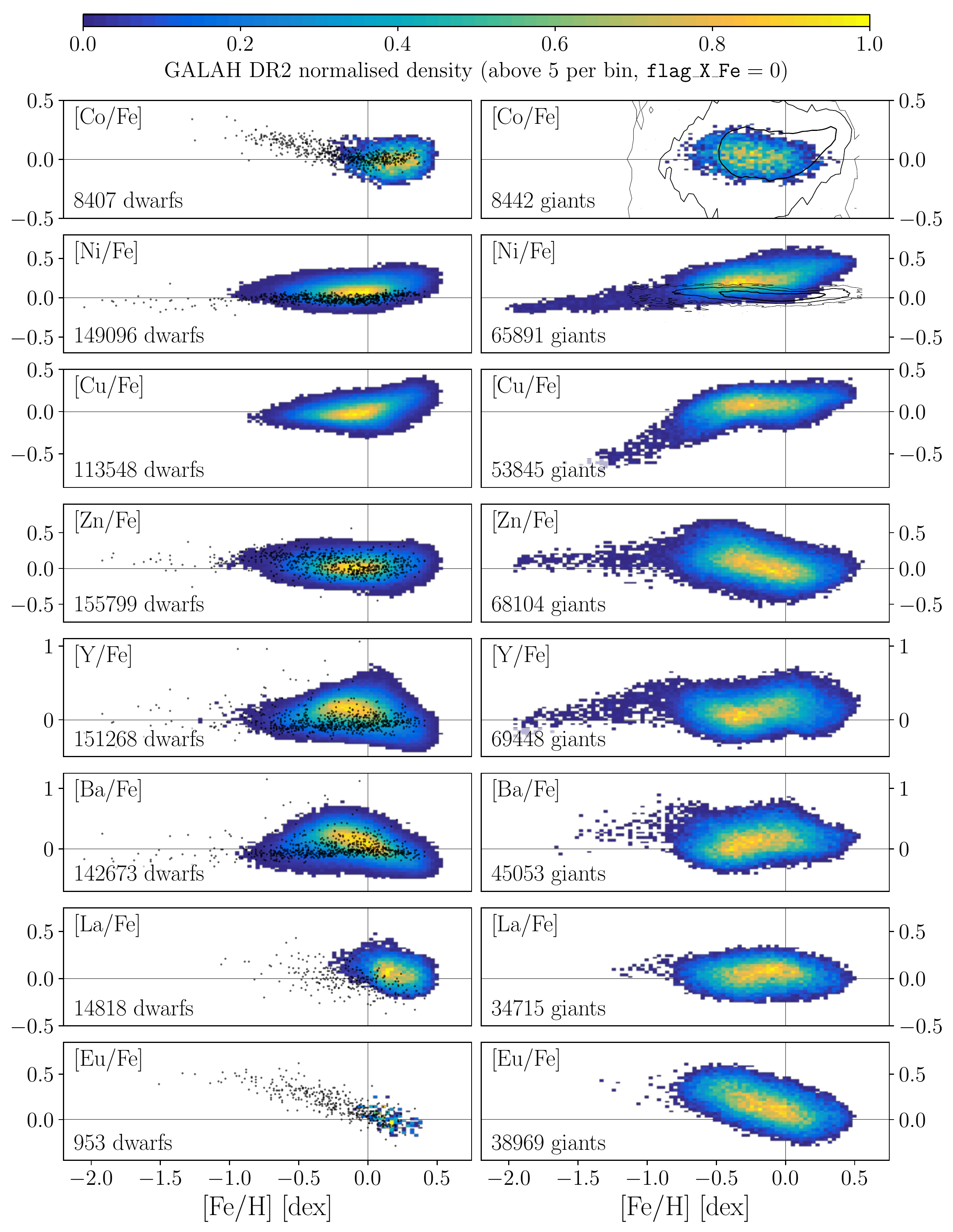}
\caption{Continuation of Figures~\ref{fig:Dwarf_Giant_comparison1_flag0} and \ref{fig:Dwarf_Giant_comparison2_flag0} for elements Co through Eu.}
    \label{fig:Dwarf_Giant_comparison3_flag0}
\end{figure*}

\subsection{Systematic trends in the parameter space}
\label{sec:behind_the_flags}

\begin{figure*}
\includegraphics[width=\textwidth]{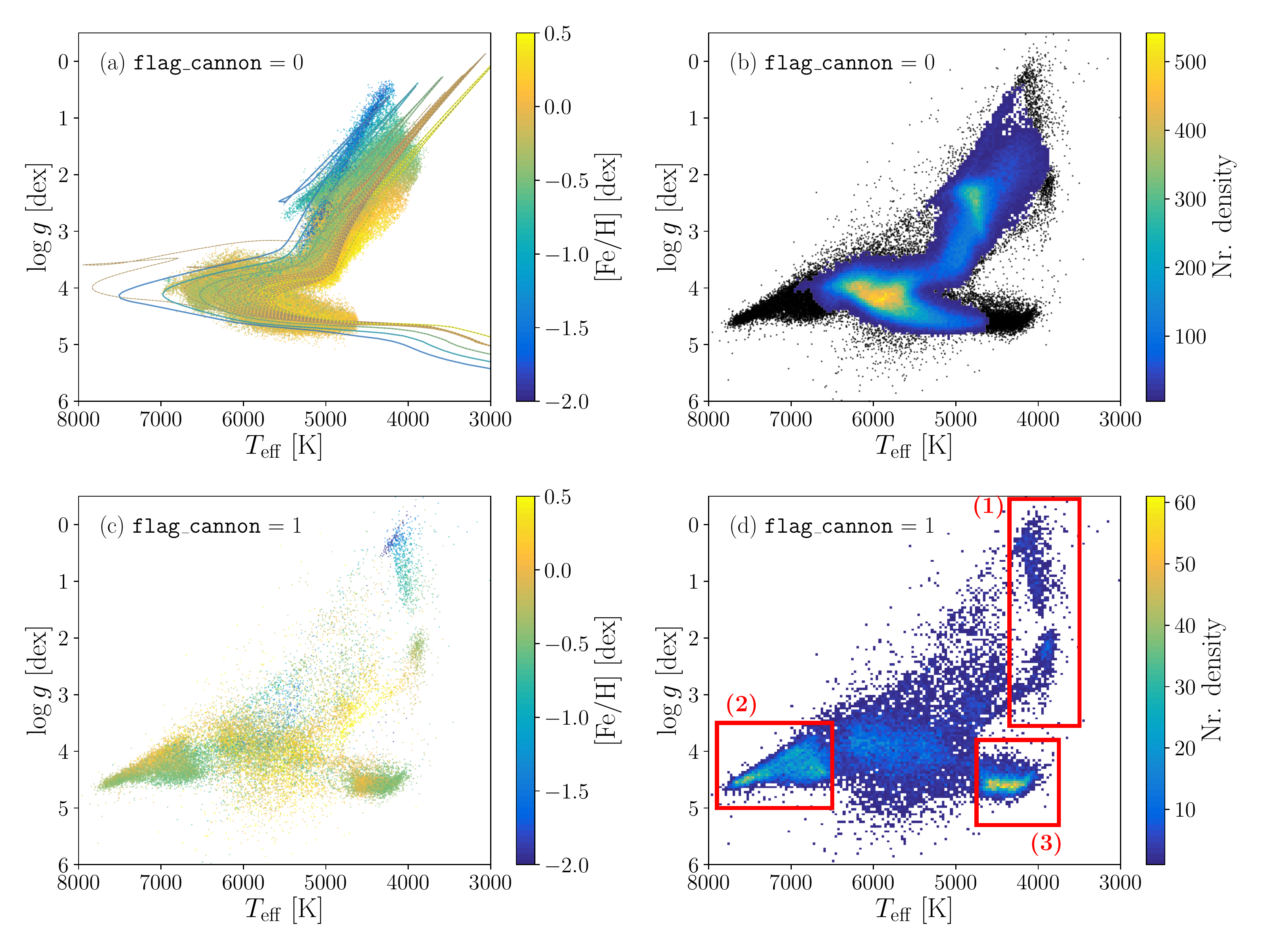}
\caption{Kiel diagrams of the GALAH data release 2. (a) stars with $\texttt{flag\_cannon} = 0$ colored by their iron abundance, (b) stars with $\texttt{flag\_cannon} = 0$ as density plot with stars with $\texttt{flag\_cannon} = 1$ as black dots in the background for perspective, (c) stars with $\texttt{flag\_cannon} = 1$ colored by their iron abundance, (d) stars with $\texttt{flag\_cannon} = 1$ as density plot.}
\label{fig:kiel_flags}
\end{figure*}

The GALAH survey data covers a large range of stars with different stellar parameters and includes spectra with peculiarities. Although the majority of stars can be described with one set of stellar labels (see the unflagged stars of this data release shown in Figure~\ref{fig:kiel_flags}), we face many challenges in the analysis, which can compromise these labels. The identification of the influence of shortcomings of our analysis is complex and an ongoing process. We want to stress that our pipeline is only tailored for non-peculiar spectra of stars of spectral type F, G and K, which consist the vast majority of the survey targets. We are also mindful that a quadratic model might not describe all spectra perfectly within the parameter space and imperfect training labels can introduce systematic trends in the final results. We hence identify and point out several systematic trends in the parameter space of this data release, which may contribute to these shortcomings of the analysis:

\begin{landscape}
\begin{figure}
\includegraphics[width=\columnwidth]{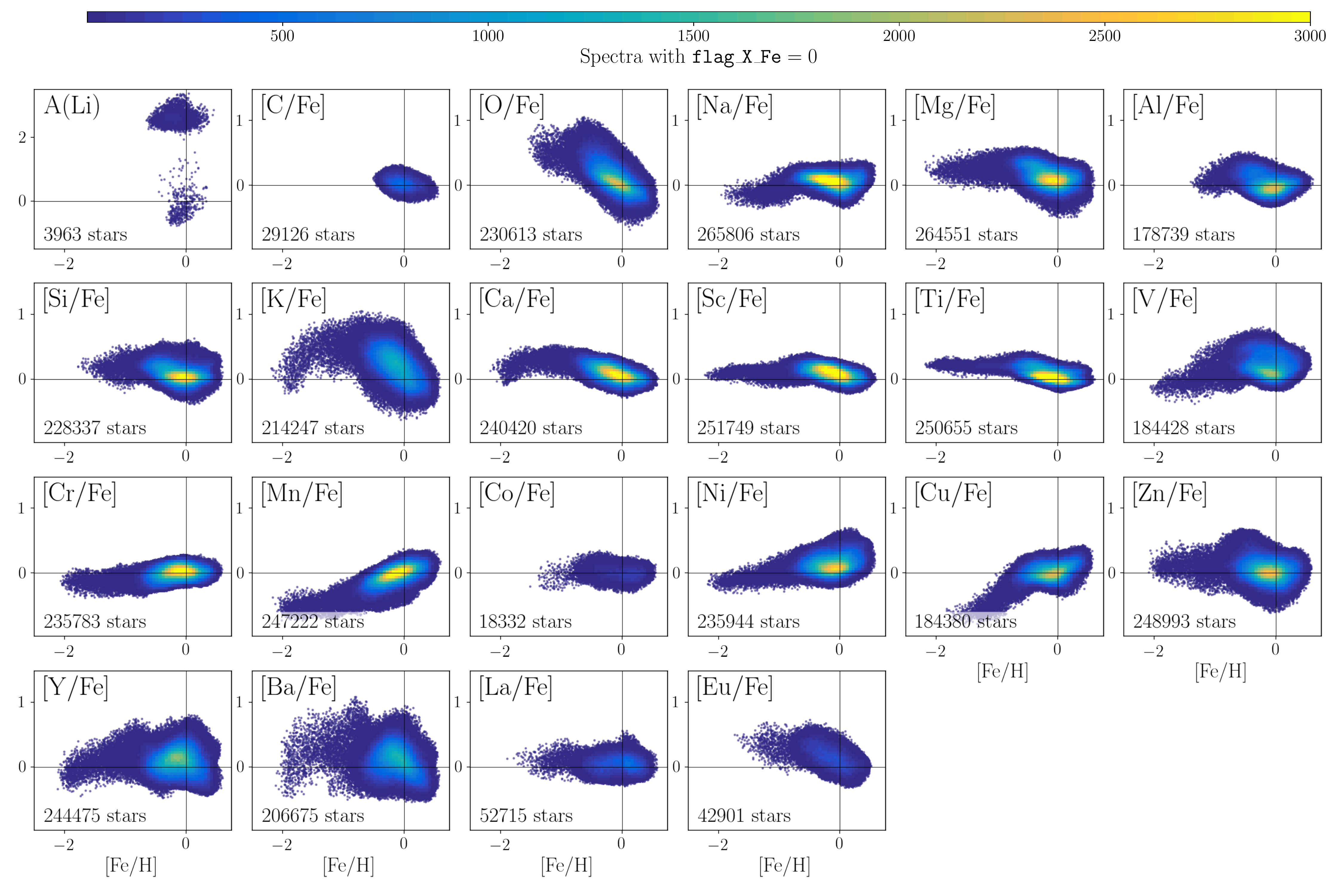}
\caption{Distribution of the element abundances included in GALAH Data Release 2 over the iron abundance [Fe/H]. Shown are relative abundances [X/Fe] for stars with $\texttt{X\_FE\_FLAG} = 0$ (with exception of Li, for which we show the absolute abundance). Colours indicate the stellar density, truncated at a maximum of 3500 per density bin.}
\label{fig:abundance_overview}
\end{figure}
\end{landscape}

\begin{itemize}
\item Red clump stars: According to stellar evolution models, red clump stars should be concentrated at a certain locus in the Kiel diagram due to their similar stellar parameters ($T_{\rm eff} \approx 4800\pm200\,\mathrm{K}$ and $\log g \approx 2.5\pm0.3\,\mathrm{dex}$). While there is an over-density in the expected locus, with the DR2 stellar parameters the shape of the red clump is not entirely as expected, but extending to too high surface gravities (Figure~\ref{fig:kiel_flags}b). Although we note that this density structure might partially originate from stars at the red giant branch bump, the reason for this behaviour is not yet known. We have tested this behaviour using only asteroseismically-based estimates of the surface gravities for the training set around this region but have not achieved a significant improvement in the appearance of the red clump.
\item Cool giants: In the optical regime, strong molecular absorption lines dominate the HERMES spectra for $T_{\rm eff} \la 4500\,\mathrm{K}$. The resulting blending of our diagnostic lines (H, Sc, Ti, and Fe) introduces degeneracies and systematic trends like the overestimation of surface gravities with coincident underestimation of metallicity. Without external information to break those degeneracies, we are currently unable to estimate reliable stellar parameters for the most luminous cool giants in our sample and had to exclude this region from the training set. \textit{The Cannon} is therefore extrapolating the labels of these stars, covering the cool end of the red giant branch (group (1) in Figure~\ref{fig:kiel_flags}d.
\item Hot stars: Stars hotter than cool F types are typically dominated by extended Balmer lines and exhibit only a few, weak metal lines within the HERMES wavelength range. The spectra of these stars hence include less and degenerate information on all stellar parameters. Until now, we have neither been able to analyse these stars with our pipeline, nor include these stars in the training set. Although \textit{The Cannon} model can extrapolate the stellar parameters of these stars, their parameters show systematic trends (group (2) in Figure~\ref{fig:kiel_flags}d). The surface gravities of these stars are overestimated and stars with effective temperatures $>7000\,\mathrm{K}$ are typically underestimated, as the comparison with Gaia benchmark stars (Section~\ref{sec:GBS}) and IRFM (Section~\ref{sec:irfm}) have shown. It is also noted that the high rotational velocities of these hotter stars make the analysis more difficult both for the SME and \textit{The Cannon} steps. 
\item Cool dwarfs: Similar to cool giants, molecular absorption lines dominate the spectra of cool dwarfs. Because of the low amount of non-peculiar cool dwarfs overlapping with TGAS, we have not been able to include enough cool dwarfs in the training set to ensure a good coverage of these stars and are hence flagging them (group (3) in Figure~\ref{fig:kiel_flags}d). We note however, that the significant upturn for cool dwarfs in the previous data releases \citep{Martell2017, Sharma2018} has been largely removed by the use of astrometric information to break degeneracies with surface gravity.
\item Some over-densities in abundance space: We have tried to implement a flagging algorithm to identify unreliable data but we can not exclude that some reliable stars are flagged and vice-versa. \textbf{We recommend that only unflagged stars and abundances should be used} as far as possible. This would for example avoid the problematic groups in Figure~\ref{fig:kiel_flags}, or under-abundant [Si/Fe] or [Ti/Fe] (see Figures~\ref{fig:Dwarf_Giant_comparison1}, \ref{fig:Dwarf_Giant_comparison2}, and \ref{fig:Dwarf_Giant_comparison3}) that otherwise would lead to erroneous abundances. In some cases however it has not been possible to flag the results even if the abundances are expected to be questionable. We caution that the K textsc{i} 7699\,\AA\ line can be affected by interstellar medium absorption, which has not been taken into account; this is expected to be of particular concern at low metallicity when the stellar line is weak and for low-latitude fields where the reddening is high. We also stress that the V lines employed in GALAH are especially vulnerable to blending, which is likely causing the inferred V abundances in giant stars in particular to be susceptible to systematic errors (Figure~\ref{fig:Dwarf_Giant_comparison3_flag0}). As always when not accounting for departures from LTE, systematic errors may be present; in DR2 we have made an effort to include non-LTE calculations for key elements such as Li, O, Na, Mg, Al, Si, and Fe, but several other elements remain to be studied in detail. 

\end{itemize}

\section{Catalogue selection and content}
\label{sec:catalog}

The stars in GALAH DR2 are selected with straightforward criteria. All stars from the main GALAH survey, observed between 16th January 2014 and 12th September 2017, are considered for inclusion. These main survey fields have \texttt{field\_id} between 0 and 6545. Although we share observing and analysis infrastructure with K2-HERMES, TESS-HERMES, HERMES Open Cluster Program, and HERMES Bulge survey, stars observed for those surveys are not in this public data release. 

From the GALAH main survey we select the DR2 data set by making the following cuts and selections:
\begin{itemize}
	\item Only stars with reliable radial velocity estimates (\texttt{rv\_synt} exists and \texttt{e\_rv\_synt}$ < 3\,$km\,s$^{-1}$) are included.
	\item If fewer than five stars in a given observing plate configuration were successfully reduced, the entire configuration is excluded.
	\item Spectroscopic binaries and emission-line stars are included, but are flagged as described in Section~\ref{sec:tsne}.
	\item If stars were observed multiple times, we only report the highest SNR observation and remove duplicates.
	\item We report up to 23 element abundances (Li, C, O, Na, Mg, Al, Si, K, Ca, Sc, Ti, V, Cr, Mn, Fe, Co, Ni, Cu, Zn, Y, Ba, La, Eu) per star. An overview of these elements is shown in Figure~\ref{fig:abundance_overview}. Future releases of the GALAH survey will include abundances of additional elements, such as Rb, Sr, Zr, Mo, Ru, Ce, Nd, and Sm. These elements are not included in DR2 since they will require additional verification to ensure that the inferred abundances are trustworthy. 
\end{itemize}

\begin{figure}
  \includegraphics[width=\columnwidth]{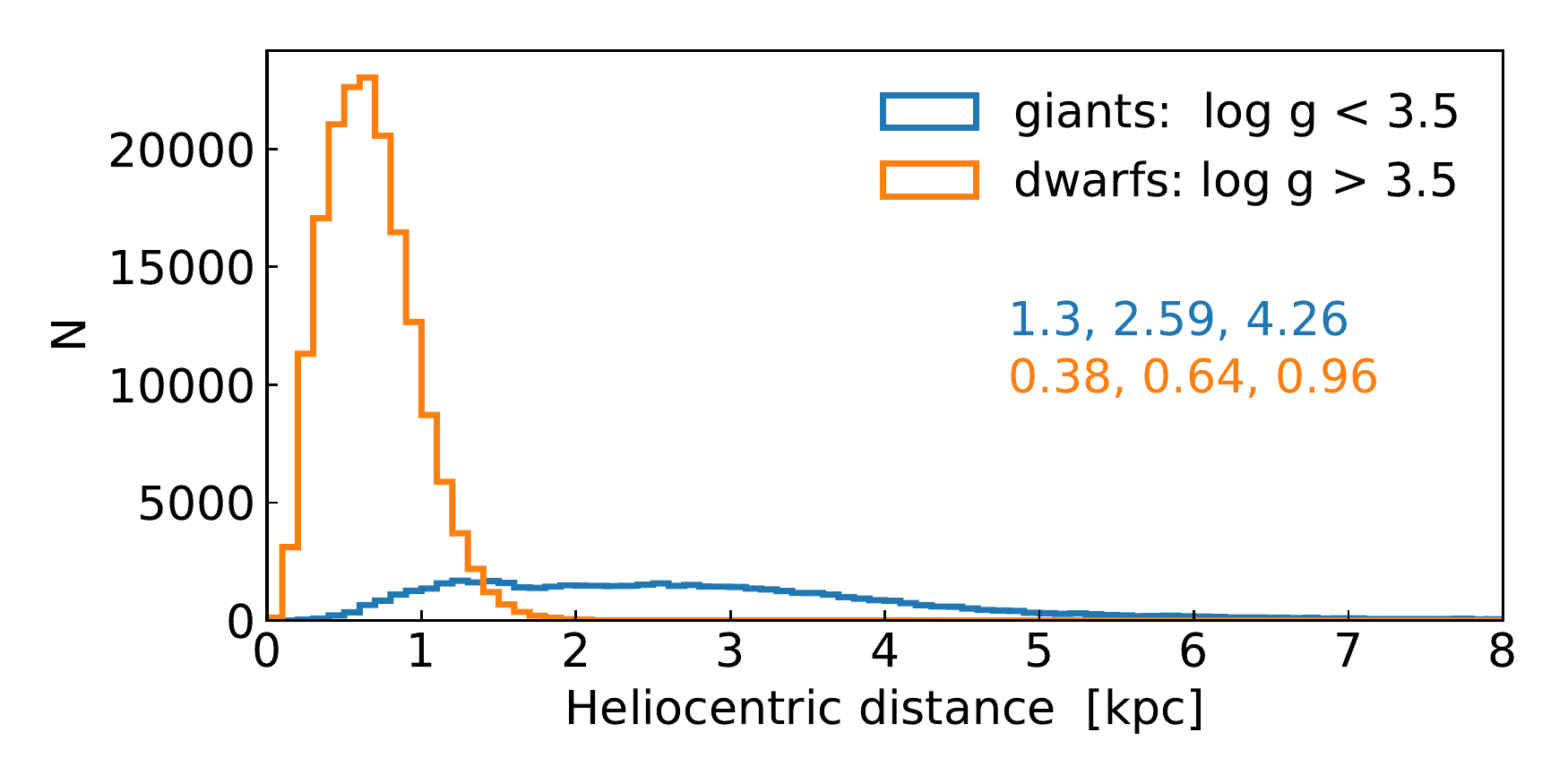}
  \caption{Distribution of distance for giants and dwarfs. The numbers denote the 16th, 50th and 84th percentile values. Stars shown are with distance error of less than 30\% and \texttt{flag\_cannon=0}.} 
  \label{fig:galah_dist_bstep}
\end{figure}

We use the isochrone-based Bayesian method described in \citet{Sharma2018} to estimate distances to all stars in DR2. Figure \ref{fig:galah_dist_bstep} shows the distance distribution for dwarfs and giants, dividing them simply at log~$g=3.5$. Dwarfs are mainly confined to the solar neighborhood (84\% are within 1 kpc), while giants extend much further (84\% are within 4 kpc).

\begin{figure}
  \includegraphics[width=\columnwidth]{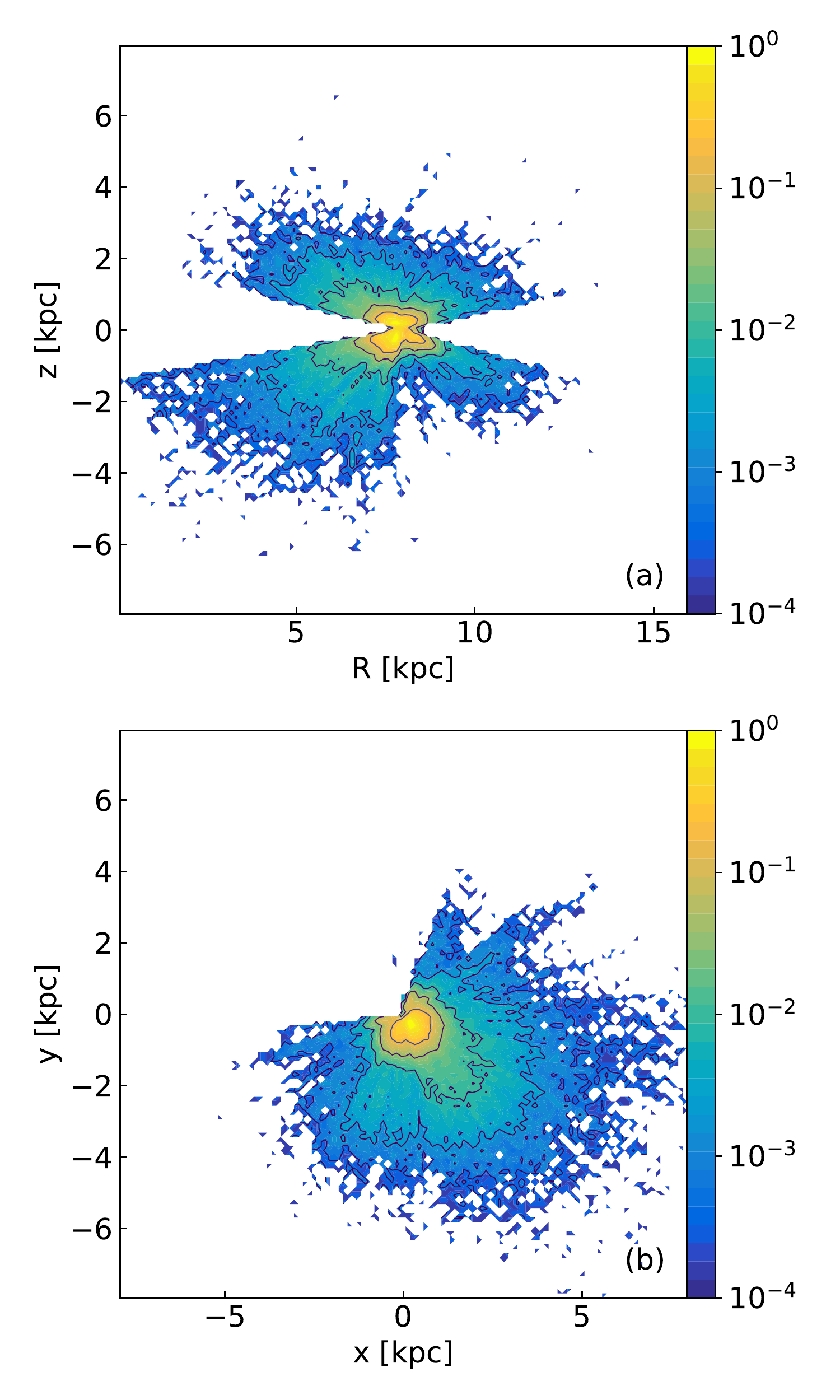}
  \caption{Distribution of stars in Galacto-centric $(R,z)$ plane and heliocentric $(x,y)$ plane (Galactic centre being at $(x,y)=(8,0)$), colour coded by normalised stellar density. Stars shown are with distance error of less than 30\% and \texttt{flag\_cannon=0}.} 
  \label{fig:galah_rzxy_bstep}
\end{figure}

The 342,682 stars in GALAH DR2 provide a very detailed sample of the Milky Way in the solar neighbourhood. To visualise the extent of DR2 in the Galaxy, we transform the coordinates and distances into (x,y,z) heliocentric Cartesian and (R,$\phi$,z) Galacto-centric cylindrical coordinates. Figure \ref{fig:galah_rzxy_bstep} shows the density of stars across the (R,z) and (x,y) planes. 

The observational selections (location of the telescope, avoiding the Galactic plane and high-latitude fields) can be seen in the spatial footprint of DR2. We have very few stars with $x<0$ and $y>0$ because we primarily observe fields with $\delta<0$ and $|b|<60$, which excludes the $90<l<180$ region (\autoref{fig:galah_fov}). 

The tSNE dimensionality reduction process discussed in Section~\ref{sec:tsne} identifies spectra in several unusual categories. These include binary and triple stars, stars with emission lines, and also spectra that showed difficulty in the data reduction. These stars are not excluded from the data set, but they are marked with the \texttt{flag\_cannon} field in the table. This is a bitmask that specifies what makes the spectrum unusual, as defined in Table \ref{tab:catalog}. We recommend using values from the DR2 catalogue for stars with \texttt{flag\_cannon}$=0$.

Table \ref{tab:catalog} lists the columns in the GALAH DR2 catalogue, along with data types, units, and a brief description of each field. The catalogue includes identifiers to enable cross-identification; GALAH observational information; J2000 astrometry; photometry from 2MASS and the $V_\mathrm{JK}$ magnitude used for target selection; SNR in each HERMES channel; radial velocity measured with the two methods described in Section~\ref{sec:reduction}; information about the quality of \textit{The Cannon} solution for the stellar parameters; stellar parameters; and abundances for 23 elements from Li to Eu. We recommend using abundances with \texttt{\texttt{flag\_x\_fe}}$=0$.

The GALAH DR2 catalogue and documentation is available at \url{https://galah-survey.org} and at 
\url{https://datacentral.aao.gov.au/docs/pages/galah/}, through a search form or ADQL query. The catalogue is also available through TAP via \url{https://datacentral.aao.gov.au/vo/tap}. Some Python tools for processing the data are available online in an open-source repository\footnote{\url{https://github.com/svenbuder/GALAH_DR2}}.

\begin{table*}
	\centering
	\caption{Column names, units, data types and descriptions for the GALAH DR2 table}
	\label{tab:catalog}
	\begin{tabular}{llll} 
		\hline\hline
		Field & Units & Data type & Description \\
		\hline
\texttt{star\_id} & & char[16] & 2MASS ID number by default, UCAC4 ID number if 2MASS unavailable (begins with UCAC4-)\\
\texttt{sobject\_id} 	& & int64 & Unique per-observation star ID \\
\texttt{gaia\_id} & & int64 &	\Gaia\ DR2 identifier\\
\texttt{ndfclass} & & char[8] & Observation type (MFOBJECT (regular observation) or MFFLX (benchmark observation))\\
\texttt{field\_id} & & int64 & GALAH field identification number\\
\texttt{raj2000} & deg & float64 & Right ascension from 2MASS, J2000\\
\texttt{dej2000} & deg & float64 & Declination from 2MASS, J2000\\
\texttt{jmag} & mag & float64 & J magnitude from 2MASS\\
\texttt{hmag} & mag & float64 & H magnitude from 2MASS\\
\texttt{kmag} & mag & float64 & K magnitude from 2MASS\\
\texttt{vmag\_jk} & mag & float64 & Synthetic V magnitude calculated from JHK, used for target selection\\
\texttt{e\_jmag} & mag & float64 & Uncertainty in J magnitude, from 2MASS\\
\texttt{e\_hmag} & mag & float64 & Uncertainty in H magnitude, from 2MASS\\
\texttt{e\_kmag} & mag & float64 & Uncertainty in K magnitude, from 2MASS\\
\texttt{snr\_c1} & & float64 & Signal to noise per pixel in the HERMES blue channel\\
\texttt{snr\_c2} & & float64 & Signal to noise per pixel in the HERMES green channel\\
\texttt{snr\_c3} & & float64 & Signal to noise per pixel in the HERMES red channel\\
\texttt{snr\_c4} & & float64 & Signal to noise per pixel in the HERMES IR channel\\
\texttt{rv\_synt} & km\,s$^{-1}$ & float64 & Radial velocity from cross-correlation against synthetic spectra\\
\texttt{rv\_obst} & km\,s$^{-1}$ & float64 & Radial velocity from internal cross-correlation against data\\
\texttt{rv\_nogr\_obst} & km\,s$^{-1}$ & float64 & Radial velocity from internal cross-correlation against data, uncorrected for gravitational redshift\\
\texttt{e\_rv\_synt} & km\,s$^{-1}$ & float64 & Uncertainty in \texttt{rv\_synt}\\
\texttt{e\_rv\_obst} & km\,s$^{-1}$ & float64 & Uncertainty in \texttt{rv\_obst}\\
\texttt{e\_rv\_nogr\_obst} & km\,s$^{-1}$ & float64 & Uncertainty in \texttt{rv\_nogr\_obst}\\
\texttt{chi2\_cannon} & & float64 & Summed chi-squared over all spectral pixels\\
\texttt{sp\_label\_distance} & & float64 & Label distance similar to \citet{Ho2017}\\
\texttt{flag\_cannon} & & int64 & \parbox[t]{11cm}{Flags for spectrum information in a bitmask format\\ 0=No flag \textbf{recommended}\\+1 ($1^{st}$ bit raised)=\textit{The Cannon} starts to extrapolate. For some stars the values could be incorrect.\\+2 ($2^{nd}$ bit raised)=The $\chi^{2}$ of the best fitting model spectrum is significantly higher or lower\\+4 ($3^{rd}$ bit raised)=Reduction flag raised\\+8 ($4^{th}$ bit raised)=Binary star\\+16 ($5^{th}$ bit raised)=Negative flux\\+32 ($6^{th}$ bit raised)=Oscillating continuum\\+64 ($7^{th}$ bit raised)=General reduction issues\\+128 ($8^{th}$ bit raised)=Emission lines}\\
\texttt{teff} & K & float64 & Effective temperature\\
\texttt{e\_teff} & K & float64 & Uncertainty of \texttt{teff}\\
\texttt{logg} & dex & float64 & Surface gravity\\
\texttt{e\_logg} & dex & float64 & Uncertainty of \texttt{logg}\\
\texttt{fe\_h} & dex & float64 & Iron abundance (not overall metallicity [M/H])\\
\texttt{e\_fe\_h} & dex & float64 & Uncertainty in \texttt{fe\_h}\\
\texttt{vmic} & km\,s$^{-1}$ & float64 & Microturbulence velocity\\
\texttt{e\_vmic} & km\,s$^{-1}$ & float64 & Uncertainty in \texttt{vmic}\\
\texttt{vsini} & km\,s$^{-1}$ & float64 & Line of sight rotational velocity\\
\texttt{e\_vsini} & km\,s$^{-1}$ & float64 & Uncertainty in \texttt{vsini}\\
\texttt{alpha\_fe} & dex & float64 & $\alpha$ enhancement, determined as an error-weighted combination of Mg, Si, Ca, Ti abundances\\
\texttt{e\_alpha\_fe} & dex & float64 & Uncertainty in \texttt{alpha\_fe}\\
\texttt{x\_fe} & dex & float64 & [X/Fe] abundance for element X. \\
\texttt{e\_x\_fe} & dex & float64 & Uncertainty in \texttt{x\_fe}\\
\texttt{flag\_x\_fe} & & int64 & \parbox[t]{11cm}{Flags indicating difficulty in abundance determination in a bitmask format\\0=No flag \textbf{recommended}\\+1 ($1^{st}$ bit raised)=Line strength below 2-$\sigma$ upper limit\\+2 ($2^{nd}$ bit raised)=\textit{The Cannon} starts to extrapolate. For some stars the values could be incorrect.\\+4  ($3^{rd}$ bit raised)=The $\chi^{2}$ of the best fitting model spectrum is significantly higher or lower.\\+8 ($4^{th}$ bit raised)=\texttt{flag\_cannon} is not 0}\\
\hline
	\end{tabular}
\end{table*}

\section{Science results accompanied by this Data Release}\label{sec:papers}

A number of science and technical papers based on GALAH DR2 results or GALAH spectra have been published recently, or are being made available along with the general description of the data release. They include: 
\begin{itemize}
     \item \citet{Quillen2018}, ``The GALAH Survey: Stellar streams and how stellar velocity distributions vary with Galactic longitude, hemisphere and metallicity'': We find that structure in the planar (u,v) velocity distribution in the disc depends on metallicity and viewing direction. We infer that there is fine structure in local velocity distributions that varies over distances of a few hundred pc in the Galaxy.
     \item \citet{Duong2018}, ``The GALAH survey: properties of the Galactic disc(s) in the solar neighbourhood'': We investigate the vertical density and abundance profiles of the chemically and kinematically defined `thick' and `thin' discs of the Galaxy. The steep vertical metallicity gradient of the low-$\alpha$ population is in agreement with models where radial migration has a major role in the evolution of the thin disc. In the high-$\alpha$ population, a negative gradient in metallicity and a small gradient in [$\alpha$/M] indicate that it experienced a settling phase, but also formed prior to the onset of major SNIa enrichment. 
     \item \citet{Kos2018}, ``The GALAH Survey: chemical tagging of star clusters and new members in the Pleiades'': The technique of chemical tagging uses the elemental abundances in stellar atmospheres to 'reconstruct' chemically homogeneous star clusters that have long since dispersed. Reliable clustering in a noisy high-dimensional space is a difficult problem that remains largely unsolved. Here, we explore t-distributed stochastic neighbour embedding (t-SNE), which identifies an optimal mapping of a high-dimensional space into fewer dimensions whilst conserving the original clustering information. We show that this method is a reliable tool for chemical tagging because it can: (i) resolve clustering in chemical space alone, (ii) recover known open and globular clusters with high efficiency and low contamination, and (iii) relate field stars to known clusters.
    \item Buder et al., ``The GALAH survey: An abundance, age, and kinematic inventory of the solar neighbourhood made with TGAS'': We investigate the age and kinematic structure in the local disc using GALAH stars in the TGAS \citep{Lindegren2016} data set. We find that age is a clearer indicator of thin disc vs thick disc membership than kinematics or chemical composition.
	\item Khanna et al., ``Velocity fluctuations in the Milky Way using red clump giants'': We investigate the possibility of streaming motions in the Galactic disc. We find no evidence for large-scale velocity fluctuations away from the Galactic plane. We also identify and remove systematic effects and distance errors from data in the midplane and find significantly less power in velocity fluctuations than previously claimed.
    \item Gao et al., ``The GALAH Survey: Verifying abundance trends in the open cluster M67 using non-LTE spectroscopy'': We carry out a careful non-LTE analysis of Li, O, Na, Mg, Al, Si and Fe abundances using GALAH spectra for 69 stars in the open cluster M67. We find that the star-to-star scatter in abundance is on the order of 0.05 dex, and we find abundance trends with temperature that are broadly consistent with the atomic diffusion models of \citet{Dotter2017}.
    \item Simpson et al., ``The GALAH survey: Co-orbiting stars and chemical tagging'': We investigate the pairs and groups of stars with very similar distances from the Sun and proper motions reported by \citet{Oh2017}. GALAH DR2 contains two stars in 15 of the apparently co-moving pairs, and we find that nine of them are truly co-orbiting. Of those nine pairs, we have reliable stellar parameters and abundances for six. Of those, three have highly similar abundance patterns and are likely to be co-natal, while three have quite different abundance patterns and are likely to be participating in a dynamical resonance. We emphasize the importance of stellar streams and co-orbiting pairs as a testbed for full chemical tagging in the Galactic disc.
	\item Zwitter et al., ``The GALAH Survey: Accurate Radial Velocities and Library of Observed Stellar Template Spectra'': We derive precise radial velocities ($\sigma \approx 100$ m~s$^{-1}$) for 306,633 stars from GALAH DR2 that fall in well-populated areas of the H-R diagram in the metallicity range $-0.6 \leq$[Fe/H]$\leq +0.3$. This is done by comparison against 1362 median spectra constructed from sets of 103 to 2276 GALAH spectra with virtually identical stellar parameters. The level of accuracy achieved is adequate for studies of dynamics {\it within\/} stellar clusters, associations and streams in the Galaxy. The library of median spectra should be useful as a reference set in the HERMES bandpasses, and the radial velocity values and their errors are included in GALAH DR2 as \texttt{\texttt{rv\_obst}} and \texttt{\texttt{e\_rv\_obst}}. For reference also the radial velocities without gravitational redshift \texttt{\texttt{rv\_nogr\_obst}} are reported, as \Gaia\ DR2 will allow a more accurate estimate of this effect. 
	\item Simpson et al., ``The GALAH and \textit{TESS}-HERMES surveys: high-resolution spectroscopy of luminous supergiants in the Magellanic Clouds and Bridge'': We report the serendipitous observation of about 560 members of the Magellanic Clouds by the Milky Way spectroscopic GALAH and TESS-HERMES surveys. We also find at least one star that appears associated with structured star formation in the Magellanic Bridge. All the observed stars in the Magellanic Clouds are intrinsically luminous supergiant stars, well outside the normal parameter range of \textit{The Cannon} analysis pipeline used by these surveys. But we find that these supergiants are located in coherent (if astrophysically incorrect) places in the label space, allowing us to identify these Magellanic Cloud stars.
    \item Kos et al., ``Holistic spectroscopy: complete reconstruction of a wide-field, multi-object spectroscopic image using a photonic comb'': We present a new method for extraction of spectroscopic data using forward modelling of the raw images, accounting for optical aberrations, scattered light, and variable fibre throughput. Using this method we can produce 1D spectra with an effective resolution over twice the nominal resolution of the HERMES spectrograph.
\end{itemize}

\section{Conclusions} \label{sec:conclusions}

The stellar parameter and abundance information contained in GALAH DR2 will enable major steps forward in Galactic Archaeology, including detailed work to identify clusters within the chemical space and characterize its structure and dimensionality. In combination with the dynamical information provided by \Gaia\ DR2, we will work toward a reliable narrative of how the Milky Way was assembled and how it has evolved since, using chemodynamics and chemical tagging.

The GALAH survey team will continue the project, collecting additional data toward the goal of one million stars and continuing to develop the analysis procedure, in particular to incorporate parallax information from \Gaia. Future GALAH data releases will include re-reductions and reanalyses of all stars from DR2 as well as new stars observed in the intervening time.

\section*{Acknowledgements}
Based on data acquired through the Australian Astronomical Observatory, under programmes: A/2013B/13 (The GALAH pilot survey); A/2014A/25, A/2015A/19, A2017A/18 (The GALAH survey). We acknowledge the traditional owners of the land on which the AAT stands, the Gamilaraay people, and pay our respects to elders past and present.

The following software and programming languages made this research possible: \textsc{IRAF} \citep{Tody1986,Tody1993}, \textsc{configure} \citep{Miszalski2006}; Python (versions 2.7 \& 3.6); Astropy \citep[version 2.0;][]{Robitaille2013,PriceWhelan2018}, a community-developed core Python package for Astronomy; pandas \citep[version 0.20.2;][]{McKinney2011}; \textsc{topcat} \citep[version 4.4;][]{Taylor2005}; \textsc{galpy} \citep[version 1.3;][]{Bovy2015}. This research made use of APLpy, an open-source plotting package for Python \citep{Robitaille2012}. This research has made use of the VizieR catalogue access tool, CDS, Strasbourg, France. The original description of the VizieR service was published in A\&AS 143, 23.

This work has made use of data from the European Space Agency (ESA) mission {\it \Gaia} (\url{http://www.cosmos.esa.int/gaia}), processed by the {\it \Gaia} Data Processing and Analysis Consortium (DPAC, \url{http://www.cosmos.esa.int/web/gaia/dpac/consortium}). Funding for the DPAC has been provided by national institutions, in particular the institutions participating in the {\it \Gaia} Multilateral Agreement.

SB and KL acknowledge funds from the Alexander von Humboldt Foundation in the framework of the Sofja Kovalevskaja Award endowed by the Federal Ministry of Education and Research. SB, MA and KL acknowledge travel support from Universities Australia and Deutsche Akademische Austauschdienst. The research by MA, LD, JL, AMA, and DN has been supported by an Australian Research Council Laureate Fellowship to MA (grant FL110100012). LD gratefully acknowledges a scholarship from Zonta International District 24 and support from ARC grant DP160103747. LD, KF and Y-ST are grateful for support from Australian Research Council grant DP160103747. JK is supported by a Discovery Project grant from the Australian Research Council (DP150104667) awarded to J. Bland-Hawthorn and T. Bedding. KL acknowledges funds from the Swedish Research Council (Grant nr. 2015-00415\_3) and Marie Sklodowska Curie Actions (Cofund Project INCA 600398). SLM acknowledges support from the Australian Research Council through grant DE140100598. LC is the recipient of an ARC Future Fellowship (project number FT160100402). GT, K\v{C}, and TZ acknowledge the financial  support  from  the  Slovenian  Research  Agency  (research core funding No. P1-0188). DMN was supported by the Allan C. and Dorothy H. Davis Fellowship. M{\v Z} acknowledges support from ARC grant DP170102233. DS is the recipient
of an ARC Future Fellowship (project number FT140100147).

Parts of this research were conducted by the Australian Research Council Centre of Excellence for All Sky Astrophysics in 3 Dimensions (ASTRO 3D), through project number CE170100013. 

Parts of the computations were performed on resources provided by the Swedish National Infrastructure for Computing (SNIC) at UPPMAX under project 2015/1-309 and 2016/1-400, as well as the MPCDF supercomputing facilities in Garching.

The authors thank D. W. Hogg, H.-W. Rix, and A. C. Eilers for useful discussions on \textit{The Cannon}.




\bibliographystyle{mnras}
\bibliography{bib} 

\begin{thebibliography}{}
\makeatletter
\relax
\def\mn@urlcharsother{\let\do\@makeother \do\$\do\&\do\#\do\^\do\_\do\%\do\~}
\def\mn@doi{\begingroup\mn@urlcharsother \@ifnextchar [ {\mn@doi@}
  {\mn@doi@[]}}
\def\mn@doi@[#1]#2{\def\@tempa{#1}\ifx\@tempa\@empty \href
  {http://dx.doi.org/#2} {doi:#2}\else \href {http://dx.doi.org/#2} {#1}\fi
  \endgroup}
\def\mn@eprint#1#2{\mn@eprint@#1:#2::\@nil}
\def\mn@eprint@arXiv#1{\href {http://arxiv.org/abs/#1} {{\tt arXiv:#1}}}
\def\mn@eprint@dblp#1{\href {http://dblp.uni-trier.de/rec/bibtex/#1.xml}
  {dblp:#1}}
\def\mn@eprint@#1:#2:#3:#4\@nil{\def\@tempa {#1}\def\@tempb {#2}\def\@tempc
  {#3}\ifx \@tempc \@empty \let \@tempc \@tempb \let \@tempb \@tempa \fi \ifx
  \@tempb \@empty \def\@tempb {arXiv}\fi \@ifundefined
  {mn@eprint@\@tempb}{\@tempb:\@tempc}{\expandafter \expandafter \csname
  mn@eprint@\@tempb\endcsname \expandafter{\@tempc}}}

\bibitem[\protect\citeauthoryear{{Abolfathi} et~al.,}{{Abolfathi}
  et~al.}{2017}]{SDSSDR14}
{Abolfathi} B.,  et~al., 2017, preprint, \href
  {http://adsabs.harvard.edu/abs/2017arXiv170709322A} {} (\mn@eprint {arXiv}
  {1707.09322})

\bibitem[\protect\citeauthoryear{{Albareti} et~al.,}{{Albareti}
  et~al.}{2017}]{SDSSDR13}
{Albareti} F.~D.,  et~al., 2017, \mn@doi [\apjs] {10.3847/1538-4365/aa8992},
  \href {http://adsabs.harvard.edu/abs/2017ApJS..233...25A} {233, 25}

\bibitem[\protect\citeauthoryear{{Aleo}, {Sobotka}  \& {Ram{\'{\i}}rez}}{{Aleo}
  et~al.}{2017}]{Aleo2017}
{Aleo} P.~D.,  {Sobotka} A.~C.,   {Ram{\'{\i}}rez} I.,  2017, \mn@doi [\apj]
  {10.3847/1538-4357/aa83b6}, \href
  {http://adsabs.harvard.edu/abs/2017ApJ...846...24A} {846, 24}

\bibitem[\protect\citeauthoryear{{Allende Prieto}, {Koesterke}, {Ludwig},
  {Freytag}  \& {Caffau}}{{Allende Prieto} et~al.}{2013}]{AllendePrieto2013}
{Allende Prieto} C.,  {Koesterke} L.,  {Ludwig} H.-G.,  {Freytag} B.,
  {Caffau} E.,  2013, \mn@doi [\aap] {10.1051/0004-6361/201220064}, \href
  {http://adsabs.harvard.edu/abs/2013A%26A...550A.103A} {550, A103}

\bibitem[\protect\citeauthoryear{{Amarsi} \& {Asplund}}{{Amarsi} \&
  {Asplund}}{2017}]{Amarsi2017}
{Amarsi} A.~M.,  {Asplund} M.,  2017, \mn@doi [\mnras] {10.1093/mnras/stw2445},
  \href {http://adsabs.harvard.edu/abs/2017MNRAS.464..264A} {464, 264}

\bibitem[\protect\citeauthoryear{{Amarsi}, {Asplund}, {Collet}  \&
  {Leenaarts}}{{Amarsi} et~al.}{2016a}]{Amarsi2016b}
{Amarsi} A.~M.,  {Asplund} M.,  {Collet} R.,   {Leenaarts} J.,  2016a, \mn@doi
  [\mnras] {10.1093/mnras/stv2608}, \href
  {http://adsabs.harvard.edu/abs/2016MNRAS.455.3735A} {455, 3735}

\bibitem[\protect\citeauthoryear{{Amarsi}, {Lind}, {Asplund}, {Barklem}  \&
  {Collet}}{{Amarsi} et~al.}{2016b}]{Amarsi2016a}
{Amarsi} A.~M.,  {Lind} K.,  {Asplund} M.,  {Barklem} P.~S.,   {Collet} R.,
  2016b, \mn@doi [\mnras] {10.1093/mnras/stw2077}, \href
  {http://adsabs.harvard.edu/abs/2016MNRAS.463.1518A} {463, 1518}

\bibitem[\protect\citeauthoryear{{Amarsi}, {Nordlander}, {Barklem}, {Asplund},
  {Collet}  \& {Lind}}{{Amarsi} et~al.}{2018}]{Amarsi2018}
{Amarsi} A.~M.,  {Nordlander} T.,  {Barklem} P.~S.,  {Asplund} M.,  {Collet}
  R.,   {Lind} K.,  2018, preprint, \href
  {http://adsabs.harvard.edu/abs/2018arXiv180402305A} {} (\mn@eprint {arXiv}
  {1804.02305})

\bibitem[\protect\citeauthoryear{{Asplund}, {Nordlund}, {Trampedach}, {Allende
  Prieto}  \& {Stein}}{{Asplund} et~al.}{2000}]{Asplund2000}
{Asplund} M.,  {Nordlund} {\AA}.,  {Trampedach} R.,  {Allende Prieto} C.,
  {Stein} R.~F.,  2000, \aap, \href
  {http://adsabs.harvard.edu/abs/2000A%26A...359..729A} {359, 729}

\bibitem[\protect\citeauthoryear{{Asplund}, {Grevesse}, {Sauval}  \&
  {Scott}}{{Asplund} et~al.}{2009}]{Asplund2009}
{Asplund} M.,  {Grevesse} N.,  {Sauval} A.~J.,   {Scott} P.,  2009, \mn@doi
  [\araa] {10.1146/annurev.astro.46.060407.145222}, \href
  {http://adsabs.harvard.edu/abs/2009ARA%26A..47..481A} {47, 481}

\bibitem[\protect\citeauthoryear{{Astraatmadja} \&
  {Bailer-Jones}}{{Astraatmadja} \& {Bailer-Jones}}{2016}]{Astraatmadja2016}
{Astraatmadja} T.~L.,  {Bailer-Jones} C.~A.~L.,  2016, \mn@doi [\apj]
  {10.3847/1538-4357/833/1/119}, \href
  {http://adsabs.harvard.edu/abs/2016ApJ...833..119A} {833, 119}

\bibitem[\protect\citeauthoryear{{Astropy Collaboration} et~al.,}{{Astropy
  Collaboration} et~al.}{2013}]{Robitaille2013}
{Astropy Collaboration} et~al., 2013, \mn@doi [\aap]
  {10.1051/0004-6361/201322068}, \href
  {http://adsabs.harvard.edu/abs/2013A%26A...558A..33A} {558, A33}

\bibitem[\protect\citeauthoryear{{Bard} \& {Kock}}{{Bard} \& {Kock}}{1994}]{BK}
{Bard} A.,  {Kock} M.,  1994, Astron. and Astrophys., 282, 1014

\bibitem[\protect\citeauthoryear{{Bard}, {Kock}  \& {Kock}}{{Bard}
  et~al.}{1991}]{BKK}
{Bard} A.,  {Kock} A.,   {Kock} M.,  1991, Astron. and Astrophys., 248, 315

\bibitem[\protect\citeauthoryear{{Barden} et~al.,}{{Barden}
  et~al.}{2010}]{Barden2010}
{Barden} S.~C.,  et~al., 2010, in Ground-based and Airborne Instrumentation for
  Astronomy III. p. 773509, \mn@doi{10.1117/12.856103}

\bibitem[\protect\citeauthoryear{{Bastian} \& {Lardo}}{{Bastian} \&
  {Lardo}}{2017}]{Bastian2017}
{Bastian} N.,  {Lardo} C.,  2017, preprint, \href
  {http://adsabs.harvard.edu/abs/2017arXiv171201286B} {} (\mn@eprint {arXiv}
  {1712.01286})

\bibitem[\protect\citeauthoryear{{Battistini} \& {Bensby}}{{Battistini} \&
  {Bensby}}{2015}]{Battistini2015}
{Battistini} C.,  {Bensby} T.,  2015, \mn@doi [\aap]
  {10.1051/0004-6361/201425327}, \href
  {http://adsabs.harvard.edu/abs/2015A%26A...577A...9B} {577, A9}

\bibitem[\protect\citeauthoryear{{Battistini} \& {Bensby}}{{Battistini} \&
  {Bensby}}{2016}]{Battistini2016}
{Battistini} C.,  {Bensby} T.,  2016, \mn@doi [\aap]
  {10.1051/0004-6361/201527385}, \href
  {http://adsabs.harvard.edu/abs/2016A%26A...586A..49B} {586, A49}

\bibitem[\protect\citeauthoryear{{Bensby} \& {Lind}}{{Bensby} \&
  {Lind}}{2018}]{Bensby2018}
{Bensby} T.,  {Lind} K.,  2018, preprint

\bibitem[\protect\citeauthoryear{{Bensby}, {Feltzing}  \& {Oey}}{{Bensby}
  et~al.}{2014}]{Bensby2014}
{Bensby} T.,  {Feltzing} S.,   {Oey} M.~S.,  2014, \mn@doi [\aap]
  {10.1051/0004-6361/201322631}, \href
  {http://adsabs.harvard.edu/abs/2014A%26A...562A..71B} {562, A71}

\bibitem[\protect\citeauthoryear{{Biemont} \& {Godefroid}}{{Biemont} \&
  {Godefroid}}{1980}]{1980A&A....84..361B}
{Biemont} E.,  {Godefroid} M.,  1980, \aap, \href
  {http://adsabs.harvard.edu/abs/1980A%26A....84..361B} {84, 361}

\bibitem[\protect\citeauthoryear{{Biemont}, {Grevesse}, {Hannaford}  \&
  {Lowe}}{{Biemont} et~al.}{1981}]{BGHL}
{Biemont} E.,  {Grevesse} N.,  {Hannaford} P.,   {Lowe} R.~M.,  1981, \mn@doi
  [\apj] {10.1086/159213}, \href
  {http://cdsads.u-strasbg.fr/abs/1981ApJ...248..867B} {248, 867}

\bibitem[\protect\citeauthoryear{{Bi{\'e}mont} et~al.,}{{Bi{\'e}mont}
  et~al.}{2011}]{BBEHL}
{Bi{\'e}mont} {\'E}.,  et~al., 2011, \mn@doi [\mnras]
  {10.1111/j.1365-2966.2011.18637.x}, \href
  {http://adsabs.harvard.edu/abs/2011MNRAS.414.3350B} {414, 3350}

\bibitem[\protect\citeauthoryear{{Blackwell}, {Petford}, {Shallis}  \&
  {Simmons}}{{Blackwell} et~al.}{1982a}]{GESB82c}
{Blackwell} D.~E.,  {Petford} A.~D.,  {Shallis} M.~J.,   {Simmons} G.~J.,
  1982a, \mn@doi [\mnras] {10.1093/mnras/199.1.43}, \href
  {http://adsabs.harvard.edu/abs/1982MNRAS.199...43B} {199, 43}

\bibitem[\protect\citeauthoryear{{Blackwell}, {Petford}  \&
  {Simmons}}{{Blackwell} et~al.}{1982b}]{GESB82d}
{Blackwell} D.~E.,  {Petford} A.~D.,   {Simmons} G.~J.,  1982b, \mn@doi
  [\mnras] {10.1093/mnras/201.3.595}, \href
  {http://adsabs.harvard.edu/abs/1982MNRAS.201..595B} {201, 595}

\bibitem[\protect\citeauthoryear{{Blackwell}, {Menon}  \&
  {Petford}}{{Blackwell} et~al.}{1983}]{1983MNRAS.204..883B}
{Blackwell} D.~E.,  {Menon} S.~L.~R.,   {Petford} A.~D.,  1983, \mn@doi
  [\mnras] {10.1093/mnras/204.3.883}, \href
  {http://adsabs.harvard.edu/abs/1983MNRAS.204..883B} {204, 883}

\bibitem[\protect\citeauthoryear{{Blackwell}, {Booth}, {Menon}  \&
  {Petford}}{{Blackwell} et~al.}{1986}]{GESB86}
{Blackwell} D.~E.,  {Booth} A.~J.,  {Menon} S.~L.~R.,   {Petford} A.~D.,  1986,
  \mn@doi [\mnras] {10.1093/mnras/220.2.289}, \href
  {http://adsabs.harvard.edu/abs/1986MNRAS.220..289B} {220, 289}

\bibitem[\protect\citeauthoryear{{Bland-Hawthorn}, {Krumholz}  \&
  {Freeman}}{{Bland-Hawthorn} et~al.}{2010}]{BlandHawthorn2010a}
{Bland-Hawthorn} J.,  {Krumholz} M.~R.,   {Freeman} K.,  2010, \mn@doi [\apj]
  {10.1088/0004-637X/713/1/166}, \href
  {http://adsabs.harvard.edu/abs/2010ApJ...713..166B} {713, 166}

\bibitem[\protect\citeauthoryear{Bland-Hawthorn, Kos, Betters, Silva, O'Byrne,
  Patterson  \& Leon-Saval}{Bland-Hawthorn et~al.}{2017}]{Bland-Hawthorn2017}
Bland-Hawthorn J.,  Kos J.,  Betters C.~H.,  Silva G.~D.,  O'Byrne J.,
  Patterson R.,   Leon-Saval S.~G.,  2017, \mn@doi [Optics Express]
  {10.1364/OE.25.015614}, 25, 15614

\bibitem[\protect\citeauthoryear{{Bovy}}{{Bovy}}{2015}]{Bovy2015}
{Bovy} J.,  2015, \mn@doi [\apjs] {10.1088/0067-0049/216/2/29}, \href
  {http://adsabs.harvard.edu/abs/2015ApJS..216...29B} {216, 29}

\bibitem[\protect\citeauthoryear{Brown et~al.,}{Brown et~al.}{2016}]{Brown2016}
Brown A. G.~A.,  et~al., 2016, \mn@doi [Astronomy {\&} Astrophysics]
  {10.1051/0004-6361/201629512}, 595, A2

\bibitem[\protect\citeauthoryear{{Brzeski}, {Case}  \& {Gers}}{{Brzeski}
  et~al.}{2011}]{Brzeski2011}
{Brzeski} J.,  {Case} S.,   {Gers} L.,  2011, in Optomechanics 2011:
  Innovations and Solutions. p. 812504, \mn@doi{10.1117/12.896389}

\bibitem[\protect\citeauthoryear{{Buder} et~al.,}{{Buder}
  et~al.}{2018}]{Buder2018a}
{Buder} S.,  et~al., 2018, \aap, submitted

\bibitem[\protect\citeauthoryear{{Cardon}, {Smith}, {Scalo}, {Testerman}  \&
  {Whaling}}{{Cardon} et~al.}{1982}]{1982ApJ...260..395C}
{Cardon} B.~L.,  {Smith} P.~L.,  {Scalo} J.~M.,  {Testerman} L.,   {Whaling}
  W.,  1982, \mn@doi [\apj] {10.1086/160264}, \href
  {http://adsabs.harvard.edu/abs/1982ApJ...260..395C} {260, 395}

\bibitem[\protect\citeauthoryear{{Carlsson}, {Sturesson}  \&
  {Svanberg}}{{Carlsson} et~al.}{1989}]{CSS89}
{Carlsson} J.,  {Sturesson} L.,   {Svanberg} S.,  1989, \mn@doi [Zeitschrift
  fur Physik D Atoms Molecules Clusters] {10.1007/BF01438501}, \href
  {http://adsabs.harvard.edu/abs/1989ZPhyD..11..287C} {11, 287}

\bibitem[\protect\citeauthoryear{{Casagrande} \& {VandenBerg}}{{Casagrande} \&
  {VandenBerg}}{2014}]{Casagrande2014}
{Casagrande} L.,  {VandenBerg} D.~A.,  2014, \mn@doi [\mnras]
  {10.1093/mnras/stu1476}, \href
  {http://adsabs.harvard.edu/abs/2014MNRAS.444..392C} {444, 392}

\bibitem[\protect\citeauthoryear{{Casagrande}, {Ram{\'{\i}}rez},
  {Mel{\'e}ndez}, {Bessell}  \& {Asplund}}{{Casagrande}
  et~al.}{2010}]{Casagrande2010}
{Casagrande} L.,  {Ram{\'{\i}}rez} I.,  {Mel{\'e}ndez} J.,  {Bessell} M.,
  {Asplund} M.,  2010, \mn@doi [\aap] {10.1051/0004-6361/200913204}, \href
  {http://adsabs.harvard.edu/abs/2010A%26A...512A..54C} {512, A54}

\bibitem[\protect\citeauthoryear{{Casagrande} et~al.,}{{Casagrande}
  et~al.}{2014}]{Casagrande2014b}
{Casagrande} L.,  et~al., 2014, \mn@doi [\mnras] {10.1093/mnras/stu089}, \href
  {http://adsabs.harvard.edu/abs/2014MNRAS.439.2060C} {439, 2060}

\bibitem[\protect\citeauthoryear{{Casey}, {Hogg}, {Ness}, {Rix}, {Ho}  \&
  {Gilmore}}{{Casey} et~al.}{2016}]{Casey2016}
{Casey} A.~R.,  {Hogg} D.~W.,  {Ness} M.,  {Rix} H.-W.,  {Ho} A.~Q.,
  {Gilmore} G.,  2016, preprint, \href
  {http://adsabs.harvard.edu/abs/2016arXiv160303040C} {} (\mn@eprint {arXiv}
  {1603.03040})

\bibitem[\protect\citeauthoryear{{Chambers} et~al.,}{{Chambers}
  et~al.}{2016}]{Chambers2016}
{Chambers} K.~C.,  et~al., 2016, preprint, \href
  {http://adsabs.harvard.edu/abs/2016arXiv161205560C} {} (\mn@eprint {arXiv}
  {1612.05560})

\bibitem[\protect\citeauthoryear{{Chang} \& {Tang}}{{Chang} \&
  {Tang}}{1990}]{1990JQSRT..43..207C}
{Chang} T.~N.,  {Tang} X.,  1990, \mn@doi [\jqsrt]
  {10.1016/0022-4073(90)90053-9}, \href
  {http://adsabs.harvard.edu/abs/1990JQSRT..43..207C} {43, 207}

\bibitem[\protect\citeauthoryear{{Chiavassa}, {Casagrande}, {Collet}, {Magic},
  {Bigot}, {Th{\'e}venin}  \& {Asplund}}{{Chiavassa}
  et~al.}{2018}]{Chiavassa2018}
{Chiavassa} A.,  {Casagrande} L.,  {Collet} R.,  {Magic} Z.,  {Bigot} L.,
  {Th{\'e}venin} F.,   {Asplund} M.,  2018, \mn@doi [\aap]
  {10.1051/0004-6361/201732147}, \href
  {http://adsabs.harvard.edu/abs/2018A%26A...611A..11C} {611, A11}

\bibitem[\protect\citeauthoryear{{Dalton} et~al.,}{{Dalton}
  et~al.}{2014}]{Dalton2014}
{Dalton} G.,  et~al., 2014, in Ground-based and Airborne Instrumentation for
  Astronomy V. p. 91470L (\mn@eprint {arXiv} {1412.0843}),
  \mn@doi{10.1117/12.2055132}

\bibitem[\protect\citeauthoryear{{Davidson}, {Snoek}, {Volten}  \&
  {Doenszelmann}}{{Davidson} et~al.}{1992}]{DSVD92}
{Davidson} M.~D.,  {Snoek} L.~C.,  {Volten} H.,   {Doenszelmann} A.,  1992,
  \aap, \href {http://adsabs.harvard.edu/abs/1992A%26A...255..457D} {255, 457}

\bibitem[\protect\citeauthoryear{{De Laverny}, {Recio-Blanco}, {Worley}  \&
  {Plez}}{{De Laverny} et~al.}{2012}]{Laverny2012}
{De Laverny} P.,  {Recio-Blanco} A.,  {Worley} C.~C.,   {Plez} B.,  2012,
  \mn@doi [\aap] {10.1051/0004-6361/201219330}, \href
  {http://adsabs.harvard.edu/abs/2012A%26A...544A.126D} {544, A126}

\bibitem[\protect\citeauthoryear{{De Silva}, {Freeman}  \&
  {Bland-Hawthorn}}{{De Silva} et~al.}{2009}]{DeSilva2009}
{De Silva} G.~M.,  {Freeman} K.~C.,   {Bland-Hawthorn} J.,  2009, \mn@doi
  [\pasa] {10.1071/AS08019}, \href
  {http://adsabs.harvard.edu/abs/2009PASA...26...11D} {26, 11}

\bibitem[\protect\citeauthoryear{{De Silva} et~al.,}{{De Silva}
  et~al.}{2015}]{DeSilva2015}
{De Silva} G.~M.,  et~al., 2015, \mn@doi [\mnras] {10.1093/mnras/stv327}, \href
  {http://adsabs.harvard.edu/abs/2015MNRAS.449.2604D} {449, 2604}

\bibitem[\protect\citeauthoryear{{Den Hartog}, {Lawler}, {Sneden}  \&
  {Cowan}}{{Den Hartog} et~al.}{2003}]{HLSC}
{Den Hartog} E.~A.,  {Lawler} J.~E.,  {Sneden} C.,   {Cowan} J.~J.,  2003,
  \mn@doi [Astrophys. J. Suppl. Ser.] {10.1086/376940}, 148, 543

\bibitem[\protect\citeauthoryear{{Den Hartog}, {Lawler}, {Sobeck}, {Sneden}  \&
  {Cowan}}{{Den Hartog} et~al.}{2011}]{DLSSC}
{Den Hartog} E.~A.,  {Lawler} J.~E.,  {Sobeck} J.~S.,  {Sneden} C.,   {Cowan}
  J.~J.,  2011, \mn@doi [\apjs] {10.1088/0067-0049/194/2/35}, \href
  {http://adsabs.harvard.edu/abs/2011ApJS..194...35D} {194, 35}

\bibitem[\protect\citeauthoryear{{Den Hartog}, {Ruffoni}, {Lawler},
  {Pickering}, {Lind}  \& {Brewer}}{{Den Hartog}
  et~al.}{2014}]{2014ApJS..215...23D}
{Den Hartog} E.~A.,  {Ruffoni} M.~P.,  {Lawler} J.~E.,  {Pickering} J.~C.,
  {Lind} K.,   {Brewer} N.~R.,  2014, \mn@doi [\apjs]
  {10.1088/0067-0049/215/2/23}, \href
  {http://adsabs.harvard.edu/abs/2014ApJS..215...23D} {215, 23}

\bibitem[\protect\citeauthoryear{{Dotter}, {Conroy}, {Cargile}  \&
  {Asplund}}{{Dotter} et~al.}{2017}]{Dotter2017}
{Dotter} A.,  {Conroy} C.,  {Cargile} P.,   {Asplund} M.,  2017, \mn@doi [\apj]
  {10.3847/1538-4357/aa6d10}, \href
  {http://adsabs.harvard.edu/abs/2017ApJ...840...99D} {840, 99}

\bibitem[\protect\citeauthoryear{{Duong} et~al.,}{{Duong}
  et~al.}{2018}]{Duong2018}
{Duong} L.,  et~al., 2018, MNRAS

\bibitem[\protect\citeauthoryear{Ester, Kriegel, Jorg  \& Xu}{Ester
  et~al.}{1996}]{Ester1996}
Ester M.,  Kriegel H.-p.,  Jorg S.,   Xu X.,  1996, in Proceedings of 2nd
  International Conference on KDD. pp 226--231, \url
  {http://citeseerx.ist.psu.edu/viewdoc/summary?doi=10.1.1.71.1980}

\bibitem[\protect\citeauthoryear{{Farrell}, {Birchall}, {Heald}, {Shortridge},
  {Vuong}  \& {Sheinis}}{{Farrell} et~al.}{2014}]{Farrell2014}
{Farrell} T.~J.,  {Birchall} M.~N.,  {Heald} R.~W.,  {Shortridge} K.,  {Vuong}
  M.~V.,   {Sheinis} A.~I.,  2014, in Software and Cyberinfrastructure for
  Astronomy III. p. 915223, \mn@doi{10.1117/12.2054805}

\bibitem[\protect\citeauthoryear{{Forbes} \& {Bridges}}{{Forbes} \&
  {Bridges}}{2010}]{Forbes2010}
{Forbes} D.~A.,  {Bridges} T.,  2010, \mn@doi [\mnras]
  {10.1111/j.1365-2966.2010.16373.x}, \href
  {http://adsabs.harvard.edu/abs/2010MNRAS.404.1203F} {404, 1203}

\bibitem[\protect\citeauthoryear{{Freeman} \& {Bland-Hawthorn}}{{Freeman} \&
  {Bland-Hawthorn}}{2002}]{FreemanBlandHawthorn2002}
{Freeman} K.,  {Bland-Hawthorn} J.,  2002, \mn@doi [\araa]
  {10.1146/annurev.astro.40.060401.093840}, \href
  {http://adsabs.harvard.edu/abs/2002ARA%26A..40..487F} {40, 487}

\bibitem[\protect\citeauthoryear{{Froese Fischer} \& {Tachiev}}{{Froese
  Fischer} \& {Tachiev}}{2012}]{GESMCHF}
{Froese Fischer} C.,  {Tachiev} G.,  2012, Multiconfiguration Hartree-Fock and
  Multiconfiguration Dirac-Hartree-Fock Collection, Version 2

\bibitem[\protect\citeauthoryear{{Fuhr}, {Martin}  \& {Wiese}}{{Fuhr}
  et~al.}{1988}]{FMW}
{Fuhr} J.~R.,  {Martin} G.~A.,   {Wiese} W.~L.,  1988, Journal of Physical and
  Chemical Reference Data, Volume 17, Suppl.~4.~New York: American Institute of
  Physics (AIP) and American Chemical Society, 1988, \href
  {http://cdsads.u-strasbg.fr/abs/1988JPCRD..17S....F} {17}

\bibitem[\protect\citeauthoryear{{Gao}, {Lind}  \& {Amarsi}}{{Gao}
  et~al.}{2018}]{Gao2018}
{Gao} X.~D.,  {Lind} K.,   {Amarsi} A.~M.,  2018, MNRAS

\bibitem[\protect\citeauthoryear{{Garz}}{{Garz}}{1973}]{GARZ}
{Garz} T.,  1973, \aap, \href
  {http://cdsads.u-strasbg.fr/abs/1973A%26A....26..471G} {26, 471}

\bibitem[\protect\citeauthoryear{{Gilmore} et~al.,}{{Gilmore}
  et~al.}{2012}]{Gilmore2012}
{Gilmore} G.,  et~al., 2012, The Messenger, \href
  {http://adsabs.harvard.edu/abs/2012Msngr.147...25G} {147, 25}

\bibitem[\protect\citeauthoryear{{Gray}}{{Gray}}{2008}]{Gray2008}
{Gray} D.~F.,  2008, {The Observation and Analysis of Stellar Photospheres}.
Cambridge University Press

\bibitem[\protect\citeauthoryear{{Grevesse}, {Blackwell}  \&
  {Petford}}{{Grevesse} et~al.}{1989}]{1989A&A...208..157G}
{Grevesse} N.,  {Blackwell} D.~E.,   {Petford} A.~D.,  1989, \aap, \href
  {http://adsabs.harvard.edu/abs/1989A%26A...208..157G} {208, 157}

\bibitem[\protect\citeauthoryear{{Grevesse}, {Asplund}  \& {Sauval}}{{Grevesse}
  et~al.}{2007}]{Grevesse2007}
{Grevesse} N.,  {Asplund} M.,   {Sauval} A.~J.,  2007, \mn@doi [\ssr]
  {10.1007/s11214-007-9173-7}, \href
  {http://adsabs.harvard.edu/abs/2007SSRv..130..105G} {130, 105}

\bibitem[\protect\citeauthoryear{{Grevesse}, {Scott}, {Asplund}  \&
  {Sauval}}{{Grevesse} et~al.}{2015}]{Grevesse2015}
{Grevesse} N.,  {Scott} P.,  {Asplund} M.,   {Sauval} A.~J.,  2015, \mn@doi
  [\aap] {10.1051/0004-6361/201424111}, \href
  {http://adsabs.harvard.edu/abs/2015A%26A...573A..27G} {573, A27}

\bibitem[\protect\citeauthoryear{{Gustafsson}, {Edvardsson}, {Eriksson},
  {J{\o}rgensen}, {Nordlund}  \& {Plez}}{{Gustafsson}
  et~al.}{2008}]{Gustafsson2008}
{Gustafsson} B.,  {Edvardsson} B.,  {Eriksson} K.,  {J{\o}rgensen} U.~G.,
  {Nordlund} {\AA}.,   {Plez} B.,  2008, \mn@doi [\aap]
  {10.1051/0004-6361:200809724}, \href
  {http://adsabs.harvard.edu/abs/2008A%26A...486..951G} {486, 951}

\bibitem[\protect\citeauthoryear{{Harris}}{{Harris}}{1996}]{Harris1996}
{Harris} W.~E.,  1996, \mn@doi [\aj] {10.1086/118116}, \href
  {http://adsabs.harvard.edu/abs/1996AJ....112.1487H} {112, 1487}

\bibitem[\protect\citeauthoryear{{Hawkins}, {Masseron}, {Jofr{\'e}}, {Gilmore},
  {Elsworth}  \& {Hekker}}{{Hawkins} et~al.}{2016}]{Hawkins2016b}
{Hawkins} K.,  {Masseron} T.,  {Jofr{\'e}} P.,  {Gilmore} G.,  {Elsworth} Y.,
  {Hekker} S.,  2016, \mn@doi [\aap] {10.1051/0004-6361/201628812}, \href
  {http://adsabs.harvard.edu/abs/2016A%26A...594A..43H} {594, A43}

\bibitem[\protect\citeauthoryear{{Heijmans} et~al.,}{{Heijmans}
  et~al.}{2012}]{Heijmans2012}
{Heijmans} J.,  et~al., 2012, in Ground-based and Airborne Instrumentation for
  Astronomy IV. p. 84460W, \mn@doi{10.1117/12.925806}

\bibitem[\protect\citeauthoryear{{Heiter} et~al.,}{{Heiter}
  et~al.}{2015a}]{Heiter2015b}
{Heiter} U.,  et~al., 2015a, \mn@doi [\physscr]
  {10.1088/0031-8949/90/5/054010}, \href
  {http://adsabs.harvard.edu/abs/2015PhyS...90e4010H} {90, 054010}

\bibitem[\protect\citeauthoryear{{Heiter}, {Jofr{\'e}}, {Gustafsson}, {Korn},
  {Soubiran}  \& {Th{\'e}venin}}{{Heiter} et~al.}{2015b}]{Heiter2015}
{Heiter} U.,  {Jofr{\'e}} P.,  {Gustafsson} B.,  {Korn} A.~J.,  {Soubiran} C.,
   {Th{\'e}venin} F.,  2015b, \mn@doi [\aap] {10.1051/0004-6361/201526319},
  \href {http://adsabs.harvard.edu/abs/2015A%26A...582A..49H} {582, A49}

\bibitem[\protect\citeauthoryear{{Hibbert}, {Biemont}, {Godefroid}  \&
  {Vaeck}}{{Hibbert} et~al.}{1993}]{1993A&AS...99..179H}
{Hibbert} A.,  {Biemont} E.,  {Godefroid} M.,   {Vaeck} N.,  1993, \aaps, \href
  {http://adsabs.harvard.edu/abs/1993A%26AS...99..179H} {99, 179}

\bibitem[\protect\citeauthoryear{{Hinkle}, {Wallace}, {Valenti}  \&
  {Harmer}}{{Hinkle} et~al.}{2000}]{Hinkle2000}
{Hinkle} K.,  {Wallace} L.,  {Valenti} J.,   {Harmer} D.,  2000, {Visible and
  Near Infrared Atlas of the Arcturus Spectrum 3727-9300 A}.
ASP

\bibitem[\protect\citeauthoryear{{Ho} et~al.,}{{Ho} et~al.}{2017}]{Ho2017}
{Ho} A.~Y.~Q.,  et~al., 2017, \mn@doi [\apj] {10.3847/1538-4357/836/1/5}, \href
  {http://adsabs.harvard.edu/abs/2017ApJ...836....5H} {836, 5}

\bibitem[\protect\citeauthoryear{{Jofr{\'e}}, {Panter}, {Hansen}  \&
  {Weiss}}{{Jofr{\'e}} et~al.}{2010}]{Jofre2010}
{Jofr{\'e}} P.,  {Panter} B.,  {Hansen} C.~J.,   {Weiss} A.,  2010, \mn@doi
  [\aap] {10.1051/0004-6361/201014013}, \href
  {http://adsabs.harvard.edu/abs/2010A%26A...517A..57J} {517, A57}

\bibitem[\protect\citeauthoryear{{Jofr{\'e}} et~al.,}{{Jofr{\'e}}
  et~al.}{2014}]{Jofre2014}
{Jofr{\'e}} P.,  et~al., 2014, \mn@doi [\aap] {10.1051/0004-6361/201322440},
  \href {http://adsabs.harvard.edu/abs/2014A%26A...564A.133J} {564, A133}

\bibitem[\protect\citeauthoryear{{Johnson} \& {Pilachowski}}{{Johnson} \&
  {Pilachowski}}{2010}]{Johnson2010}
{Johnson} C.~I.,  {Pilachowski} C.~A.,  2010, \mn@doi [\apj]
  {10.1088/0004-637X/722/2/1373}, \href
  {http://adsabs.harvard.edu/abs/2010ApJ...722.1373J} {722, 1373}

\bibitem[\protect\citeauthoryear{{Kausch} et~al.,}{{Kausch}
  et~al.}{2015}]{Kausch2015}
{Kausch} W.,  et~al., 2015, \mn@doi [\aap] {10.1051/0004-6361/201423909}, \href
  {http://adsabs.harvard.edu/abs/2015A%26A...576A..78K} {576, A78}

\bibitem[\protect\citeauthoryear{{Kelleher} \& {Podobedova}}{{Kelleher} \&
  {Podobedova}}{2008}]{Kelleher2008}
{Kelleher} D.~E.,  {Podobedova} L.~I.,  2008, \mn@doi [Journal of Physical and
  Chemical Reference Data] {10.1063/1.2734564}, \href
  {http://adsabs.harvard.edu/abs/2008JPCRD..37..709K} {37, 709}

\bibitem[\protect\citeauthoryear{{Kerkhoff}, {Schmidt}  \&
  {Zimmermann}}{{Kerkhoff} et~al.}{1980}]{1980ZPhyA.298..249K}
{Kerkhoff} H.,  {Schmidt} M.,   {Zimmermann} P.,  1980, \mn@doi [Zeitschrift
  fur Physik A Hadrons and Nuclei] {10.1007/BF01425154}, \href
  {http://adsabs.harvard.edu/abs/1980ZPhyA.298..249K} {298, 249}

\bibitem[\protect\citeauthoryear{{Kharchenko}, {Piskunov}, {Schilbach},
  {R{\"o}ser}  \& {Scholz}}{{Kharchenko} et~al.}{2013}]{Kharchenko2013}
{Kharchenko} N.~V.,  {Piskunov} A.~E.,  {Schilbach} E.,  {R{\"o}ser} S.,
  {Scholz} R.-D.,  2013, \mn@doi [\aap] {10.1051/0004-6361/201322302}, \href
  {http://adsabs.harvard.edu/abs/2013A%26A...558A..53K} {558, A53}

\bibitem[\protect\citeauthoryear{Kjeldsen \& Bedding}{Kjeldsen \&
  Bedding}{1995}]{Kjeldsen1995}
Kjeldsen H.,  Bedding T.~R.,  1995, \mn@doi [Astronomy \& Astrophysics]
  {10.1007/s13398-014-0173-7.2}, 293, 87

\bibitem[\protect\citeauthoryear{{Kock} \& {Richter}}{{Kock} \&
  {Richter}}{1968}]{KR}
{Kock} M.,  {Richter} J.,  1968, \zap, \href
  {http://cdsads.u-strasbg.fr/abs/1968ZA.....69..180K} {69, 180}

\bibitem[\protect\citeauthoryear{{Kollmeier} et~al.,}{{Kollmeier}
  et~al.}{2017}]{Kollmeier2017}
{Kollmeier} J.~A.,  et~al., 2017, preprint, \href
  {http://adsabs.harvard.edu/abs/2017arXiv171103234K} {} (\mn@eprint {arXiv}
  {1711.03234})

\bibitem[\protect\citeauthoryear{{Kos} \& {Zwitter}}{{Kos} \&
  {Zwitter}}{2013}]{KosZwitter2013}
{Kos} J.,  {Zwitter} T.,  2013, \mn@doi [\apj] {10.1088/0004-637X/774/1/72},
  \href {http://adsabs.harvard.edu/abs/2013ApJ...774...72K} {774, 72}

\bibitem[\protect\citeauthoryear{{Kos} et~al.,}{{Kos} et~al.}{2017}]{Kos2017}
{Kos} J.,  et~al., 2017, \mn@doi [\mnras] {10.1093/mnras/stw2064}, \href
  {http://adsabs.harvard.edu/abs/2017MNRAS.464.1259K} {464, 1259}

\bibitem[\protect\citeauthoryear{{Kos} et~al.,}{{Kos} et~al.}{2018}]{Kos2018}
{Kos} J.,  et~al., 2018, \mn@doi [\mnras] {10.1093/mnras/stx2637}, \href
  {http://adsabs.harvard.edu/abs/2018MNRAS.473.4612K} {473, 4612}

\bibitem[\protect\citeauthoryear{{Kunder} et~al.,}{{Kunder}
  et~al.}{2017}]{Kunder2017}
{Kunder} A.,  et~al., 2017, \mn@doi [\aj] {10.3847/1538-3881/153/2/75}, \href
  {http://adsabs.harvard.edu/abs/2017AJ....153...75K} {153, 75}

\bibitem[\protect\citeauthoryear{{Kurucz}}{{Kurucz}}{1975}]{K75}
{Kurucz} R.~L.,  1975, Robert L. Kurucz on-line database of observed and
  predicted atomic transitions, (K75)

\bibitem[\protect\citeauthoryear{{Kurucz}}{{Kurucz}}{2006}]{K06}
{Kurucz} R.~L.,  2006, Robert L. Kurucz on-line database of observed and
  predicted atomic transitions

\bibitem[\protect\citeauthoryear{{Kurucz}}{{Kurucz}}{2007}]{K07}
{Kurucz} R.~L.,  2007, Robert L. Kurucz on-line database of observed and
  predicted atomic transitions

\bibitem[\protect\citeauthoryear{{Kurucz}}{{Kurucz}}{2008}]{K08}
{Kurucz} R.~L.,  2008, Robert L. Kurucz on-line database of observed and
  predicted atomic transitions

\bibitem[\protect\citeauthoryear{{Kurucz}}{{Kurucz}}{2009}]{K09}
{Kurucz} R.~L.,  2009, Robert L. Kurucz on-line database of observed and
  predicted atomic transitions

\bibitem[\protect\citeauthoryear{{Kurucz}}{{Kurucz}}{2013}]{K13}
{Kurucz} R.~L.,  2013, Robert L. Kurucz on-line database of observed and
  predicted atomic transitions

\bibitem[\protect\citeauthoryear{{Kurucz}}{{Kurucz}}{2014}]{K14}
{Kurucz} R.~L.,  2014, Robert L. Kurucz on-line database of observed and
  predicted atomic transitions

\bibitem[\protect\citeauthoryear{{Kurucz} \& {Peytremann}}{{Kurucz} \&
  {Peytremann}}{1975}]{KP}
{Kurucz} R.~L.,  {Peytremann} E.,  1975, SAO Special Report, \href
  {http://cdsads.u-strasbg.fr/abs/1975SAOSR.362.....K} {362, 1}

\bibitem[\protect\citeauthoryear{{Lawler} \& {Dakin}}{{Lawler} \&
  {Dakin}}{1989}]{LD}
{Lawler} J.~E.,  {Dakin} J.~T.,  1989, \mn@doi [Journal of the Optical Society
  of America B Optical Physics] {10.1364/JOSAB.6.001457}, \href
  {http://adsabs.harvard.edu/abs/1989JOSAB...6.1457L} {6, 1457}

\bibitem[\protect\citeauthoryear{{Lawler}, {Bonvallet}  \& {Sneden}}{{Lawler}
  et~al.}{2001a}]{LBS}
{Lawler} J.~E.,  {Bonvallet} G.,   {Sneden} C.,  2001a, \mn@doi [Astrophys. J.]
  {10.1086/321549}, 556, 452

\bibitem[\protect\citeauthoryear{{Lawler}, {Wickliffe}, {den Hartog}  \&
  {Sneden}}{{Lawler} et~al.}{2001b}]{LWHS}
{Lawler} J.~E.,  {Wickliffe} M.~E.,  {den Hartog} E.~A.,   {Sneden} C.,  2001b,
  \mn@doi [Astrophys. J.] {10.1086/323407}, 563, 1075

\bibitem[\protect\citeauthoryear{{Lawler}, {Den Hartog}, {Sneden}  \&
  {Cowan}}{{Lawler} et~al.}{2006}]{LD-HS}
{Lawler} J.~E.,  {Den Hartog} E.~A.,  {Sneden} C.,   {Cowan} J.~J.,  2006,
  \mn@doi [Astrophys. J. Suppl. Ser.] {10.1086/498213}, 162, 227

\bibitem[\protect\citeauthoryear{{Lawler}, {Sneden}, {Cowan}, {Ivans}  \& {Den
  Hartog}}{{Lawler} et~al.}{2009}]{LSCI}
{Lawler} J.~E.,  {Sneden} C.,  {Cowan} J.~J.,  {Ivans} I.~I.,   {Den Hartog}
  E.~A.,  2009, \mn@doi [Astrophys. J. Suppl. Ser.]
  {10.1088/0067-0049/182/1/51}, 182, 51

\bibitem[\protect\citeauthoryear{{Lawler}, {Guzman}, {Wood}, {Sneden}  \&
  {Cowan}}{{Lawler} et~al.}{2013a}]{LGWSC}
{Lawler} J.~E.,  {Guzman} A.,  {Wood} M.~P.,  {Sneden} C.,   {Cowan} J.~J.,
  2013a, \mn@doi [\apjs] {10.1088/0067-0049/205/2/11}, \href
  {http://adsabs.harvard.edu/abs/2013ApJS..205...11L} {205, 11}

\bibitem[\protect\citeauthoryear{{Lawler}, {Guzman}, {Wood}, {Sneden}  \&
  {Cowan}}{{Lawler} et~al.}{2013b}]{2013ApJS..205...11L}
{Lawler} J.~E.,  {Guzman} A.,  {Wood} M.~P.,  {Sneden} C.,   {Cowan} J.~J.,
  2013b, \mn@doi [\apjs] {10.1088/0067-0049/205/2/11}, \href
  {http://adsabs.harvard.edu/abs/2013ApJS..205...11L} {205, 11}

\bibitem[\protect\citeauthoryear{{Lawler}, {Wood}, {Den Hartog}, {Feigenson},
  {Sneden}  \& {Cowan}}{{Lawler} et~al.}{2014}]{2014ApJS..215...20L}
{Lawler} J.~E.,  {Wood} M.~P.,  {Den Hartog} E.~A.,  {Feigenson} T.,  {Sneden}
  C.,   {Cowan} J.~J.,  2014, \mn@doi [\apjs] {10.1088/0067-0049/215/2/20},
  \href {http://adsabs.harvard.edu/abs/2014ApJS..215...20L} {215, 20}

\bibitem[\protect\citeauthoryear{{Lennard}, {Whaling}, {Scalo}  \&
  {Testerman}}{{Lennard} et~al.}{1975}]{LWST}
{Lennard} W.~N.,  {Whaling} W.,  {Scalo} J.~M.,   {Testerman} L.,  1975,
  \mn@doi [\apj] {10.1086/153538}, \href
  {http://cdsads.u-strasbg.fr/abs/1975ApJ...197..517L} {197, 517}

\bibitem[\protect\citeauthoryear{{Lewis} et~al.,}{{Lewis}
  et~al.}{2002}]{Lewis2002}
{Lewis} I.~J.,  et~al., 2002, \mn@doi [\mnras]
  {10.1046/j.1365-8711.2002.05333.x}, \href
  {http://adsabs.harvard.edu/abs/2002MNRAS.333..279L} {333, 279}

\bibitem[\protect\citeauthoryear{{Lin}, {Dotter}, {Ting}  \& {Asplund}}{{Lin}
  et~al.}{2018}]{Lin2018}
{Lin} J.,  {Dotter} A.,  {Ting} Y.-S.,   {Asplund} M.,  2018, \mn@doi [\mnras]
  {10.1093/mnras/sty709}, \href
  {http://adsabs.harvard.edu/abs/2018MNRAS.tmp..697L} {}

\bibitem[\protect\citeauthoryear{{Lind}, {Asplund}  \& {Barklem}}{{Lind}
  et~al.}{2009}]{Lind2009}
{Lind} K.,  {Asplund} M.,   {Barklem} P.~S.,  2009, \mn@doi [\aap]
  {10.1051/0004-6361/200912221}, \href
  {http://adsabs.harvard.edu/abs/2009A%26A...503..541L} {503, 541}

\bibitem[\protect\citeauthoryear{{Lind}, {Asplund}, {Barklem}  \&
  {Belyaev}}{{Lind} et~al.}{2011}]{Lind2011}
{Lind} K.,  {Asplund} M.,  {Barklem} P.~S.,   {Belyaev} A.~K.,  2011, \mn@doi
  [\aap] {10.1051/0004-6361/201016095}, \href
  {http://adsabs.harvard.edu/abs/2011A%26A...528A.103L} {528, A103}

\bibitem[\protect\citeauthoryear{{Lindegren} et~al.,}{{Lindegren}
  et~al.}{2016}]{Lindegren2016}
{Lindegren} L.,  et~al., 2016, \mn@doi [\aap] {10.1051/0004-6361/201628714},
  \href {http://adsabs.harvard.edu/abs/2016A%26A...595A...4L} {595, A4}

\bibitem[\protect\citeauthoryear{{Luo} et~al.,}{{Luo} et~al.}{2015}]{Luo2015}
{Luo} A.-L.,  et~al., 2015, \mn@doi [Research in Astronomy and Astrophysics]
  {10.1088/1674-4527/15/8/002}, \href
  {http://adsabs.harvard.edu/abs/2015RAA....15.1095L} {15, 1095}

\bibitem[\protect\citeauthoryear{{Magic}, {Collet}, {Asplund}, {Trampedach},
  {Hayek}, {Chiavassa}, {Stein}  \& {Nordlund}}{{Magic}
  et~al.}{2013}]{Magic2013}
{Magic} Z.,  {Collet} R.,  {Asplund} M.,  {Trampedach} R.,  {Hayek} W.,
  {Chiavassa} A.,  {Stein} R.~F.,   {Nordlund} {\AA}.,  2013, \mn@doi [\aap]
  {10.1051/0004-6361/201321274}, \href
  {http://adsabs.harvard.edu/abs/2013A%26A...557A..26M} {557, A26}

\bibitem[\protect\citeauthoryear{{Majewski}, {Zasowski}  \&
  {Nidever}}{{Majewski} et~al.}{2011}]{Majewski2011}
{Majewski} S.~R.,  {Zasowski} G.,   {Nidever} D.~L.,  2011, \mn@doi [\apj]
  {10.1088/0004-637X/739/1/25}, \href
  {http://adsabs.harvard.edu/abs/2011ApJ...739...25M} {739, 25}

\bibitem[\protect\citeauthoryear{{Marigo} et~al.,}{{Marigo}
  et~al.}{2017}]{Marigo2017}
{Marigo} P.,  et~al., 2017, \mn@doi [\apj] {10.3847/1538-4357/835/1/77}, \href
  {http://adsabs.harvard.edu/abs/2017ApJ...835...77M} {835, 77}

\bibitem[\protect\citeauthoryear{{Martell} et~al.,}{{Martell}
  et~al.}{2017}]{Martell2017}
{Martell} S.~L.,  et~al., 2017, \mn@doi [\mnras] {10.1093/mnras/stw2835}, \href
  {http://adsabs.harvard.edu/abs/2017MNRAS.465.3203M} {465, 3203}

\bibitem[\protect\citeauthoryear{{May}, {Richter}  \& {Wichelmann}}{{May}
  et~al.}{1974}]{MRW}
{May} M.,  {Richter} J.,   {Wichelmann} J.,  1974, \aaps, \href
  {http://cdsads.u-strasbg.fr/abs/1974A%26AS...18..405M} {18, 405}

\bibitem[\protect\citeauthoryear{Mckinney}{Mckinney}{2011}]{McKinney2011}
Mckinney W.,  2011, in Python High Performance Science Computer.

\bibitem[\protect\citeauthoryear{{Meggers}, {Corliss}  \& {Scribner}}{{Meggers}
  et~al.}{1975}]{MC}
{Meggers} W.~F.,  {Corliss} C.~H.,   {Scribner} B.~F.,  1975, {Tables of
  spectral-line intensities. Part I, II\_- arranged by elements.}.
NBS

\bibitem[\protect\citeauthoryear{{Mel{\'e}ndez} \& {Barbuy}}{{Mel{\'e}ndez} \&
  {Barbuy}}{2009}]{2009A&A...497..611M}
{Mel{\'e}ndez} J.,  {Barbuy} B.,  2009, \mn@doi [\aap]
  {10.1051/0004-6361/200811508}, \href
  {http://adsabs.harvard.edu/abs/2009A%26A...497..611M} {497, 611}

\bibitem[\protect\citeauthoryear{{Mendoza}, {Eissner}, {LeDourneuf}  \&
  {Zeippen}}{{Mendoza} et~al.}{1995}]{1995JPhB...28.3485M}
{Mendoza} C.,  {Eissner} W.,  {LeDourneuf} M.,   {Zeippen} C.~J.,  1995,
  \mn@doi [Journal of Physics B Atomic Molecular Physics]
  {10.1088/0953-4075/28/16/006}, \href
  {http://adsabs.harvard.edu/abs/1995JPhB...28.3485M} {28, 3485}

\bibitem[\protect\citeauthoryear{{Miglio} et~al.,}{{Miglio}
  et~al.}{2017}]{Miglio2017}
{Miglio} A.,  et~al., 2017, \mn@doi [Astronomische Nachrichten]
  {10.1002/asna.201713385}, \href
  {http://adsabs.harvard.edu/abs/2017AN....338..644M} {338, 644}

\bibitem[\protect\citeauthoryear{{Miszalski}, {Shortridge}, {Saunders},
  {Parker}  \& {Croom}}{{Miszalski} et~al.}{2006}]{Miszalski2006}
{Miszalski} B.,  {Shortridge} K.,  {Saunders} W.,  {Parker} Q.~A.,   {Croom}
  S.~M.,  2006, \mn@doi [\mnras] {10.1111/j.1365-2966.2006.10777.x}, \href
  {http://adsabs.harvard.edu/abs/2006MNRAS.371.1537M} {371, 1537}

\bibitem[\protect\citeauthoryear{{Nahar}}{{Nahar}}{1993}]{1993PhyS...48..297N}
{Nahar} S.~N.,  1993, \mn@doi [\physscr] {10.1088/0031-8949/48/3/008}, \href
  {http://adsabs.harvard.edu/abs/1993PhyS...48..297N} {48, 297}

\bibitem[\protect\citeauthoryear{{Nataf} et~al.,}{{Nataf}
  et~al.}{2016}]{Nataf2016}
{Nataf} D.~M.,  et~al., 2016, \mn@doi [\mnras] {10.1093/mnras/stv2843}, \href
  {http://adsabs.harvard.edu/abs/2016MNRAS.456.2692N} {456, 2692}

\bibitem[\protect\citeauthoryear{{Ness}, {Hogg}, {Rix}, {Ho}  \&
  {Zasowski}}{{Ness} et~al.}{2015}]{Ness2015}
{Ness} M.,  {Hogg} D.~W.,  {Rix} H.-W.,  {Ho} A.~Y.~Q.,   {Zasowski} G.,  2015,
  \mn@doi [\apj] {10.1088/0004-637X/808/1/16}, \href
  {http://adsabs.harvard.edu/abs/2015ApJ...808...16N} {808, 16}

\bibitem[\protect\citeauthoryear{{Nissen}, {Hoeg}  \& {Schuster}}{{Nissen}
  et~al.}{1997}]{Nissen1997}
{Nissen} P.~E.,  {Hoeg} E.,   {Schuster} W.~J.,  1997, in {Bonnet} R.~M.,
  et~al., eds,  ESA Special Publication Vol. 402, Hipparcos - Venice '97. pp
  225--230

\bibitem[\protect\citeauthoryear{{Nordlander} \& {Lind}}{{Nordlander} \&
  {Lind}}{2017}]{Nordlander2017}
{Nordlander} T.,  {Lind} K.,  2017, \mn@doi [\aap]
  {10.1051/0004-6361/201730427}, \href
  {http://adsabs.harvard.edu/abs/2017A%26A...607A..75N} {607, A75}

\bibitem[\protect\citeauthoryear{{O'Brian}, {Wickliffe}, {Lawler}, {Whaling}
  \& {Brault}}{{O'Brian} et~al.}{1991}]{BWL}
{O'Brian} T.~R.,  {Wickliffe} M.~E.,  {Lawler} J.~E.,  {Whaling} W.,   {Brault}
  J.~W.,  1991, Journal of the Optical Society of America B Optical Physics, 8,
  1185

\bibitem[\protect\citeauthoryear{{O'brian} \& {Lawler}}{{O'brian} \&
  {Lawler}}{1991}]{BL}
{O'brian} T.~R.,  {Lawler} J.~E.,  1991, \mn@doi [\pra]
  {10.1103/PhysRevA.44.7134}, \href
  {http://adsabs.harvard.edu/abs/1991PhRvA..44.7134O} {44, 7134}

\bibitem[\protect\citeauthoryear{{Oh}, {Price-Whelan}, {Hogg}, {Morton}  \&
  {Spergel}}{{Oh} et~al.}{2017}]{Oh2017}
{Oh} S.,  {Price-Whelan} A.~M.,  {Hogg} D.~W.,  {Morton} T.~D.,   {Spergel}
  D.~N.,  2017, \mn@doi [\aj] {10.3847/1538-3881/aa6ffd}, \href
  {http://adsabs.harvard.edu/abs/2017AJ....153..257O} {153, 257}

\bibitem[\protect\citeauthoryear{{Osorio}, {Barklem}, {Lind}, {Belyaev},
  {Spielfiedel}, {Guitou}  \& {Feautrier}}{{Osorio} et~al.}{2015}]{Osorio2015}
{Osorio} Y.,  {Barklem} P.~S.,  {Lind} K.,  {Belyaev} A.~K.,  {Spielfiedel} A.,
   {Guitou} M.,   {Feautrier} N.,  2015, \mn@doi [\aap]
  {10.1051/0004-6361/201525846}, \href
  {http://adsabs.harvard.edu/abs/2015A%26A...579A..53O} {579, A53}

\bibitem[\protect\citeauthoryear{{Palmeri}, {Quinet}, {Wyart}  \&
  {Bi{\'e}mont}}{{Palmeri} et~al.}{2000}]{PQWB}
{Palmeri} P.,  {Quinet} P.,  {Wyart} J.,   {Bi{\'e}mont} E.,  2000, \mn@doi
  [Physica Scripta] {10.1238/Physica.Regular.061a00323}, 61, 323

\bibitem[\protect\citeauthoryear{Perryman et~al.,}{Perryman
  et~al.}{1997}]{Perryman1997}
Perryman M. A.~C.,  et~al., 1997, Astronomy \& Astrophysics, 323, L49

\bibitem[\protect\citeauthoryear{{Perryman} et~al.,}{{Perryman}
  et~al.}{2001}]{Perryman2001}
{Perryman} M.~A.~C.,  et~al., 2001, \mn@doi [\aap]
  {10.1051/0004-6361:20010085}, \href
  {http://adsabs.harvard.edu/abs/2001A%26A...369..339P} {369, 339}

\bibitem[\protect\citeauthoryear{{Piskunov} \& {Valenti}}{{Piskunov} \&
  {Valenti}}{2017}]{Piskunov2017}
{Piskunov} N.,  {Valenti} J.~A.,  2017, \mn@doi [\aap]
  {10.1051/0004-6361/201629124}, \href
  {http://adsabs.harvard.edu/abs/2017A%26A...597A..16P} {597, A16}

\bibitem[\protect\citeauthoryear{{Quillen} et~al.,}{{Quillen}
  et~al.}{2018}]{Quillen2018}
{Quillen} A.~C.,  et~al., 2018, preprint, \href
  {http://adsabs.harvard.edu/abs/2018arXiv180202924Q} {} (\mn@eprint {arXiv}
  {1802.02924})

\bibitem[\protect\citeauthoryear{{Raassen} \& {Uylings}}{{Raassen} \&
  {Uylings}}{1998}]{RU}
{Raassen} A.~J.~J.,  {Uylings} P.~H.~M.,  1998, \aap, \href
  {http://adsabs.harvard.edu/abs/1998A%26A...340..300R} {340, 300}

\bibitem[\protect\citeauthoryear{{Ralchenko}, {Kramida}, {Reader}  \& {NIST ASD
  Team}}{{Ralchenko} et~al.}{2010}]{NIST10}
{Ralchenko} Y.,  {Kramida} A.,  {Reader} J.,   {NIST ASD Team} 2010, NIST
  Atomic Spectra Database (ver. 4.0.0), [Online].

\bibitem[\protect\citeauthoryear{{Randich}, {Gilmore}  \& {Gaia-ESO
  Consortium}}{{Randich} et~al.}{2013}]{Randich2013}
{Randich} S.,  {Gilmore} G.,   {Gaia-ESO Consortium} 2013, The Messenger, \href
  {http://adsabs.harvard.edu/abs/2013Msngr.154...47R} {154, 47}

\bibitem[\protect\citeauthoryear{{Rauer} et~al.,}{{Rauer}
  et~al.}{2014}]{Rauer2014}
{Rauer} H.,  et~al., 2014, \mn@doi [Experimental Astronomy]
  {10.1007/s10686-014-9383-4}, \href
  {http://adsabs.harvard.edu/abs/2014ExA....38..249R} {38, 249}

\bibitem[\protect\citeauthoryear{{Ricker} et~al.,}{{Ricker}
  et~al.}{2015}]{Ricker2015}
{Ricker} G.~R.,  et~al., 2015, \mn@doi [Journal of Astronomical Telescopes,
  Instruments, and Systems] {10.1117/1.JATIS.1.1.014003}, \href
  {http://adsabs.harvard.edu/abs/2015JATIS...1a4003R} {1, 014003}

\bibitem[\protect\citeauthoryear{{Robitaille} \& {Bressert}}{{Robitaille} \&
  {Bressert}}{2012}]{Robitaille2012}
{Robitaille} T.,  {Bressert} E.,  2012, {APLpy: Astronomical Plotting Library
  in Python}, Astrophysics Source Code Library (\mn@eprint {ascl} {1208.017})

\bibitem[\protect\citeauthoryear{{Ruffoni}, {Den Hartog}, {Lawler}, {Brewer},
  {Lind}, {Nave}  \& {Pickering}}{{Ruffoni} et~al.}{2014}]{2014MNRAS.441.3127R}
{Ruffoni} M.~P.,  {Den Hartog} E.~A.,  {Lawler} J.~E.,  {Brewer} N.~R.,  {Lind}
  K.,  {Nave} G.,   {Pickering} J.~C.,  2014, \mn@doi [\mnras]
  {10.1093/mnras/stu780}, \href
  {http://adsabs.harvard.edu/abs/2014MNRAS.441.3127R} {441, 3127}

\bibitem[\protect\citeauthoryear{{Ryabchikova}, {Piskunov}, {Stempels}, {Kupka}
   \& {Weiss}}{{Ryabchikova} et~al.}{1999}]{T83av}
{Ryabchikova} T.~A.,  {Piskunov} N.~E.,  {Stempels} H.~C.,  {Kupka} F.,
  {Weiss} W.~W.,  1999, \mn@doi [Physica Scripta Volume T]
  {10.1238/Physica.Topical.083a00162}, 83, 162

\bibitem[\protect\citeauthoryear{{Saito} et~al.,}{{Saito}
  et~al.}{2012}]{Saito2012}
{Saito} R.~K.,  et~al., 2012, \mn@doi [\aap] {10.1051/0004-6361/201219448},
  \href {http://adsabs.harvard.edu/abs/2012A%26A...544A.147S} {544, A147}

\bibitem[\protect\citeauthoryear{{Sharma}, {Stello}, {Bland-Hawthorn}, {Huber}
  \& {Bedding}}{{Sharma} et~al.}{2016}]{Sharma2016}
{Sharma} S.,  {Stello} D.,  {Bland-Hawthorn} J.,  {Huber} D.,   {Bedding}
  T.~R.,  2016, \mn@doi [\apj] {10.3847/0004-637X/822/1/15}, \href
  {http://adsabs.harvard.edu/abs/2016ApJ...822...15S} {822, 15}

\bibitem[\protect\citeauthoryear{{Sharma} et~al.,}{{Sharma}
  et~al.}{2018}]{Sharma2018}
{Sharma} S.,  et~al., 2018, \mn@doi [\mnras] {10.1093/mnras/stx2582}, \href
  {http://adsabs.harvard.edu/abs/2018MNRAS.473.2004S} {473, 2004}

\bibitem[\protect\citeauthoryear{{Sheinis} et~al.,}{{Sheinis}
  et~al.}{2015}]{Sheinis2015}
{Sheinis} A.,  et~al., 2015, \mn@doi [Journal of Astronomical Telescopes,
  Instruments, and Systems] {10.1117/1.JATIS.1.3.035002}, \href
  {http://adsabs.harvard.edu/abs/2015JATIS...1c5002S} {1, 035002}

\bibitem[\protect\citeauthoryear{{Simpson} et~al.,}{{Simpson}
  et~al.}{2017}]{Simpson2017}
{Simpson} J.~D.,  et~al., 2017, \mn@doi [\mnras] {10.1093/mnras/stx1892}, \href
  {http://adsabs.harvard.edu/abs/2017MNRAS.471.4087S} {471, 4087}

\bibitem[\protect\citeauthoryear{Skrutskie et~al.,}{Skrutskie
  et~al.}{2006}]{Skrutskie2006}
Skrutskie M.~F.,  et~al., 2006, \mn@doi [The Astronomical Journal]
  {10.1086/498708}, 131, 1163

\bibitem[\protect\citeauthoryear{{Smette, A.} et~al.,}{{Smette, A.}
  et~al.}{2015}]{Smette2015}
{Smette, A.} et~al., 2015, \mn@doi [A&A] {10.1051/0004-6361/201423932}, 576,
  A77

\bibitem[\protect\citeauthoryear{{Smiljanic} et~al.,}{{Smiljanic}
  et~al.}{2014}]{Smiljanic2014}
{Smiljanic} R.,  et~al., 2014, \mn@doi [\aap] {10.1051/0004-6361/201423937},
  \href {http://adsabs.harvard.edu/abs/2014A%26A...570A.122S} {570, A122}

\bibitem[\protect\citeauthoryear{{Smith}}{{Smith}}{1988}]{S}
{Smith} G.,  1988, \mn@doi [Journal of Physics B Atomic Molecular Physics]
  {10.1088/0953-4075/21/16/008}, 21, 2827

\bibitem[\protect\citeauthoryear{{Smith} \& {Raggett}}{{Smith} \&
  {Raggett}}{1981}]{SR}
{Smith} G.,  {Raggett} D.~S.~J.,  1981, \mn@doi [Journal of Physics B Atomic
  Molecular Physics] {10.1088/0022-3700/14/21/016}, 14, 4015

\bibitem[\protect\citeauthoryear{{Sobeck}, {Lawler}  \& {Sneden}}{{Sobeck}
  et~al.}{2007}]{SLS}
{Sobeck} J.~S.,  {Lawler} J.~E.,   {Sneden} C.,  2007, \mn@doi [Astrophys. J.]
  {10.1086/519987}, 667, 1267

\bibitem[\protect\citeauthoryear{{Spite} \& {Spite}}{{Spite} \&
  {Spite}}{1982}]{Spite1982}
{Spite} F.,  {Spite} M.,  1982, \aap, \href
  {http://adsabs.harvard.edu/abs/1982A%26A...115..357S} {115, 357}

\bibitem[\protect\citeauthoryear{{Stello} et~al.,}{{Stello}
  et~al.}{2015}]{Stello2015}
{Stello} D.,  et~al., 2015, \mn@doi [\apjl] {10.1088/2041-8205/809/1/L3}, \href
  {http://adsabs.harvard.edu/abs/2015ApJ...809L...3S} {809, L3}

\bibitem[\protect\citeauthoryear{Stello et~al.,}{Stello
  et~al.}{2017}]{Stello2017}
Stello D.,  et~al., 2017, \mn@doi [The Astrophysical Journal]
  {10.3847/1538-4357/835/1/83}, 835, 83

\bibitem[\protect\citeauthoryear{{Taylor}}{{Taylor}}{2005}]{Taylor2005}
{Taylor} M.~B.,  2005, in {Shopbell} P.,  {Britton} M.,   {Ebert} R.,  eds,
  Astronomical Society of the Pacific Conference Series Vol. 347, Astronomical
  Data Analysis Software and Systems XIV. p.~29

\bibitem[\protect\citeauthoryear{{The Astropy Collaboration} et~al.,}{{The
  Astropy Collaboration} et~al.}{2018}]{PriceWhelan2018}
{The Astropy Collaboration} et~al., 2018, preprint, \href
  {http://adsabs.harvard.edu/abs/2018arXiv180102634T} {} (\mn@eprint {arXiv}
  {1801.02634})

\bibitem[\protect\citeauthoryear{{Ting}, {Conroy}, {Rix}  \& {Asplund}}{{Ting}
  et~al.}{2018}]{Ting2018}
{Ting} Y.-S.,  {Conroy} C.,  {Rix} H.-W.,   {Asplund} M.,  2018, preprint,
  \href {http://adsabs.harvard.edu/abs/2018arXiv180107370T} {} (\mn@eprint
  {arXiv} {1801.07370})

\bibitem[\protect\citeauthoryear{{Tody}}{{Tody}}{1986}]{Tody1986}
{Tody} D.,  1986, in {Crawford} D.~L.,  ed.,  \procspie Vol. 627,
  Instrumentation in astronomy VI. p.~733, \mn@doi{10.1117/12.968154}

\bibitem[\protect\citeauthoryear{{Tody}}{{Tody}}{1993}]{Tody1993}
{Tody} D.,  1993, in {Hanisch} R.~J.,  {Brissenden} R.~J.~V.,   {Barnes} J.,
  eds,  Astronomical Society of the Pacific Conference Series Vol. 52,
  Astronomical Data Analysis Software and Systems II. p.~173

\bibitem[\protect\citeauthoryear{{Traven} et~al.,}{{Traven}
  et~al.}{2017}]{Traven2017}
{Traven} G.,  et~al., 2017, \mn@doi [\apjs] {10.3847/1538-4365/228/2/24}, \href
  {http://adsabs.harvard.edu/abs/2017ApJS..228...24T} {228, 24}

\bibitem[\protect\citeauthoryear{Ulyanov}{Ulyanov}{2016}]{Ulyanov2016}
Ulyanov D.,  2016, Multicore-TSNE,
  \url{https://github.com/DmitryUlyanov/Multicore-TSNE}

\bibitem[\protect\citeauthoryear{{Vaeck}, {Godefroid}  \& {Hansen}}{{Vaeck}
  et~al.}{1988}]{VGH}
{Vaeck} N.,  {Godefroid} M.,   {Hansen} J.~E.,  1988, \mn@doi [\pra]
  {10.1103/PhysRevA.38.2830}, \href
  {http://adsabs.harvard.edu/abs/1988PhRvA..38.2830V} {38, 2830}

\bibitem[\protect\citeauthoryear{{Valenti} \& {Piskunov}}{{Valenti} \&
  {Piskunov}}{1996}]{Valenti1996}
{Valenti} J.~A.,  {Piskunov} N.,  1996, \aaps, \href
  {http://adsabs.harvard.edu/abs/1996A%26AS..118..595V} {118, 595}

\bibitem[\protect\citeauthoryear{{Van Dokkum}}{{Van Dokkum}}{2001}]{Dokkum2001}
{Van Dokkum} P.~G.,  2001, \mn@doi [\pasp] {10.1086/323894}, \href
  {http://adsabs.harvard.edu/abs/2001PASP..113.1420V} {113, 1420}

\bibitem[\protect\citeauthoryear{{Wang}, {Li}, {Wang}, {Williams}, {Gould}  \&
  {Stwalley}}{{Wang} et~al.}{1997}]{Wang1997}
{Wang} H.,  {Li} J.,  {Wang} X.~T.,  {Williams} C.~J.,  {Gould} P.~L.,
  {Stwalley} W.~C.,  1997, \mn@doi [\pra] {10.1103/PhysRevA.55.R1569}, \href
  {http://adsabs.harvard.edu/abs/1997PhRvA..55.1569W} {55, R1569}

\bibitem[\protect\citeauthoryear{{Whaling} \& {Brault}}{{Whaling} \&
  {Brault}}{1988}]{WBb}
{Whaling} W.,  {Brault} J.~W.,  1988, \mn@doi [\physscr]
  {10.1088/0031-8949/38/5/010}, \href
  {http://cdsads.u-strasbg.fr/abs/1988PhyS...38..707W} {38, 707}

\bibitem[\protect\citeauthoryear{{Whaling}, {Hannaford}, {Lowe}, {Biemont}  \&
  {Grevesse}}{{Whaling} et~al.}{1985}]{1985A&A...153..109W}
{Whaling} W.,  {Hannaford} P.,  {Lowe} R.~M.,  {Biemont} E.,   {Grevesse} N.,
  1985, \aap, \href {http://adsabs.harvard.edu/abs/1985A%26A...153..109W} {153,
  109}

\bibitem[\protect\citeauthoryear{{Wickliffe}, {Salih}  \& {Lawler}}{{Wickliffe}
  et~al.}{1994}]{WSL}
{Wickliffe} M.~E.,  {Salih} S.,   {Lawler} J.~E.,  1994, \mn@doi [\jqsrt]
  {10.1016/0022-4073(94)90108-2}, \href
  {http://adsabs.harvard.edu/abs/1994JQSRT..51..545W} {51, 545}

\bibitem[\protect\citeauthoryear{{Wittenmyer} et~al.,}{{Wittenmyer}
  et~al.}{2018}]{Wittenmyer2018}
{Wittenmyer} R.~A.,  et~al., 2018, \mn@doi [\aj] {10.3847/1538-3881/aaa3e4},
  \href {http://adsabs.harvard.edu/abs/2018AJ....155...84W} {155, 84}

\bibitem[\protect\citeauthoryear{{Wojno} et~al.,}{{Wojno}
  et~al.}{2017}]{Wojno2017}
{Wojno} J.,  et~al., 2017, \mn@doi [\mnras] {10.1093/mnras/stx606}, \href
  {http://adsabs.harvard.edu/abs/2017MNRAS.468.3368W} {468, 3368}

\bibitem[\protect\citeauthoryear{{Wolf} et~al.,}{{Wolf}
  et~al.}{2018}]{Wolf2018}
{Wolf} C.,  et~al., 2018, \mn@doi [\pasa] {10.1017/pasa.2018.5}, \href
  {http://adsabs.harvard.edu/abs/2018PASA...35...10W} {35, e010}

\bibitem[\protect\citeauthoryear{{Wood}, {Lawler}, {Sneden}  \& {Cowan}}{{Wood}
  et~al.}{2013}]{WLSC}
{Wood} M.~P.,  {Lawler} J.~E.,  {Sneden} C.,   {Cowan} J.~J.,  2013, \mn@doi
  [\apjs] {10.1088/0067-0049/208/2/27}, \href
  {http://adsabs.harvard.edu/abs/2013ApJS..208...27W} {208, 27}

\bibitem[\protect\citeauthoryear{{Wood}, {Lawler}, {Sneden}  \& {Cowan}}{{Wood}
  et~al.}{2014}]{2014ApJS..211...20W}
{Wood} M.~P.,  {Lawler} J.~E.,  {Sneden} C.,   {Cowan} J.~J.,  2014, \mn@doi
  [\apjs] {10.1088/0067-0049/211/2/20}, \href
  {http://adsabs.harvard.edu/abs/2014ApJS..211...20W} {211, 20}

\bibitem[\protect\citeauthoryear{{Yan}, {Tambasco}  \& {Drake}}{{Yan}
  et~al.}{1998}]{1998PhRvA..57.1652Y}
{Yan} Z.-C.,  {Tambasco} M.,   {Drake} G.~W.~F.,  1998, \mn@doi [\pra]
  {10.1103/PhysRevA.57.1652}, \href
  {http://adsabs.harvard.edu/abs/1998PhRvA..57.1652Y} {57, 1652}

\bibitem[\protect\citeauthoryear{{Yong}, {Lambert}, {Allende Prieto}  \&
  {Paulson}}{{Yong} et~al.}{2004}]{Yong2004}
{Yong} D.,  {Lambert} D.~L.,  {Allende Prieto} C.,   {Paulson} D.~B.,  2004,
  \mn@doi [\apj] {10.1086/381701}, \href
  {http://adsabs.harvard.edu/abs/2004ApJ...603..697Y} {603, 697}

\bibitem[\protect\citeauthoryear{{Zacharias}, {Finch}, {Girard}, {Henden},
  {Bartlett}, {Monet}  \& {Zacharias}}{{Zacharias}
  et~al.}{2013}]{Zacharias2013}
{Zacharias} N.,  {Finch} C.~T.,  {Girard} T.~M.,  {Henden} A.,  {Bartlett}
  J.~L.,  {Monet} D.~G.,   {Zacharias} M.~I.,  2013, \mn@doi [\aj]
  {10.1088/0004-6256/145/2/44}, \href
  {http://adsabs.harvard.edu/abs/2013AJ....145...44Z} {145, 44}

\bibitem[\protect\citeauthoryear{{Zacharias}, {Finch}  \&
  {Frouard}}{{Zacharias} et~al.}{2017}]{Zacharias2017}
{Zacharias} N.,  {Finch} C.,   {Frouard} J.,  2017, \mn@doi [\aj]
  {10.3847/1538-3881/aa6196}, \href
  {http://adsabs.harvard.edu/abs/2017AJ....153..166Z} {153, 166}

\bibitem[\protect\citeauthoryear{Zhang \& Zhao}{Zhang \&
  Zhao}{2005}]{Zhang2005}
Zhang H.~W.,  Zhao G.,  2005, \mn@doi [Monthly Notices of the Royal
  Astronomical Society] {10.1111/j.1365-2966.2005.09599.x}, 364, 712

\bibitem[\protect\citeauthoryear{{de Jong} et~al.,}{{de Jong}
  et~al.}{2014}]{DeJong2014}
{de Jong} R.~S.,  et~al., 2014, in Ground-based and Airborne Instrumentation
  for Astronomy V. p. 91470M, \mn@doi{10.1117/12.2055826}

\bibitem[\protect\citeauthoryear{{van Leeuwen}}{{van
  Leeuwen}}{2007}]{vanLeeuwen2007}
{van Leeuwen} F.,  2007, \mn@doi [\aap] {10.1051/0004-6361:20078357}, \href
  {http://adsabs.harvard.edu/abs/2007A%26A...474..653V} {474, 653}

\bibitem[\protect\citeauthoryear{van~der Maaten \& Hinton}{van~der Maaten \&
  Hinton}{2008}]{vanderMaaten2008}
van~der Maaten L.,  Hinton G.,  2008, Journal of Machine Learning Research, 9,
  2579

\makeatother
\end{thebibliography}


\newpage
\noindent \rule{8.5cm}{1pt}

\noindent
$^{1}$Max Planck Institute  for Astronomy (MPIA), Koenigstuhl 17, 69117 Heidelberg, Germany\\
$^{2}$Fellow of the International Max Planck Research School for Astronomy \& Cosmic Physics at the University of Heidelberg, Germany\\
$^{3}$Research School of Astronomy \& Astrophysics, Australian National University, ACT 2611, Australia\\
$^{4}$Center of Excellence for Astrophysics in Three Dimensions (ASTRO-3D), Australia\\
$^{5}$Sydney Institute for Astronomy, School of Physics, A28, The University of Sydney, NSW 2006, Australia\\
$^{6}$Department of Physics and Astronomy, Uppsala University, Box 516, SE-751 20 Uppsala, Sweden\\
$^{7}$Department of Astronomy, Columbia University, Pupin Physics Laboratories, New York, NY 10027, USA\\
$^{8}$Center for Computational Astrophysics, Flatiron Institute, 162 Fifth Avenue, New York, NY 10010, USA\\
$^{9}$Miller Professor, Miller Institute, University of California Berkeley, CA 94720, USA\\
$^{10}$Monash Centre for Astrophysics, Monash University, Australia \\
$^{11}$School of Physics and Astronomy, Monash University, Australia\\
$^{12}$Australian Astronomical Observatory, 105 Delhi Rd, North Ryde, NSW 2113, Australia\\
$^{13}$Istituto Nazionale di Astrofisica, Osservatorio Astronomico di Padova, vicolo dell'Osservatorio 5, 35122, Padova, Italy \\
$^{14}$School of Physics, UNSW, Sydney, NSW 2052, Australia\\
$^{15}$Department of Physics and Astronomy, Macquarie University, Sydney, NSW 2109, Australia \\
$^{16}$Faculty of Mathematics and Physics, University of Ljubljana, Jadranska 19, 1000 Ljubljana, Slovenia\\
$^{17}$Department of Astronomy, University of Virginia, Charlottesville, VA 22904-4325, USA \\
$^{18}$INAF, Astrophysical Observatory of Turin, Torino, Italy \\
$^{19}$School of Physical and Chemical Sciences, University of Canterbury, New Zealand \\
$^{20}$University of Southern Queensland, Toowoomba, Queensland 4350, Australia\\
$^{21}$ICRAR, The Uni. of Western Australia, 35 Stirling Highway, Crawley, WA 6009, Australia \\
$^{22}$INAF Astronomical Observatory of Padova, 36012 Asiago, Italy \\
$^{23}$Department of Physics and Astronomy, The Johns Hopkins University, Baltimore, MD 21218, USA \\
$^{24}$Stellar Astrophysics Centre, Department of Physics and Astronomy, Aarhus University, DK-8000, Aarhus C, Denmark\\
$^{25}$Institute for Advanced Study, Princeton, NJ 08540, USA \\
$^{26}$Department of Astrophysical Sciences, Princeton University, Princeton, NJ 08544, USA \\
$^{27}$Observatories of the Carnegie Institution of Washington, 813 Santa Barbara Street, Pasadena, CA 91101, USA \\
$^{28}$Western Sydney University, Locked Bag 1797, Penrith, NSW 2751, Australia \\
$^{29}$University of Southern Queensland, Computational Engineering and Science Research Centre, Toowoomba, Queensland 4350, Australia \\
$^{30}$Department of Astronomy, The Ohio State University, 140 West 18th Avenue, Columbus, OH 43210, USA \\


\appendix

\section{Linelist}\label{sec:linelist}

\begin{table*}
\caption{Selected lines for the elemental abundance analysis.}\label{linelist}
  \label{tab:linelist}
\centering
\begin{tabular}{l l l l l p{2.8cm} l l}
\hline
Elem. & Ion & Wavelength [\AA] & LEP [eV] & $\log(g f)$ & Reference & Line mask [\AA] & Segment mask [\AA] \\
\hline
Li & 1 & 6707.7635 & 0.0000 & -0.00200 & YTD09 & 6707.650-6707.981 & 6705.761-6709.761 \\
Li & 1 & 6707.9145 & 0.0000 & -0.30300 & YTD09 & 6707.650-6707.981 & 6705.761-6709.761 \\
C  & 1 & 6587.610 & 8.5370 & -1.0210 & HBG93 & 6587.461-6587.786 & 6585.610-6589.610 \\
O  & 1 & 7771.944 & 9.1460 & 0.36900 & NIST1 & 7771.559-7772.309 & 7769.500-7777.500 \\
O  & 1 & 7774.166 & 9.1460 & 0.22300 & NIST1 & 7773.722-7774.582 & 7769.500-7777.500 \\
O  & 1 & 7775.388 & 9.1460 & 0.00200 & NIST1 & 7775.112-7775.762 & 7769.500-7777.500 \\
Na & 1 & 4751.8218 & 2.1040 & -2.0780 & GESMCHF & 4751.689-4751.944 & 4750.822-4752.822 \\
Na & 1 & 5682.6333 & 2.1020 & -0.70600 & GESMCHF & 5682.517-5682.997 & 5680.633-5684.633 \\
Na & 1 & 5688.205 & 2.1040 & -0.40400 & GESMCHF & 5687.917-5688.392 & 5686.200-5690.200 \\
Mg & 1 & 4730.0286 & 4.3460 & -2.3470 & NIST1 & 4729.908-4730.232 & 4728.500-4732.029 \\
Mg & 1 & 5711.088 & 4.3460 & -1.7240 & CT9 & 5710.857-5711.328 & 5709.090-5713.090 \\
Mg & 1 & 7691.550 & 5.7530 & -0.78300 & NIST1 & 7691.204-7691.779 & 7689.550-7695.550 \\
Al & 1 & 6696.023 & 3.1430 & -1.5690 & MEL95 & 6695.778-6696.173 & 6695.000-6697.000 \\
Al & 1 & 6698.673 & 3.1430 & -1.8700 & MEL95 & 6698.392-6698.895 & 6697.673-6699.673 \\
Al & 1 & 7835.309 & 4.0220 & -0.6890 & Kelleher & 7834.984-7835.472 & 7834.000-7837.500 \\
Al & 1 & 7836.134 & 4.0220 & -0.4940 & K75 & 7835.813-7836.431 & 7834.000-7837.500 \\
Si & 1 & 5665.5545 & 4.9200 & -1.9400 & GARZ|BL & 5665.200-5665.800 & 5663.550-5667.550 \\
Si & 1 & 5690.425 & 4.9300 & -1.7730 & GARZ|BL & 5690.180-5690.683 & 5688.430-5692.430 \\
Si & 1 & 5793.0726 & 4.9300 & -1.9630 & GARZ|BL & 5792.719-5793.293 & 5791.073-5795.073 \\
Si & 1 & 6721.8481 & 5.8630 & -1.0620 & N93 & 6721.476-6722.683 & 6719.848-6723.848 \\
K & 1 & 7698.9643 &  0.0000 & -0.1760 & Wang & 7698.573-7699.296 & 7696.960-7700.960 \\
Ca & 1 & 5857.451 & 2.9330 & 0.24000 & S & 5857.018-5857.625 & 5855.451-5859.451 \\
Ca & 1 & 5867.562 & 2.9330 & -1.5700 & S & 5867.307-5867.743 & 5865.500-5869.800 \\
\dots & \dots & \dots & \dots & \dots & \dots & \dots & \dots \\
\hline \\
\end{tabular}
\newline
\footnotetext{p{2.8cm}}{References:  AP: astrophysical, BBEHL: \cite{BBEHL}, BG: \cite{1980A&A....84..361B}, BGHL: \cite{BGHL}, BK: \cite{BK}, BKK: \cite{BKK}, BL: \cite{BL}, BMP83: \cite{1983MNRAS.204..883B}, BWL: \cite{BWL}, CT09: \cite{1990JQSRT..43..207C}, CSS89: \cite{CSS89}, CSSTW: \cite{1982ApJ...260..395C}, DLSSC: \cite{DLSSC}, DSVD92: \cite{DSVD92}, FMW: \cite{FMW}, GARZ: \cite{GARZ}, GBP89: \cite{1989A&A...208..157G}, GESB82c: \cite{GESB82c}, GESB82d: \cite{GESB82d}, GESB86: \cite{GESB86}, GESHRL14b: \cite{2014MNRAS.441.3127R}, GESHRL14d\cite{2014ApJS..215...23D}, GESMCHF: \cite{GESMCHF}, Grevesse: \cite{Grevesse2015}, HBG93: \cite{1993A&AS...99..179H}, HLSC: \cite{HLSC}, K06: \cite{K06}, K07: \cite{K07}, K08: \cite{K08}, K09: \cite{K09}, K13: \cite{K13}, K14: \cite{K14}, K75: \cite{K75}, Kelleher: \cite{Kelleher2008}, KP: \cite{KP}, KR: \cite{KR}, KSZ: \cite{1980ZPhyA.298..249K}, LBS: \cite{LBS}, LD: \cite{LD}, LD-HS: \cite{LD-HS}, LGWSC: \cite{LGWSC}, LWHFSC: \cite{2014ApJS..215...20L}, LWHS: \cite{LWHS}, LWST: \cite{LWST}, LGW13: \cite{2013ApJS..205...11L}, LSCI: \cite{LSCI}, MB09: \cite{2009A&A...497..611M}, MC: \cite{MC}, MEL95: \cite{1995JPhB...28.3485M}, MRW: \cite{MRW}, N93: \cite{1993PhyS...48..297N}, NIST10: \cite{NIST10}, PQWB: \cite{PQWB}, RU: \cite{RU}, S: \cite{S}, SLS: \cite{SLS}, SR: \cite{SR}, T83av: \cite{T83av}, VGH: \cite{VGH}, Wang: \cite{Wang1997}, WBb: \cite{WBb}, WHLBG: \cite{1985A&A...153..109W}, WLSC: \cite{WLSC}, WLSCb: \cite{2014ApJS..211...20W}, WSL: \cite{WSL}, YTD09: \cite{1998PhRvA..57.1652Y}
}
\end{table*}

\newpage $\,$
\newpage

\section{Abundance overview of extrapolated abundances}\label{sec:extrapolated_abundances}

\begin{figure*}
  \includegraphics[width=0.975\textwidth]{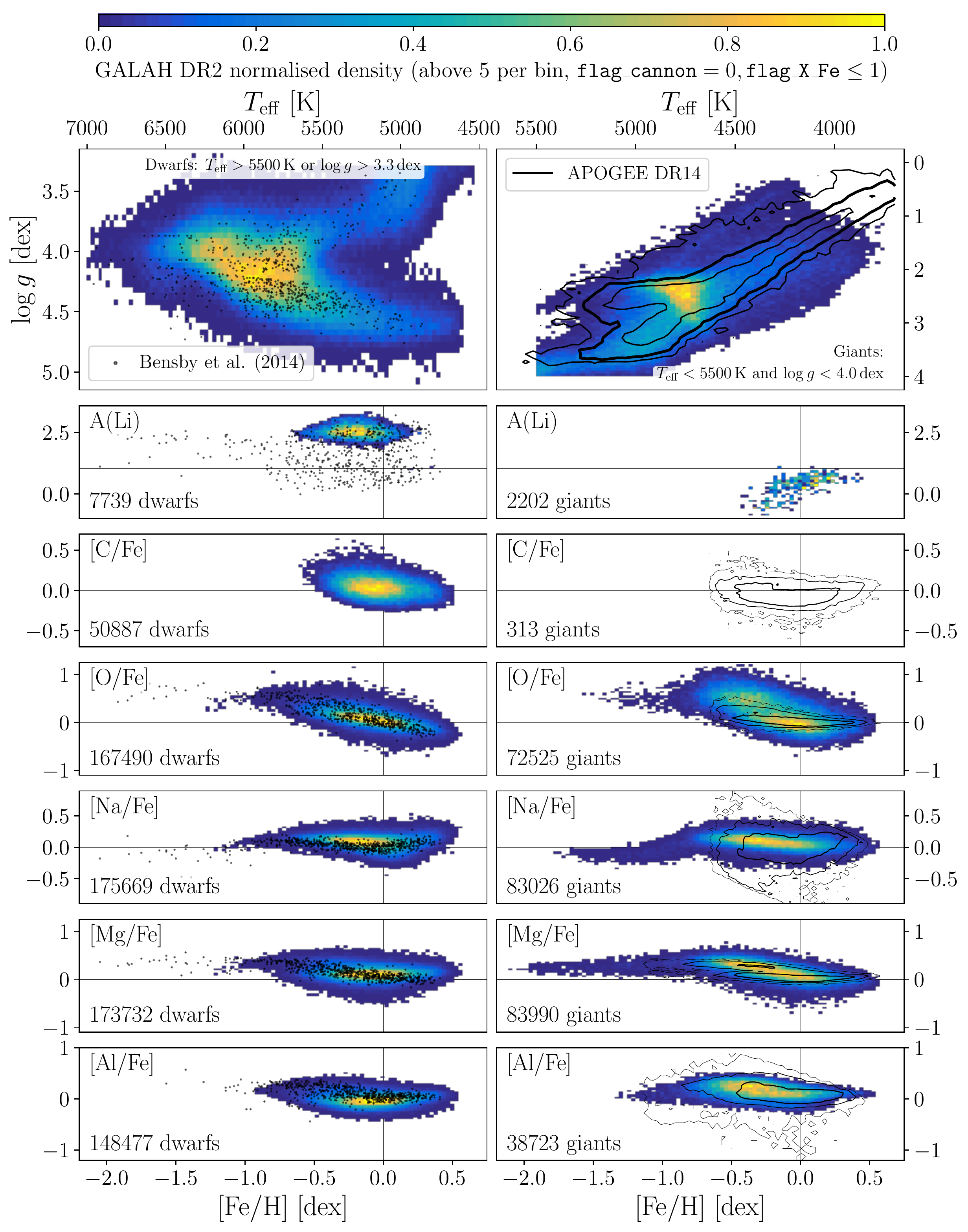}
  \caption{Comparison of the Kiel diagrams (top panel) and individual abundances (Li through Al) with the sample of 714 dwarfs \citep{Bensby2014, Battistini2015, Battistini2016, Bensby2018} on the left hand side as well as APOGEE DR14 giants \citep{SDSSDR14} on the right hand side. GALAH DR2 data (with $\texttt{flag\_cannon} = 0$ for stellar parameters and $\texttt{flag\_X\_fe} \leq 1$ for the respective element X) are plotted as colored density with a minimum of 5 stars per bin. The literature values for dwarfs are overplotted as black dots, while the APOGEE giants (with finite values, $\texttt{ASPCAPFLAG} = 0$, and $\texttt{STARFLAG} = 0$ for stellar parameters as well as $\texttt{X\_FE\_FLAG} = 0$ for the respective element X).}
  \label{fig:Dwarf_Giant_comparison1}
\end{figure*}

\begin{figure*}
  \includegraphics[width=0.975\textwidth]{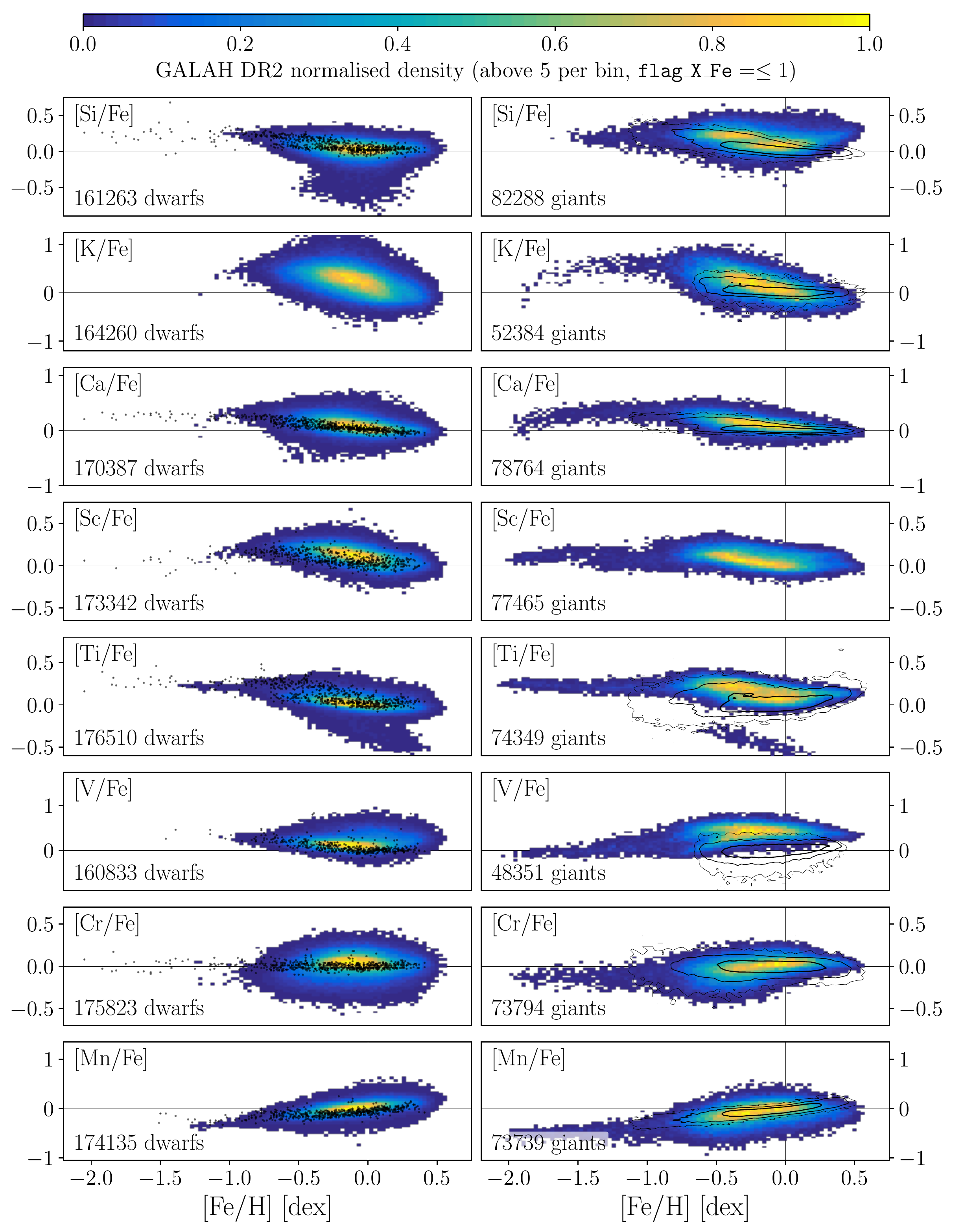}
\caption{Continuation of Figure~\ref{fig:Dwarf_Giant_comparison1} for elements Si through Mn.}
  \label{fig:Dwarf_Giant_comparison2}
\end{figure*}

\begin{figure*}
  \includegraphics[width=0.975\textwidth]{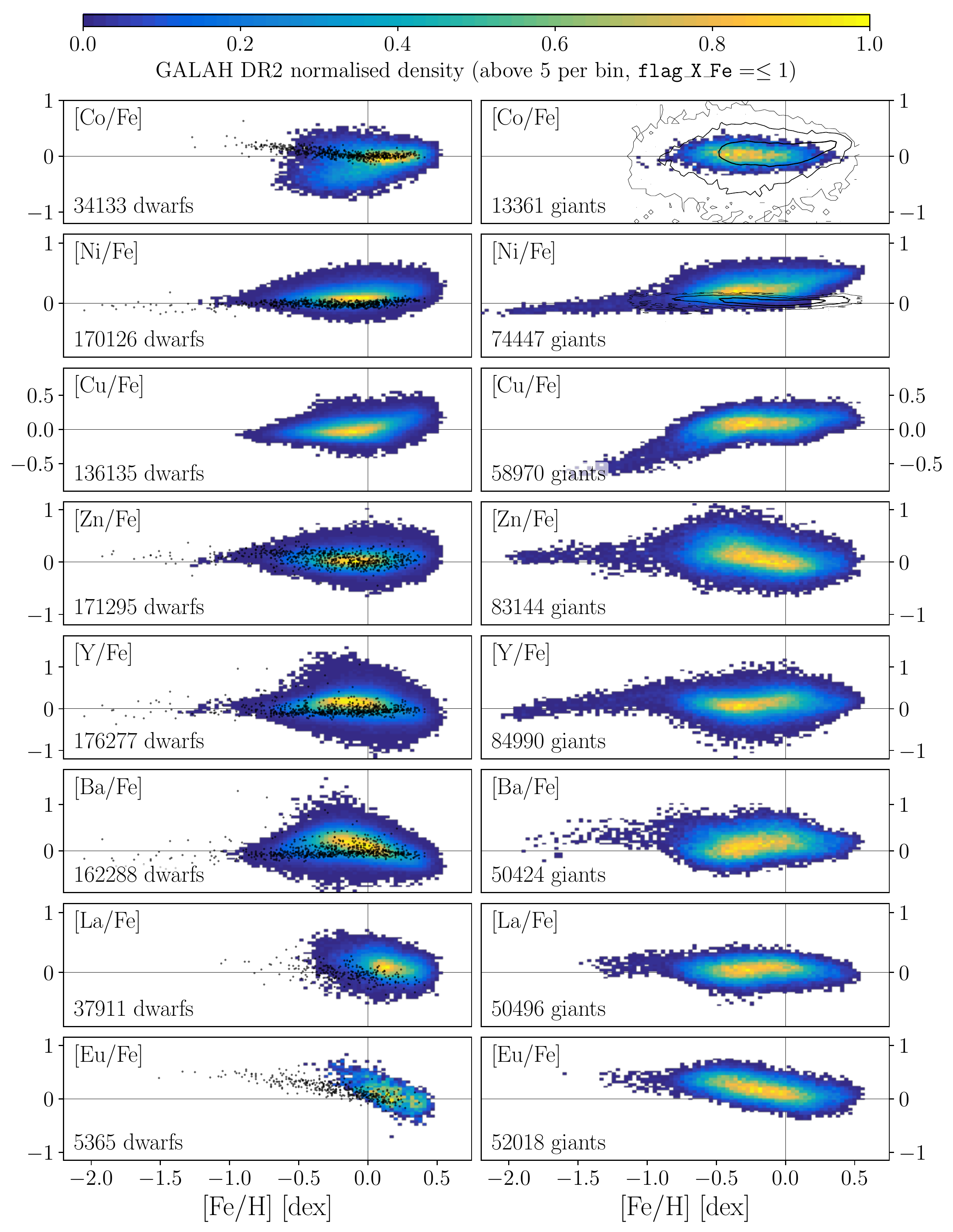}
\caption{Continuation of Figures~\ref{fig:Dwarf_Giant_comparison1} and \ref{fig:Dwarf_Giant_comparison2} for elements Co through Eu.}
    \label{fig:Dwarf_Giant_comparison3}
\end{figure*}


\bsp	
\label{lastpage}
\end{document}